\documentclass[journal,onecolumn]{IEEEtran}

\usepackage{amsmath}
\usepackage{amssymb} 
\usepackage{color} 
\usepackage{colortbl}
\usepackage{cite}
\usepackage{hyperref}
\usepackage{cleveref} 
\usepackage[pdftex]{graphicx}
\usepackage{multirow}
\usepackage{array}
\usepackage{graphicx}
\usepackage{arydshln} 
\usepackage[]{algorithm2e}
\usepackage{pdflscape}
\usepackage{scrhack}
\usepackage{graphicx}
\usepackage{wrapfig}
\usepackage[utf8]{inputenc}
\usepackage{multicol, blindtext}
\usepackage{etoolbox}
\usepackage{neuralnetwork}
\usepackage{tikz}
\usepackage{booktabs}
\usepackage{setspace}
\usepackage{amsthm}
 
\newtheorem*{remark}{Remark}

\usepackage[symbol]{footmisc}

\newcommand\undermat[2]{%
\makebox[0pt][l]{$\smash{\underbrace{\phantom{%
\begin{matrix}#2\end{matrix}}}_{\text{$#1$}}}$}#2}

\ifCLASSINFOpdf
\else
\fi

\begin{document}



%
\title{Optimizing electrode positions {in} 2D Electrical Impedance Tomography using deep learning}

%
\author{\IEEEauthorblockN{Danny~Smyl\IEEEauthorrefmark{1} and Dong~Liu\IEEEauthorrefmark{2}\IEEEauthorrefmark{3}\IEEEauthorrefmark{4}}\\
\IEEEauthorblockA{\IEEEauthorrefmark{1}Department of Civil and Structural Engineering, University of Sheffield, Sheffield, UK\\
\IEEEauthorrefmark{2}Hefei National Laboratory for Physical Sciences
at the Microscale and Department of Modern Physics, University of Science
and Technology of China, Hefei 230026, China\\
\IEEEauthorrefmark{3}CAS Key Laboratory of Microscale Magnetic Resonance, University of Science and Technology of China, Hefei 230026, China\\
\IEEEauthorrefmark{4}Synergetic Innovation Center of Quantum Information and Quantum Physics, University of Science and Technology of China, Hefei 230026, China}
\thanks{Corresponding authors: D. Smyl (d.smyl@sheffield.ac.uk) and Dong Liu (dong2016@ustc.edu.cn).\protect\\
}
\thanks{Digital Object Identifier 10.1109/IEEE.2019.0000000}}
%
%
%


\maketitle
\begin{abstract}
\normalsize
Electrical Impedance Tomography (EIT) is a powerful tool for non-destructive evaluation, state estimation, and process tomography --  {among numerous other use cases}.
For these applications, and in order to reliably reconstruct images of a given process using EIT, we must obtain high-quality voltage measurements from the  {target} of interest.
 {As such}, it is  {obvious} that the locations of electrodes used for measuring plays a key role in this task.
Yet, to date, methods for optimally placing electrodes either require knowledge on the EIT target (which is, in practice, never fully known) or are computationally difficult to implement numerically.
In this paper, we circumvent these challenges and present a straightforward deep learning based approach for optimizing electrodes positions.
It is found that the optimized electrode positions outperformed ``standard'' uniformly-distributed electrode layouts in all test cases.
 {Further, it is found that the use of optimized electrode positions computed using the approach derived herein can reduce errors in EIT reconstructions as well as improve the distinguishability of EIT measurements.}
\end{abstract}


\doublespacing

\section{Introduction}
In Electrical Impedance Tomography (EIT), the aim is to reconstruct a conductivity distribution inside an object using boundary measurements.
Information gained from the reconstructed conductivity is useful in a number of engineering applications, such as damage detection \cite{hassan2019failure,tallman2016} and stress/strain localization \cite{tallman2017inverse,loh2009} in composites, crack detection using conductive sensors \cite{Smyl2019SHM,smyl2018detection}, process tomography \cite{seppanen2009state}, and more.
For biomedical applications, such as thoracic and lung imaging, the use of EIT is also widespread nowadays \cite{liuMMC2019,mellenthin2018ace1,liu2019nonstationary}.
Owing to increases in computational resources and knowledge transfer on EIT imaging methods (e.g. the EIDORS project \cite{adler2006uses}), the use of EIT in industrial applications has increased over the past 20 years  {(for additional information on industrial applications cf. \cite{yao2017application,wei2016super,holder2004electrical})}.

We are certainly inspired by the recent advances and increased usage of EIT -- especially from modern hybrid applications merging EIT and deep learning \cite{tan2019image,wei2019dominant,hamilton2018deep}.
However, there remain significant challenges in EIT imaging making it difficult to reliably apply in many engineering applications.
The first is challenge is fundamental to EIT -- its generally low spatial resolution resulting from the diffusive nature of electric fields.
The second challenge is the sensitivity of the conductivity on the boundary measurements, which results from the severely ill-posed and nonlinear EIT inverse problem \cite{mueller2012}.
Assuming a fixed target, i.e. we can not reasonably change the geometry or constitution of the target, our best \textit{physical} options for mitigating the second challenge is to optimize the current injection and/or electrode configuration.
While the former is well addressed in the literature \cite{kaipio2004posterior,lionheart2001generalized}, regimes for optimizing electrode positions are scarce.
Perhaps the most relevant works addressing large degree of freedom systems common in engineering applications include the use of elegant optimization regimes which (i) require prior knowledge on the target conductivity distribution or (ii) can be computationally demanding or difficult for new users to implement \cite{hyvonen2014optimizing,horesh2010optimal}.

In this work, we circumvent the need to implement optimization algorithms with limitations (i) and (ii).
To do this, we present a straightforward and fast deep learning-based approach for electrode position optimization which aims to minimize the ill-conditioning of the EIT Hessian matrix and the fit between the least-squares solution and true (training) distributions.
We begin this paper by providing motivation for the optimization problem and outline the electrode position optimization algorithm.
Following, we provide representative demonstrations of optimized electrode configurations using different geometries and derive metric {s} to evaluate  {the effectiveness of the optimized electrode positions}.
We then use these metrics to illustrate potential improvements in using optimized electrode positioning relative to ``standard'' electrode positions  {and explore whether EIT reconstructions can be improved by employing the optimized electrode positions}.
Lastly, a discussion and concluding remarks are provided.

\section{Optimizing electrode positions using deep learning}
The use of deep learning/neural networks in EIT applications is not unprecedented.
In fact, deep learning has shown significant promise for both direct reconstruction using EIT data and emulating the numerical EIT forward model \cite{FAN2019109119}.
Specific applications of deep learning in EIT have included use cases in, for example, D-Bar reconstructions \cite{hamilton2019beltrami,hamilton2018deep}, convolutional neural network-based reconstruction approaches \cite{wei2019dominant,tan2018image}, {three-dimensional EIT \cite{sLiu2019}}, and tracking moisture migration in concrete specimens \cite{rymarczyk2018non}.
{In addition to these, the use of deep learning for solving inverse problems in general is an emerging field; example applications include linear inverse problems \cite{borgerding2016onsager}, photoacoustic tomography \cite{antholzer2019deep}, and magnetic resonance imaging \cite{han2019k}, and the broad use of convolutional neural networks for imaging \cite{mccann2017convolutional}.}
In this work, we do not focus on using deep learning for any of the former applications (direct reconstruction or forward model emulation); rather, we utilize deep learning as a tool for use in optimizing the position of electrodes.

We begin  {the technical portion of} this section by first describing the purpose of optimizing electrode positions in EIT.
Following, we outline the proposed deep learning based optimization approach and the motivation for selecting training parameters.
Lastly, we detail the electrode position optimization algorithm and practical aspects on implementation.

\subsection{Background, nonlinearity, ill-conditioning, and ill-posedness}
\label{BNII}
The purposes of optimizing electrode positions are to (a) maximize the quality of information contained in EIT measurements, (b) improve the conditioning of the EIT inverse problem, and (c) ultimately improve the reliability of EIT measurements and quality of reconstructions.
As a whole, accomplishing (a) - (c) will improve the robustness and quantitative nature of EIT information used for engineering and physical science applications -- which is the primary motivation for this article.
In this subsection, we will discuss and derive key variables including information from points (a) - (c) to drive training for the electrode position optimization algorithm described in the next subsection.

The initial challenge arising in electrode position optimization stems from the realization that the EIT forward problem, computing boundary voltage measurements $V$ from conductivity $\sigma$,  {has a nonlinear dependence of $V$ as a function of $\sigma$} \cite{mueller2012}.
In this work, the nonlinearity is numerically manifested via the the finite element version of the complete electrode model \cite{vauhkonen99}.
If we write down the so-called `noiseless observation model'

\begin{equation}
V = U(\sigma)
\label{obeqr}
\end{equation}

\noindent we immediately observe that solving the EIT inverse problem, estimating $\sigma$ from $V$, is highly dependent on the properties of this numerical forward model, $U$.
Importantly, the  {diffusive physical nature of the EIT problem} results in a severely ill-conditioned sensitivity matrix $J = \frac{\partial U(\sigma)}{\partial \sigma}$ and Hessian approximation $H = J^\mathrm{T} J$ \cite{borcea2002electrical}.
In particular, the ill-conditioning of $H$ has a large impact on the ill-posedness of least-squares based solutions to the EIT reconstruction problem, which is usually addressed using regularization techniques.
However, we can also reduce the ill-conditioning of $H$ to mitigate ill-posedness by reducing the Hessian's condition number $\kappa$, where $\kappa(H) = || H^{-1} || || H ||$ is a scalar value.
Practically speaking, reducing $\kappa(H)$ lowers the sensitivity of outputs computed using $H$ on the small changes to input values -- in this case the sensitivity of $\sigma$ to small changes in $V$.
Based on these reasons, the selection of $\kappa(H)$ as the first input parameter used in training the optimization algorithm is a clear choice and addresses points (b) and (c).

The choice of a second input parameter, required to address point (a), is less immediately obvious.
However, we can begin by first writing down the minimizer to the EIT problem about some reasonable initial guess $\sigma_0$ (in the least-squares sense)

\begin{equation}
\Delta \sigma = (H(\sigma_0) + \Gamma^{-1})^{-1}J(\sigma_0) ^\mathrm{T}(V_\mathrm{t}- U(\sigma_0))
\label{delsig}
\end{equation}

\noindent where $\Gamma$ is a given covariance matrix (specified in section \ref{tlro}) and $V_\mathrm{t}$ is a ``true" measurement computed from the prescribed training data $\sigma_\mathrm{t}$.
 {Note that the lack of a gradient term in Eq. \ref{delsig} results from the fact that the prior mean and $\sigma_0$ are the same in this one-step minimizer; therefore, the gradient term vanishes.}
We quickly observe that information on the quality of EIT data generated from a specific electrode configuration is included in Eq. \ref{delsig} via the forward model $ U(\sigma_0)$, sensitivity matrix $J(\sigma_0)$, and the Hessian $H(\sigma_0) $ computed at $\sigma_0$.
In other words, Eq. \ref{delsig} includes information on the data forward model and its 1$^\mathrm{st}$ and 2$^\mathrm{nd}$ derivatives.
However, for this information to be usefully implemented in a training regime, we need to first write the full one-step Gauss-Newton estimate as

\begin{equation}
\widehat{\sigma} = \sigma_0 + \Delta \sigma.
\label{GN}
\end{equation}

\noindent In this form, the information in Eq. \ref{GN} can then be directly included as a training input parameter by computing the misfit between the Gauss-Newton estimate and the true training data by taking the norm of their difference

\begin{equation}
\beta= || \sigma_t - \widehat{\sigma} ||^2.
\label{mf}
\end{equation}

\noindent It is worth noting that, while Eq. \ref{mf} does address the data quality requirement, it also directly includes information on the reconstruction quality by measuring the mismatch between reconstructed and true conductivities.

To summarize this subsection, we have derived two input parameters to drive training of the neural network detailed in the next section.
These parameters, $\kappa$ and $\beta$, include information on EIT data quality, ill-conditioning, and reconstruction quality -- key variables needed to improve EIT imaging.
Moreover, $\kappa$ and $\beta$ are both scalar quantities, containing a significant amount of information in a small amount of memory, which is a desirable feature when using them as training data for deep learning.
Finally, when we concatenate these parameters into an objective vector $\Theta = [\kappa, \beta]^\mathrm{T}$, we surmise that an optimal electrode configuration has (theoretically) been attained when the following criterion is reached

\begin{equation}
\Theta =
\begin{bmatrix}
1 \\
0
\end{bmatrix}.
\label{ob}
\end{equation}

\noindent In practice, the condition number and data mismatch will always be greater than $\Theta = [1, 0]^\mathrm{T}$; but, Eq. \ref{ob} can be used as a simple objective for the trained network described in the following subsection.

\subsection{A  {supervised} deep learning-based algorithm for optimizing electrode positions}
In the proposed approach, we aim to optimize 2D electrode configurations using trained deep neural networks.
Generally speaking, deep learning aims to develop a mapping $Q: X \to Y$ between the elements of parameters $X$ and $Y$ \cite{Arridge2019}.
In our approach, we aim to train a mapping to predict a vector of  {Cartesian ($x,y$)} electrode locations  {$E = [x_1, x_2, \dots x_k, y_1, y_2, \dots y_k]^\mathrm{T}$} as a function of the objective vector $\Theta$

\begin{equation}
E = Q_{\bar{\Theta}, \bar{E}}(\Theta)
\label{dl}
\end{equation}

\noindent using training data $\bar{\Theta}(\kappa,\beta) \in \mathbb{R}^{2 \times N_{\bar{E}} N_\sigma}$ and randomized electrode positions $\bar{E} \in \mathbb{R}^{ {2}k \times N_{\bar{E}} N_\sigma}$ consisting of $N_{\bar{E}}$ electrode configuration samples, $N_\sigma$ conductivity samples, and $k$ electrode midpoint coordinates.
Here, we converted the  {electrodes'} $x,y$ midpoint coordinates to a single-column vector by simply stacking the $x$ and $y$ components (i.e. $[x,y] \to [x;y]$). 
 {This representation results in a model output vector $E$ consisting of $2k$ elements (double the number of prescribed electrodes, $k$).}
Given the generality of the of the approach outlined in Eq. \ref{dl}, it was found that several practical constraints need to be prescribed to ensure a reliable $Q_{\bar{\Theta}, \bar{E}}$.
Herein, we assume $Q_{\bar{\Theta}, \bar{E}}$ is only valid:
\begin{itemize}
\item For a fixed 2D geometry $\Omega$.
\item For a fixed $k$.
\item When number of electrodes on a given side of $\Omega$ is fixed\footnote{ {A function was used to ensure electrodes do not overlap. In this function, random electrodes center point locations are trialed within the width of the side, unique electrode center point locations at least one element width apart are determined, and this process is iterated until all electrode center points were at least one element width apart.}}.
\item When forward model function and method for computing $J$ are fixed.
\item When electrode contact impedances $z$ are fixed.
\item When the current injection $I$ and measurement patterns $M$ are fixed.
\item When the maximum element size $\delta_e$ is fixed.
\end{itemize}

\noindent These constraints were adopted throughout this work and were determined after a preliminary error study using a suite of different neural networks and training data sizes/constitutions.
From the preliminary study, we selected a feed forward neural network with two hidden layers.
To compute the number of neurons per hidden layer, we used the fixed criteria described in \cite{huang2003learning} which proved robust in this application.
To train $Q_{\bar{\Theta}, \bar{E}}$, we selected a regularized conjugate gradient backpropagation method with Fletcher-Reeves updates \cite{nawi2007}.
 {The training was stopped when either the loss function (provided in the next subsection)} or the gradient were below $10^{-7}$.
This training architecture was chosen based on its speed (all examples trained in less than 5 minutes on a 8-core processor with 32 Gb RAM running computations in parallel in MATLAB) and ability to consistently produce physically realistic solutions (i.e. no overlapping electrodes and electrodes mapped outside of $\Omega$).

\begin{figure}[h]
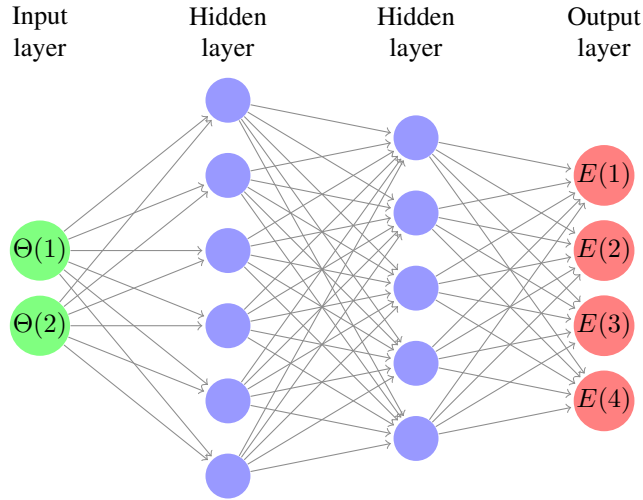

\centering
\begin{neuralnetwork}[height=6.0]
\newcommand{\nodetextclear}[2]{}
\newcommand{\nodetextx}[2]{$\Theta(#2)$}
\newcommand{\nodetexty}[2]{$E(#2)$}
\inputlayer[count=2, bias=false, title=Input\\layer, text=\nodetextx]
\hiddenlayer[count=6, bias=false, title=Hidden\\layer, text=\nodetextclear] \linklayers
\hiddenlayer[count=5, bias=false, title=Hidden\\layer, text=\nodetextclear] \linklayers
\outputlayer[count=4, title=Output\\layer, text=\nodetexty] \linklayers
\end{neuralnetwork}
\caption{Schematic illustration of a trained electrode position optimization network where the  {length of the electrode location vector} is  {arbitrarily} set to  {4 -- corresponding to 2 optimized electrode positions}. The input layer takes the optimization vector $\Theta$, transfers this data to the two hidden layers, and outputs the optimized electrode locations  {$E = [E(1), E(2), E(3), E(4)]^\mathrm{T} = [x_1, x_2, y_1, y_2]^\mathrm{T}$ }.}
\label{net}
\end{figure}

For the purposes of visualization, a schematic illustration of a trained electrode position optimization network is shown in Fig. \ref{net} where the  {length of the electrode location vector} is  {arbitrarily} set to  {4 -- corresponding to 2 optimized electrode positions}.
 {We would like to reinforce, however, that the length of the electrode location vector depends on the number of user-prescribed electrodes (equal to $k$).}
In the schematic, the input layer (taking the objective vector $\Theta$), the two hidden layers, and the output layer with  {two} optimized electrode locations  {$E = [E(1), E(2), E(3), E(4)]^\mathrm{T} = [x_1, x_2, y_1, y_2]^\mathrm{T}$ } are shown.

\subsection{Training, learning, and algorithm overview}
\label{tlro}
The training data used in this work consists of two components: $\bar{E}$ and $\bar{\Theta}$.
$\bar{E}$ is a matrix consisting of $N_{\bar{E}}$ different random electrode  {midpoint coordinates ($\bar{x}=[\bar{x}_1, \bar{x}_2 \dots \bar{x}_{N_{\bar{E}}}] \in \mathbb{R}^{k\times N_{\bar{E}} }$ and $\bar{y}=[\bar{y}_1, \bar{y}_2 \dots \bar{y}_{N_{\bar{y}}}] \in \mathbb{R}^{k\times N_{\bar{E}} }$)} subject to the constraints noted in the previous subsection.
To generate different random electrode configurations, a vector of random electrode midpoint positions is input into the EIDORS \cite{adler2006uses} unstructured meshing routine.
Following, the mesh generated using random electrode positions, prescribed current/measurement patterns, and conductivity samples are input into the numerical EIT forward model and sensitivity matrix function in order to compute $\bar{\Theta}$ parameters $\kappa$ and $\beta$.
Moreover, for each random electrode configuration, $N_\sigma$ conductivity samples are utilized and,
 {for the purposes of transparency, the training data is shown succinctly using the following matrix descriptions:}

\begin{equation}
 {
\bar{E} = 
\rotatebox[origin=c]{90}{2$k$} \Bigg{\{}
\overbrace{\Bigg[
\begin{array}{rrrrrrrrrr}
\bar{x}_1 & \dots & \bar{x}_1 & \bar{x}_2 & \dots & \bar{x}_2 & \dots & \bar{x}_{N_{\bar{E}}} & \dots & \bar{x}_{N_{\bar{E}}}\\
\undermat{N_\sigma}{\bar{y}_1 & \dots & \bar{y}_1} & \undermat{N_\sigma}{\bar{y}_2 & \dots & \bar{y}_2} & \dots & \undermat{N_\sigma}{\bar{y}_{N_{\bar{E}}} & \dots & \bar{y}_{N_{\bar{E}}}}
\end{array}
\Bigg]}^{N_{\bar{E}} \times N_\sigma}
}
\end{equation}

\noindent and

\begin{equation}
 {
\bar{\Theta} = 
\overbrace{\Bigg[
\begin{array}{rrrrrrrrr}
\kappa_1 & \kappa_2 &\dots & \kappa_{N_\sigma} & \kappa_{N_\sigma+1} & \dots & \kappa_{2N_\sigma} & \dots & \kappa_{N_{\bar{E}} N_\sigma} \\
\undermat{N_\sigma}{\beta_1 & \beta_2 &\dots & \beta_{N_\sigma}} & \undermat{N_\sigma}{\beta_{N_\sigma+1} &\dots & \beta_{2N_\sigma} } & \dots & \beta_{N_{\bar{E}} N_\sigma} 
\end{array}
\Bigg]}^{N_{\bar{E}} \times N_\sigma}
}.
\end{equation}\\

Importantly, the conductivity samples are computed only once to train $Q_{\bar{\Theta}, \bar{E}}$ and are reused in computing $\kappa$ and $\beta$ values for a given electrode configuration.
The random conductivity samples tested herein are blob-line in structure and are drawn using the Cholesky factorization of the inverse covariance matrix (i.e. $L^T L= \Gamma^{-1}$) and the resulting generator $\bar{\sigma}_\mathrm{rand} = L^{-1}r$, where $r$ is a random non-negative vector.
For completeness, $\Gamma$ is a matrix determined element wise, where the entry $(i,j)$ at locations $x_i$ and $x_j$ is given by

\begin{equation}
\label{PrCov}
\Gamma (i,j) = a \exp \Big (-\frac{||x_i - x_j||}{2b} \Big) + c \delta_{ij}
\end{equation}

\noindent where the scalars $a$, $b$, and $c$ are positive and $\delta_{ij}$ is the Kronecker delta function.
It should be noted that, since the meshing routine is unstructured, the number of elements and node locations vary slightly.
As such, we simply interpolate the conductivity samples onto different meshes using linear interpolation.
 {We would like to mention that $\Gamma$ does add structure to the random samples of $\sigma$ via the so-called ``correlation length'' $b$.
It was found that when the sample size is small, $b$ does have an effect on the optimized electrode positions.
However, when the sample size is sufficient large (greater than approximately 500 samples in this work), this effect vanished.}

Following the accumulation and storage of the training data, the data  {is randomly split into thirds and each third was designated for either network training, validation, and testing.
Next, the data is} fed into the training algorithm in order to learn the mapping $Q_{\bar{\Theta}, \bar{E}}$ using the general approach detailed in the previous subsection. 
One practical subtlety here is that regularization is required to ensure the network properly fits the data.
 {In this work, we use $L^2$ regularization with a loss function defined by}

 {
\begin{equation}
\mathcal{L} = \frac{1}{N_s}\sum_{n=1}^{N_s} (E^d_n - E_n^Q)^2 + \alpha \pi(w)
\label{NNr}
\end{equation}
}

\noindent  {where $\pi=||w||^2$, $w$ are the network weights, $\alpha$ is the regularization hyperparameter, $N_s$ is the number of samples, $E^d_n$ are the desired electrode position outputs and $E_n^Q$ are the electrode positions output from the network at sample $n$}.
If $\alpha$ is too large, the network suffers from under fitting; on the other hand if $\alpha$ is too low, the system is overly fit.
For the problems addressed in this work $\alpha = 0.01$ was used.
We remark that this is certainly not a globally optimal choice and the statistical selection of $\alpha$ will be addressed in future work.
Another important assumption requiring elaboration is the selection of the ``reasonable initial guess $\sigma_0$.''
For this, we chose the mean conductivity of a given sample.
While a more accurate guess may be the best homogeneous estimate, the use of the mean value was found to be virtually indistinguishable from the best homogeneous estimate and saved significant computing time.
To summarize the approach, training, and electrode position optimization approach proposed in this section, pseudocode is provided in Algorithm \ref{PC}.

\begin{algorithm}
\small{
\SetAlgoLined
\KwResult{Obtain optimized electrode positions, $E$}
initialize $\Omega$, $\mathrm{el width}$, $k$, $\delta_e$, $N_{\bar{E}}$, $N_\sigma$, $I$, $M$, $z$, $\alpha$\;
~~~~\%\% Generate training data $\bar{E}$ and $\bar{\Theta}$ \%\%\; 
\For{i = 1:$ N_{\bar{E}}$ }
{
Generate random electrode positions,  {$E^e= [\bar{x}_i;\bar{y}_i]$}\;
Generate mesh from meshing routine: $\Omega^e = f(E^e,\Omega, \delta_e, \mathrm{el width})$\;
Compute $\Gamma = \Gamma(\Omega^e)$\;
\If{i = 1}{
Generate $N_\sigma$ random conductivity distributions using $\bar{\sigma}_\mathrm{rand} = L^{-1}r$
}
\For{j=1:$N_\sigma$}
{
\If{j $>$ 1}
{
Interpolate conductivity sample to current mesh: $\bar{\sigma}_{\mathrm{rand},j} \to \sigma_{\mathrm{rand},j}$\;
}
$V_t = U(\sigma_{\mathrm{rand},j},\Omega^eI,M,z)$ \% ``True'' measurements\;
$J = \frac{\partial U(\sigma_{\mathrm{rand},j})}{\partial \sigma}$ \% sensitivity matrix\; 
$H = J^\mathrm{T} J$ \% Hessian\;
$\kappa= || H^{-1} || || H ||$ \% Condition number\; 
~~~~\%\% 1-step Gauss-Newton estimate \%\%\; 
$\sigma_0 = \mathrm{mean}(\sigma_{\mathrm{rand},j})$ \% Reasonable initial guess\;
$J(\sigma_0) = f(\sigma_0)$ \% sensitivity matrix about $\sigma_0$\;
$H(\sigma_0)= J(\sigma_0)^\mathrm{T} J(\sigma_0)$ \% Hessian about $\sigma_0$\;
$\Delta \sigma = (H(\sigma_0) + \Gamma^{-1})^{-1}J(\sigma_0) ^\mathrm{T}(V_\mathrm{t}- U(\sigma_0))$ \% Gauss-Newton update\;
$\widehat{\sigma} = \sigma_0 + \Delta \sigma$\% Gauss-Newton estimate\;
$\beta= || \sigma_t - \widehat{\sigma} ||^2$ \% Conductivity misfit\;
~~~~\%\% collect parameters\%\%\; 
$\theta(:,j) = [\kappa, \beta]^\mathrm{T}$\;
$\varepsilon(:,j) = E^e$\;
}
$\bar{\Theta}= [\bar{\Theta}, \theta]$,~$\bar{E} = [\bar{E}, \varepsilon]$\;

}
~~~~\%\% Network training \%\%\; 
$L_1 = \mathrm{floor}(~((k+2) \times N_{\bar{E}})^{1/2} + 2 \times (N_{\bar{E}}/(k+2))^{1/2}~)$ \% Huang \# neurons for layer 1\;
$L_2 = \mathrm{floor}(~k\times(N_{\bar{E}}/(k+2))^{1/2}~)$ \% Huang \# neurons for layer 2\;
$Q_{\bar{\Theta}, \bar{E}}$ = train($\bar{E}$, $\bar{\Theta}$, $L_1$, $L_2$, $\alpha$) ]\% CG backpropagation method with Fletcher-Reeves updates\;
~~~~\%\% Compute optimized electrode positions, $E$ \%\%\; 
$\Theta = [1, 0]^\mathrm{T}$\% cf. Eq \ref{ob} text for rationale\; 
$E = Q_{\bar{\Theta}, \bar{E}}(\Theta)$ \% Optimized electrode positions\;
}
\caption{Pseudocode for generating training data $\bar{E}$ and $\bar{\Theta}$, training the mapping $Q_{\bar{\Theta}, \bar{E}}$, and optimizing electrode positions.}
\label{PC}
\end{algorithm}

\begin{remark}
{
At this stage, an important point needs to be clarified.
Broadly speaking, an optimized electrode configuration -- as delineated herein -- depends on both the domain geometry and the conductivity distribution.
While one generally does not know the exact conductivity distribution \textit{a priori} (if one did, it would defeat the purpose of doing EIT), one can include information related to the expected structure of the target conductivity distribution to train the neural network.
For example, these structures could include sharp inclusions, anisotropic distributions, or smooth distributions.
The incorporation of such structural training data can be viewed as a form of prior information used in solving the EIT reconstruction problems  -- i.e. the inclusion of prior structural information used to improve the quality of data measured from the electrodes.
}
\end{remark}

\section{Examples and a metric for comparing the quality of electrode positioning}
In this section we evaluate the proposed deep learning method for optimizing electrode positions considering three geometries: a square, a rectangle, and a triangle.
While the former two geometries are rather common in engineering applications, the latter is used to test an irregular configuration.
We begin by outlining simple metric {s} for determining the quality of electrode configurations, which will be used for comparison in the examples.
Following, we demonstrate optimized electrode positions using the proposed algorithm.

\subsection{Simple metric for determining the quality of electrode configurations  {based on reducing discretization error}}
\label{qualmetric1}
In order to numerically quantify (potential) improvements gained when using the optimized electrode position approach, we require quality metrics.
It is tempting  {to develop a metric based on the comparison of} reconstructions of a similar target using a ``standard'' electrode configuration and an optimized one;  {however, such a metric would be biased}.
This is because simulated data from an ``electrode configuration A'' may be significantly different in quality compared to ``configuration B.'' 
Therefore, if an optimized sensor configuration has higher quality simulated data than a ``standard'' configuration, the potential quality of reconstructions is favorably biased towards the optimized configuration.
In this sense, we require an ``apples to apples'' comparison.

To make such a comparison, let us first assume we are interested in determining the reliability of ``electrode configuration A'' (ECA).
From a numerical perspective, ECA would be maximally reliable if ECA minimized the effects of discretization on computed voltages.
Therefore, in a perfect world, the maximally reliable ECA would be able to simulate voltages from an infinitely large suite of conductivity samples -- and compute the same voltages for each individual sample -- using a coarse or a fine mesh.
However, since all numerical models have error, we must quantify the reliability of electrode positioning statistically.

In order to do this, we take elements from the Bayesian approximation error approach (BAE) \cite{nissinen2007bayesian}.
In BAE, the idea is to statistically model the error between an accurate model $U_{\mathcal{A}}$ and a reduced model $U_h$ (generally represented by fine and coarse meshes).
In the context of this work and determining the reliability of a given electrode configuration,  {one} can also apply this methodology by first collecting the mean modeling error across ${N_\sigma}$ conductivity samples

\begin{equation}
\mu = \frac{1}{{N_\sigma}}\sum_{j=1}^{N_\sigma} \big [ U_{\mathcal{A}}(\sigma_j) - U_h(\sigma_j) \big ].
\label{qual}
\end{equation}

\noindent Now, assuming we have collected ample conductivity samples and have two electrode configurations (A and B) with resulting mean modeling errors $\mu_A$ and $\mu_B$: we can confidently state that ``electrode configuration A'' is more reliable than ``electrode configuration B'' when the overall magnitude of $\mu_A$ is significantly lower than $\mu_B$.
This statement can be put more plainly as, the use of ``electrode configuration A'' results less statistical modeling error than ``electrode configuration B'' when $||\mu_A ||_1 <||\mu_B||_1$.

The advantage of using this metric for quantifying the quality of electrode configurations is that it is independent from the metrics used to optimize the electrode configurations.
As such, the comparison of mean modeling errors  {(resulting from discretization)} for different electrode configurations (optimized and ``standard'' layouts) using fine and coarse meshes will be used in the examples for the purpose of quantitative comparison.

\subsection{ {Simple metrics for determining the quality of electrode configurations based on the conditioning of the Hessian and resistivity matrix}}
 {As discussed in section \ref{BNII}, the ill-conditioning of the Hessian has a central role in the ill-posedness of the EIT inverse problem.
Physically, the ill-conditioning stems from the diffusive nature of the EIT problem and is also numerically manifested in the Finite Element resistivity matrix $R(\sigma)$, which is used in computing $H$\footnote{ {Note: $U(\sigma) = R(\sigma)I$, where $I$ is a current injection matrix. Therefore $J=\frac{\partial (R(\sigma)I)}{\partial \sigma}$ and $H =[\frac{\partial (R(\sigma)I)}{\partial \sigma}]^T[\frac{\partial (R(\sigma)I)}{\partial \sigma}]$.}}.
The ``snowball effect" of ill-conditioned Finite Element matrices contributing to ill-conditioned Hessians is known in the inverse problems community \cite{smyl2019less} and also holds in EIT applications.
As a logical extension relevant to this work, electrode positions resulting in relatively lower condition numbers $\kappa$ for $H$ and $R$ are considered herein to be more optimal than electrode positions resulting in higher respective condition numbers.
}

 {Considering a large sampling of conductivity distributions ($N_\sigma$), we can write down measures for quantifying the quality of electrode configurations based on the conditioning of $H$ and $R$ as follows:
}

 {
\begin{equation}
\bar{\kappa}_{H} = \frac{1}{{N_\sigma}}\sum_{j=1}^{N_\sigma} \kappa(H(\sigma_j))
\label{qualH}
\end{equation}
\begin{equation}
\bar{\kappa}_{R} = \frac{1}{{N_\sigma}}\sum_{j=1}^{N_\sigma} \kappa(R(\sigma_j))
\label{qualR}
\end{equation}
}

\noindent  {where $\bar{\kappa}_{H}$ and $\bar{\kappa}_{R}$ are mean values of the condition numbers for $H$ and $R$, respectively, over conductivity samples $\sigma_j$.
These quality metrics are a simple representation of the average sensitivity of outputs computed using $H$ and $R$ to changes in the inputs.
In the context of this work, we aim to reduce $\bar{\kappa}_{H}$ and $\bar{\kappa}_{R}$ relative to ``standard layouts" by optimizing the electrode positions in order to promote reduction of ill-posedness of the EIT inverse problem.
We remark that reducing the ill-posedness of the EIT inverse problem by optimizing electrode positions cannot be guaranteed in all applications since, e.g. poor regularizer choices may negate the effect of decreasing $\bar{\kappa}_{H}$ and $\bar{\kappa}_{R}$.
}

\subsection{Example parameters}
For these examples demonstrated herein, we assume the maximum element size used in training is equal to the fixed electrode width, the minimum training element size is half the fixed electrode width, contact impedances $z = 10^{-5}$, 1 mA current injections ($k-1$ injections against electrode 1\footnote{ {In a preliminary analysis, it was found that injections against one electrode provided the most consistent results with respect to different randomly sampled data. Therefore this protocol was selected in lieu of, e.g., opposite current injections which results in fewer measurements.}}), and $k(k-1)$ adjacent electrode measurements.
The units of length and conductivity in all examples are cm and mS/cm, respectively.
To train the mapping $Q_{\bar{\Theta}, \bar{E}}$, $N_{\bar{E}} = 2000$ electrode configuration samples and $N_\sigma = 2000$ conductivity samples are drawn per example geometry.

In order to quantify the quality of optimized electrode configurations, coarse and fine meshes will be generated using (i) the optimized electrode positions and (ii) one ``standard'' electrode configuration (i.e. uniform spacing of electrodes along the geometry edges).
The coarse meshes will have the same maximum element size used in training and the fine meshes will have a maximum element size half of these values.
Following, Eqs. \ref{qual}  {- \ref{qualR}} will be used as the statistical metrics for comparative quality evaluation.

\subsection{Example 1: 1$\times$1 square}
\begin{figure}[h!]
\centering
\begin{tabular}{m{4.7cm} m{4.2cm} m{4.2cm}}
\includegraphics[width=45mm,trim=7.2cm 1cm 7.2cm 1cm, clip=true]{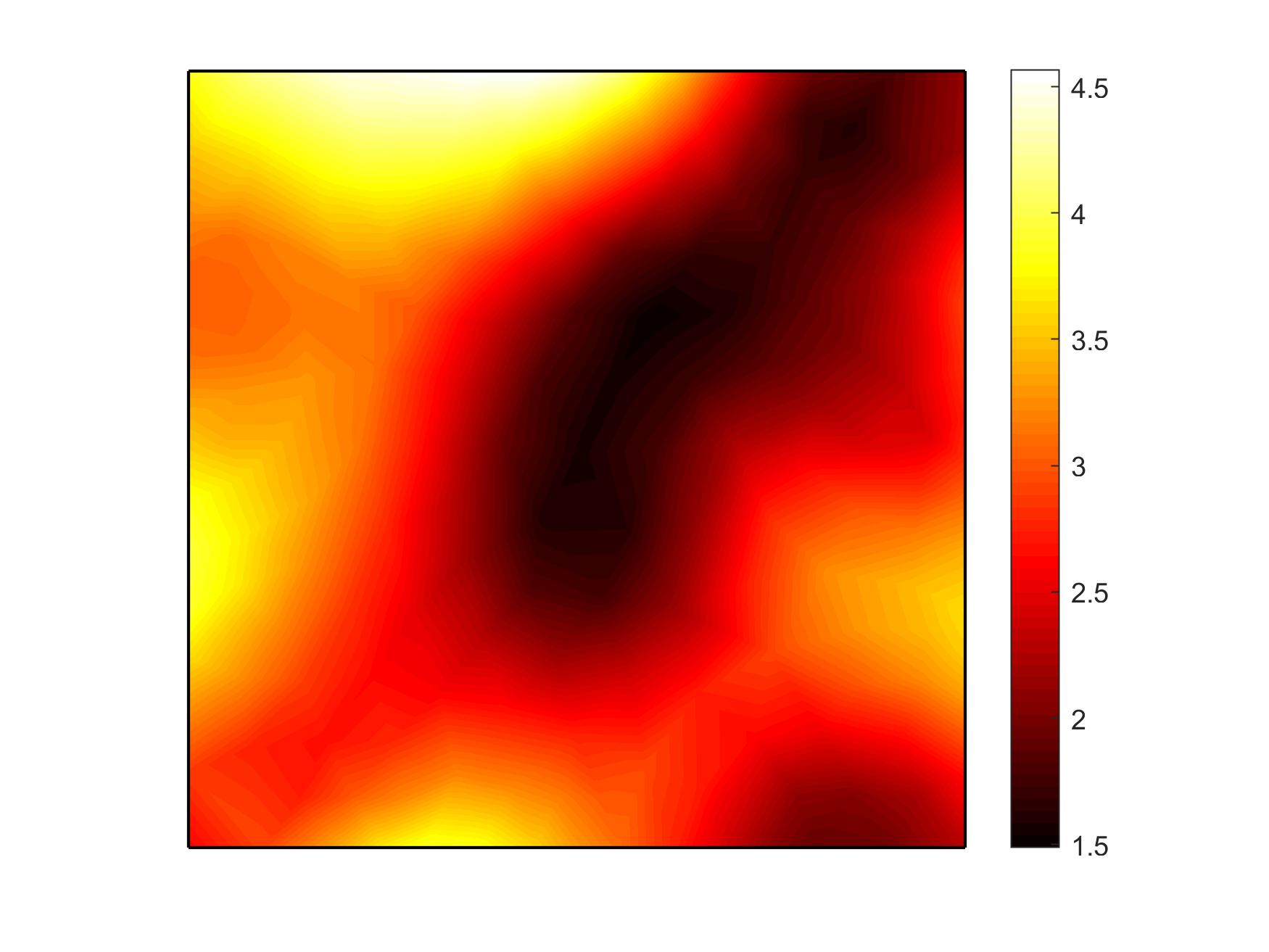} & \includegraphics[width=40mm,trim=7.2cm 1cm 6.0cm 1cm, clip=true]{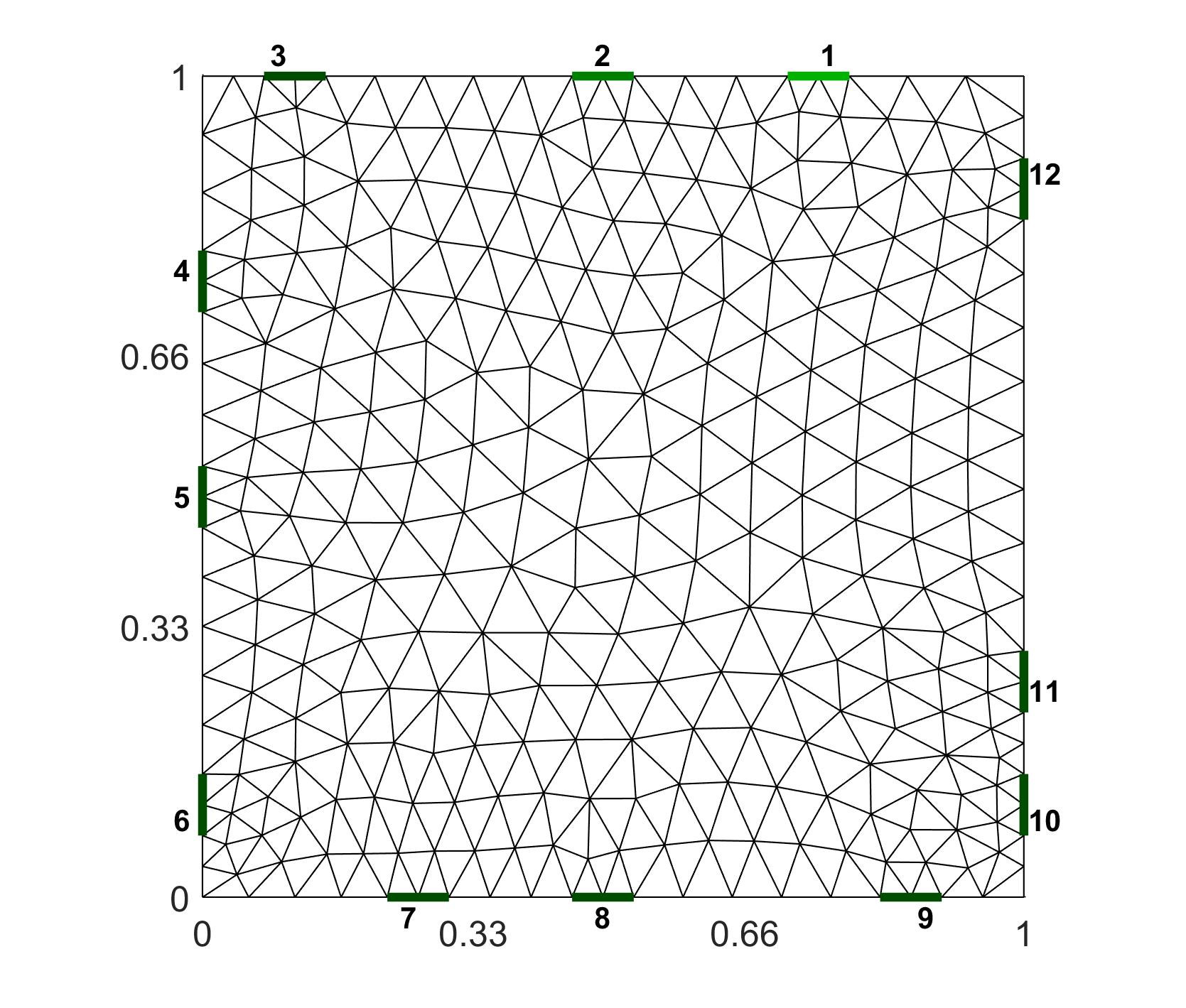} &\includegraphics[width=40mm,trim=7.2cm 1cm 6.0cm 1cm, clip=true]{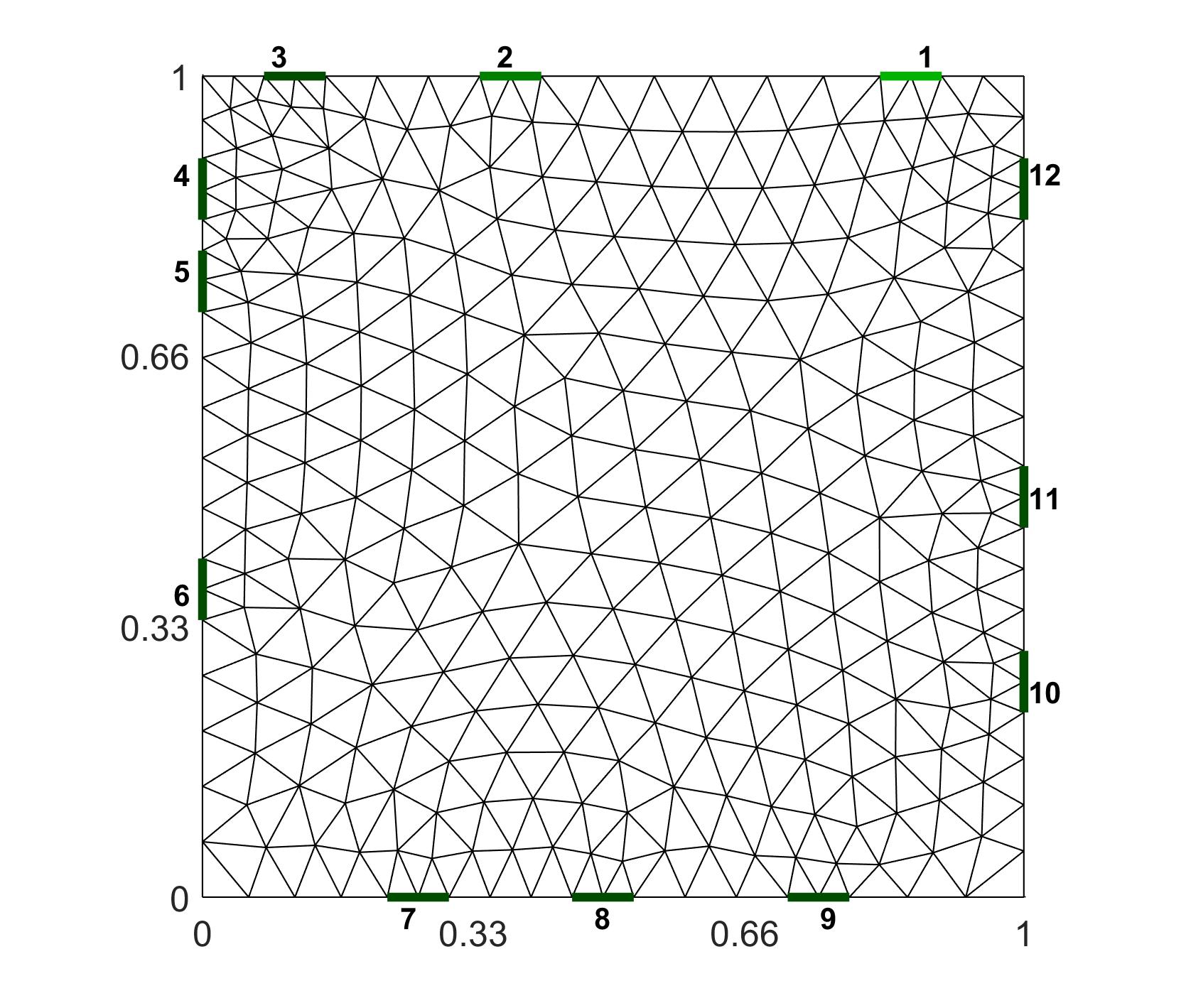} \\
\end{tabular}
\caption{Left column: one random conductivity draw from the training data (units, mS/cm). Middle column and right columns: randomized electrode locations taken from the training data.
Electrode locations are indicated numerically and are organized as counterclockwise ascending from the top right-hand corner of the geometry.}
\label{E1D}
\end{figure}

In this example, we examine the optimized electrode configuration generated for a 1$\times$1 unit square having $k=12$ electrodes.
The geometry was prescribed three electrodes per side with a fixed electrode width of 0.075.
To illustrate the example graphically, one random conductivity draw and two random electrode configurations are shown in Fig. \ref{E1D} (electrodes shown in green).
Note the counterclockwise ascending numbering of electrodes starting from the top right-hand corner of the geometry (consistent numbering is maintained through the paper).

\begin{figure}
\centering
\begin{tabular}{m{7cm} m{7cm}}
\includegraphics[width=8cm]{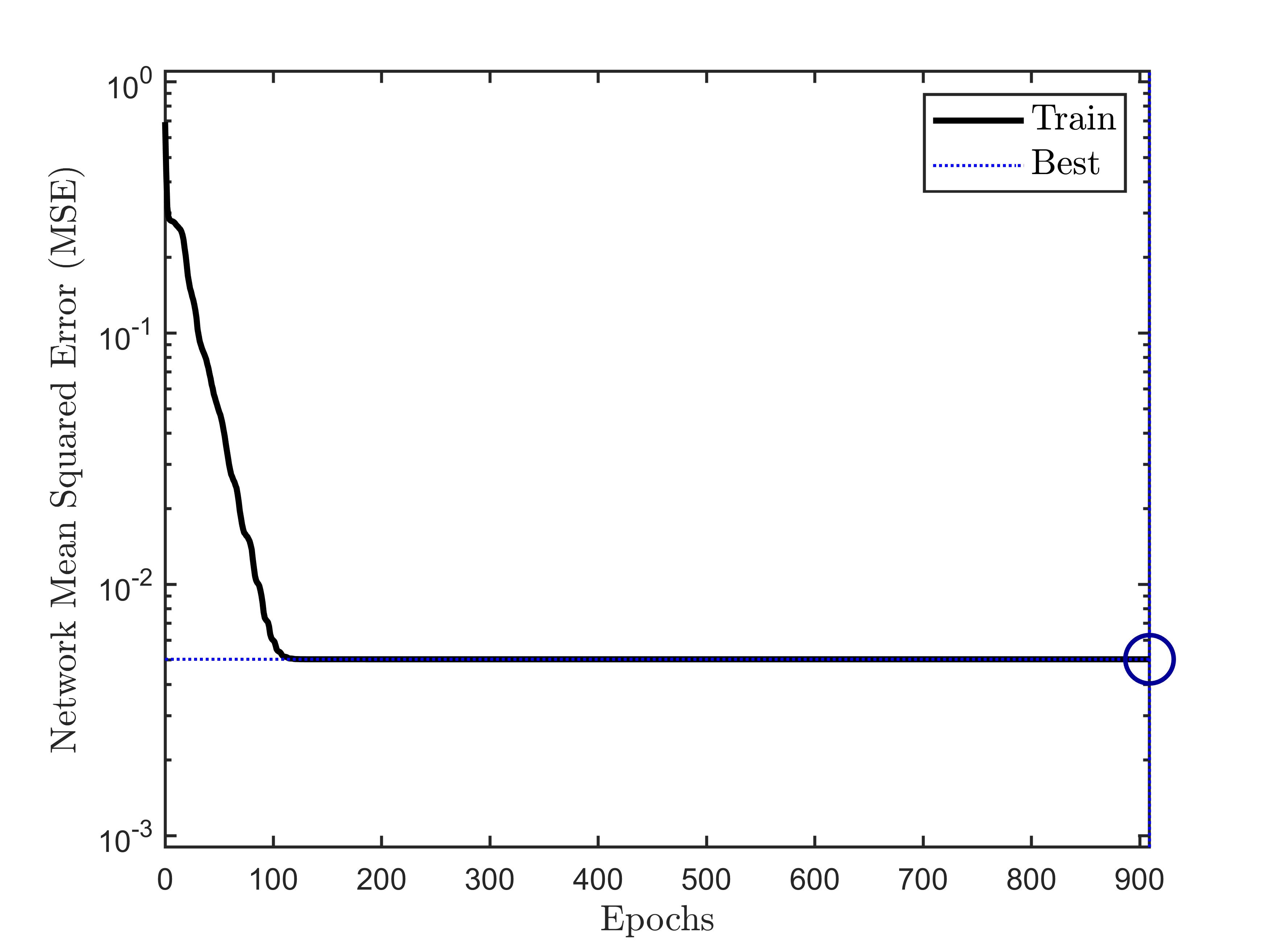} & \includegraphics[width=7.75cm]{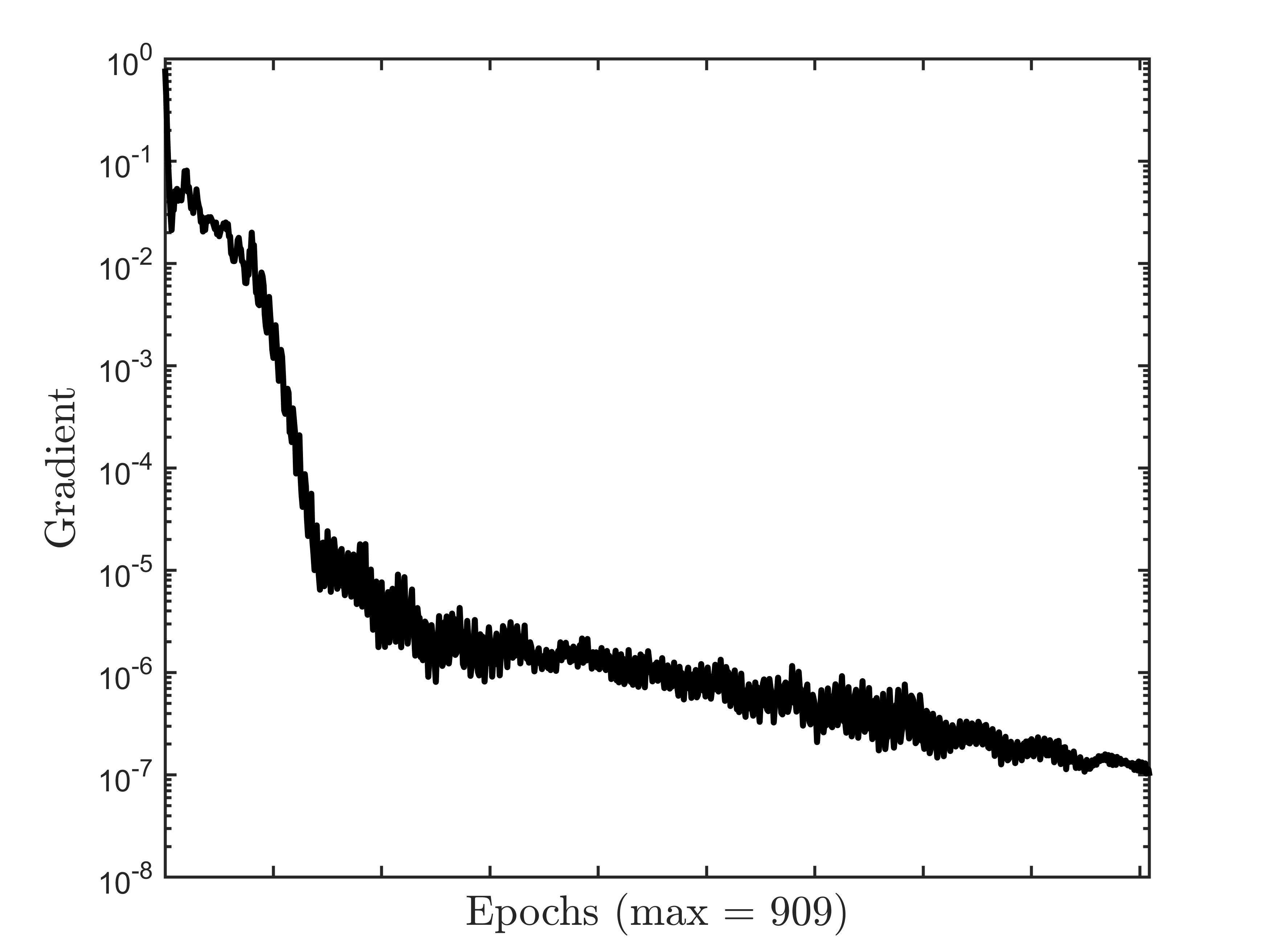} \\
\end{tabular}
\caption{Training plots for the network $Q_{\bar{\Theta}, \bar{E}}$. The left-hand side shows the MSE performance while the right-hand side shows the gradient drops across the training epochs.}
\label{E1train}
\end{figure}

After generating the training data, the neural network was trained following the procedure highlighted in the previous section.
Representative plots taken after the completion of training are provided in Fig. \ref{E1train}, where the MSE performance and gradient drops are both provided throughout the training epochs.
The total time required to train the model $Q_{\bar{\Theta}, \bar{E}}$ was approximately two minutes over a total of 909 epochs.
This performance was similar among all examples studied in this work.

\begin{figure}[h!]
\centering
\begin{tabular}{m{0.5cm} m{5.0cm} m{5.0cm}}
\textbf{\large { (a) }} & \includegraphics[width=46mm,trim=4.0cm 1cm 4.0cm 1cm, clip=true]{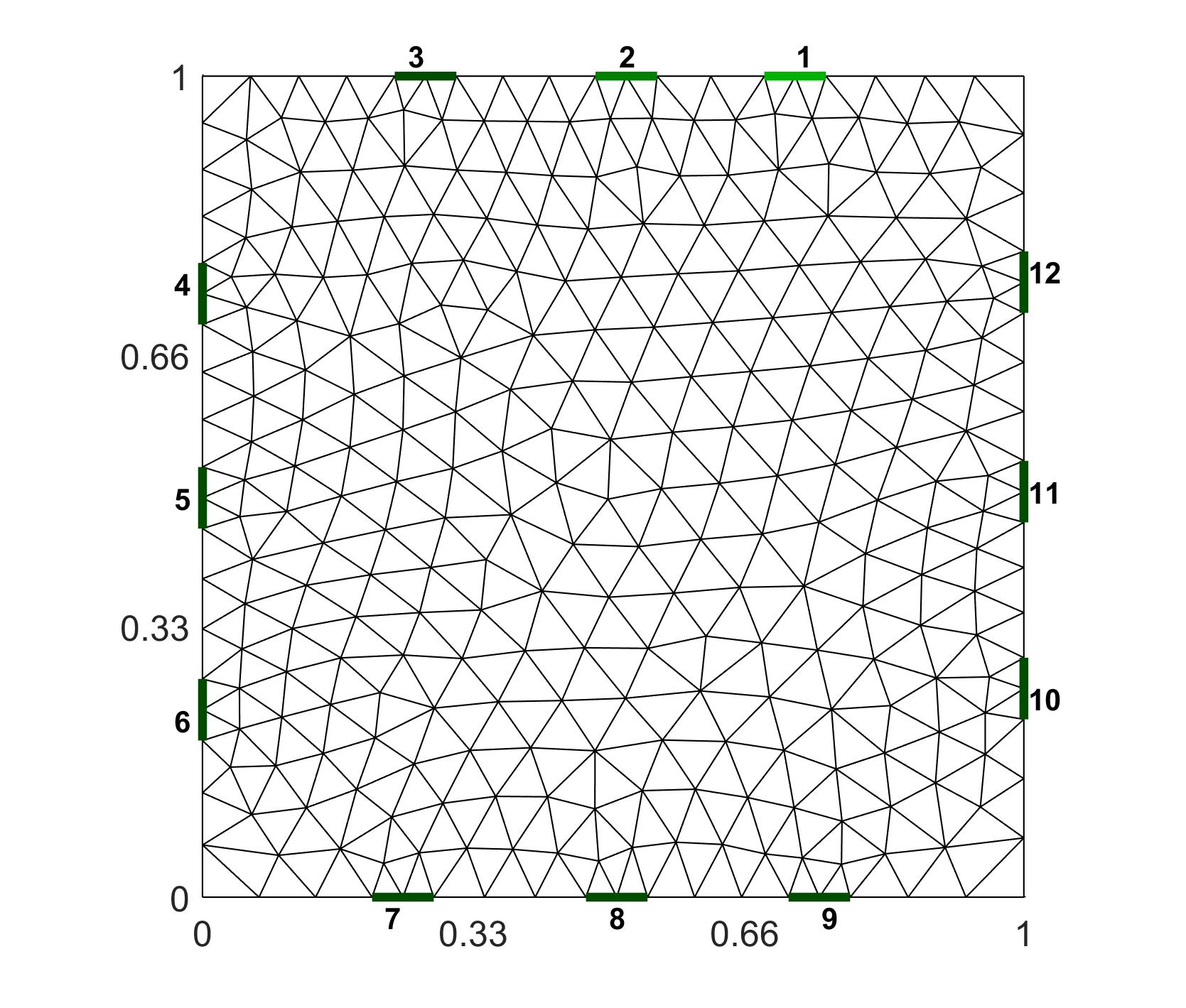} & \includegraphics[width=46mm,trim=4.0cm 1cm 4.0cm 1cm, clip=true]{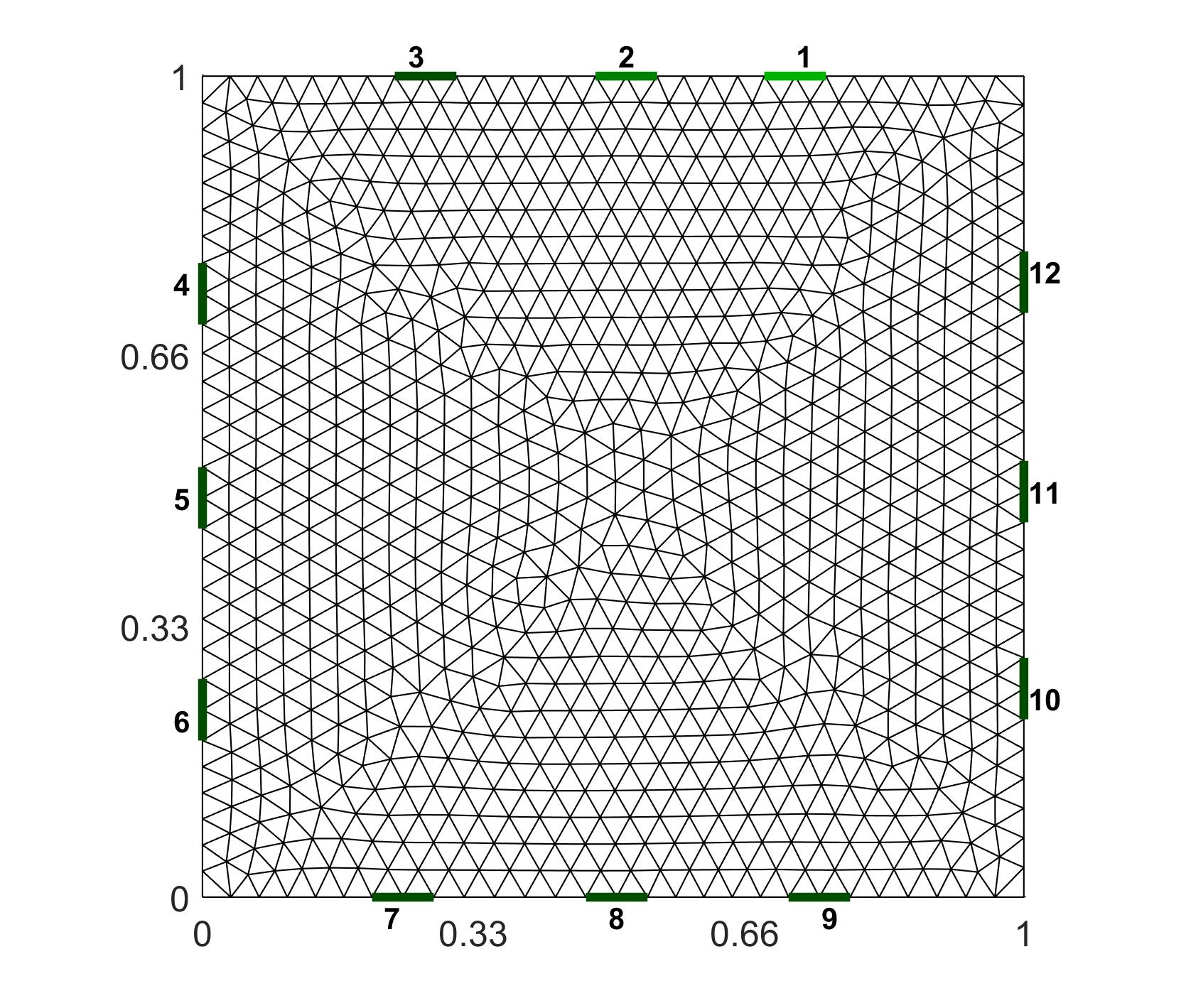} \\
\textbf{\large { (b) }} & \includegraphics[width=46mm,trim=4.0cm 1cm 4.0cm 1cm, clip=true]{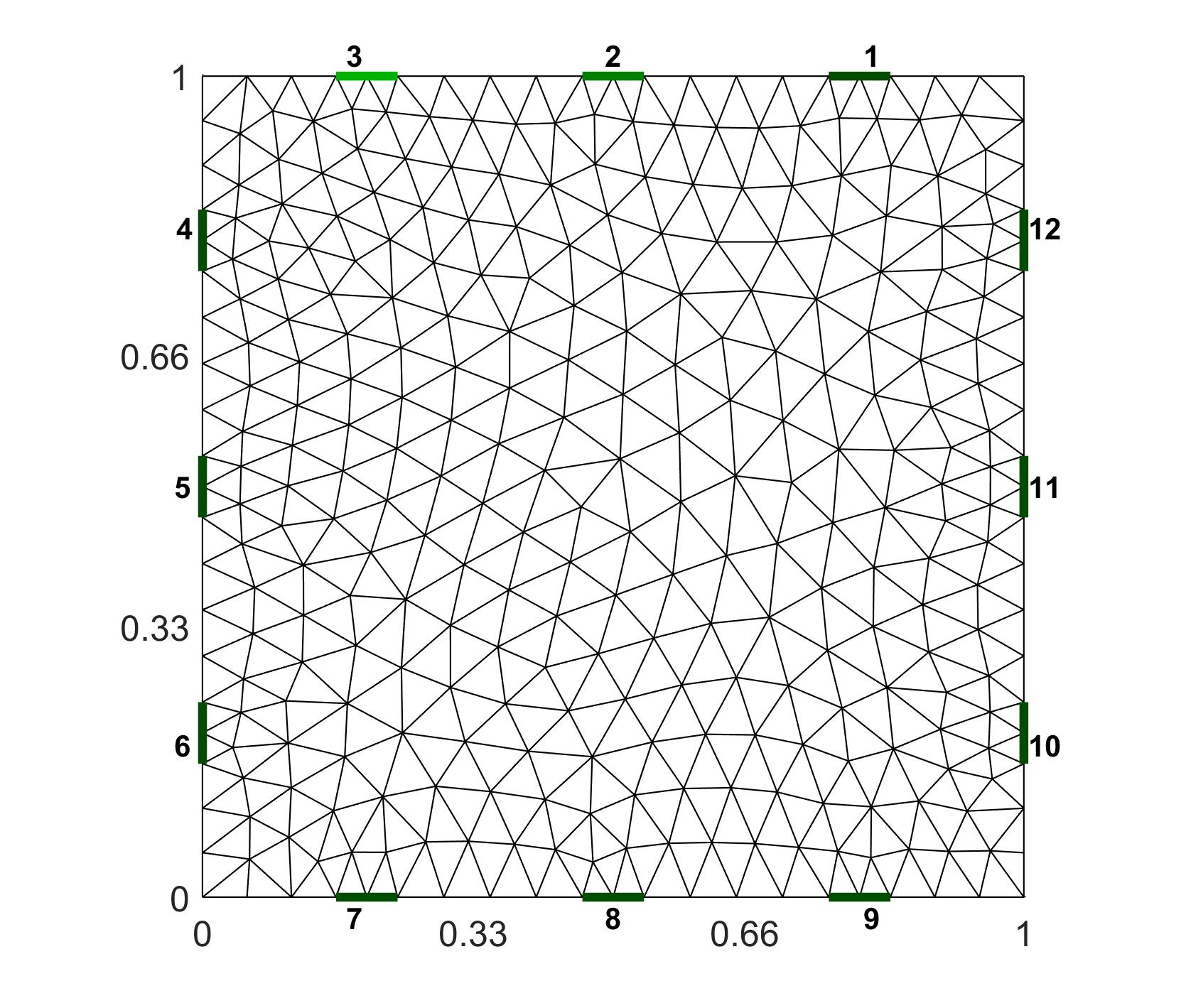} & \includegraphics[width=46mm,trim=4.0cm 1cm 4.0cm 1cm, clip=true]{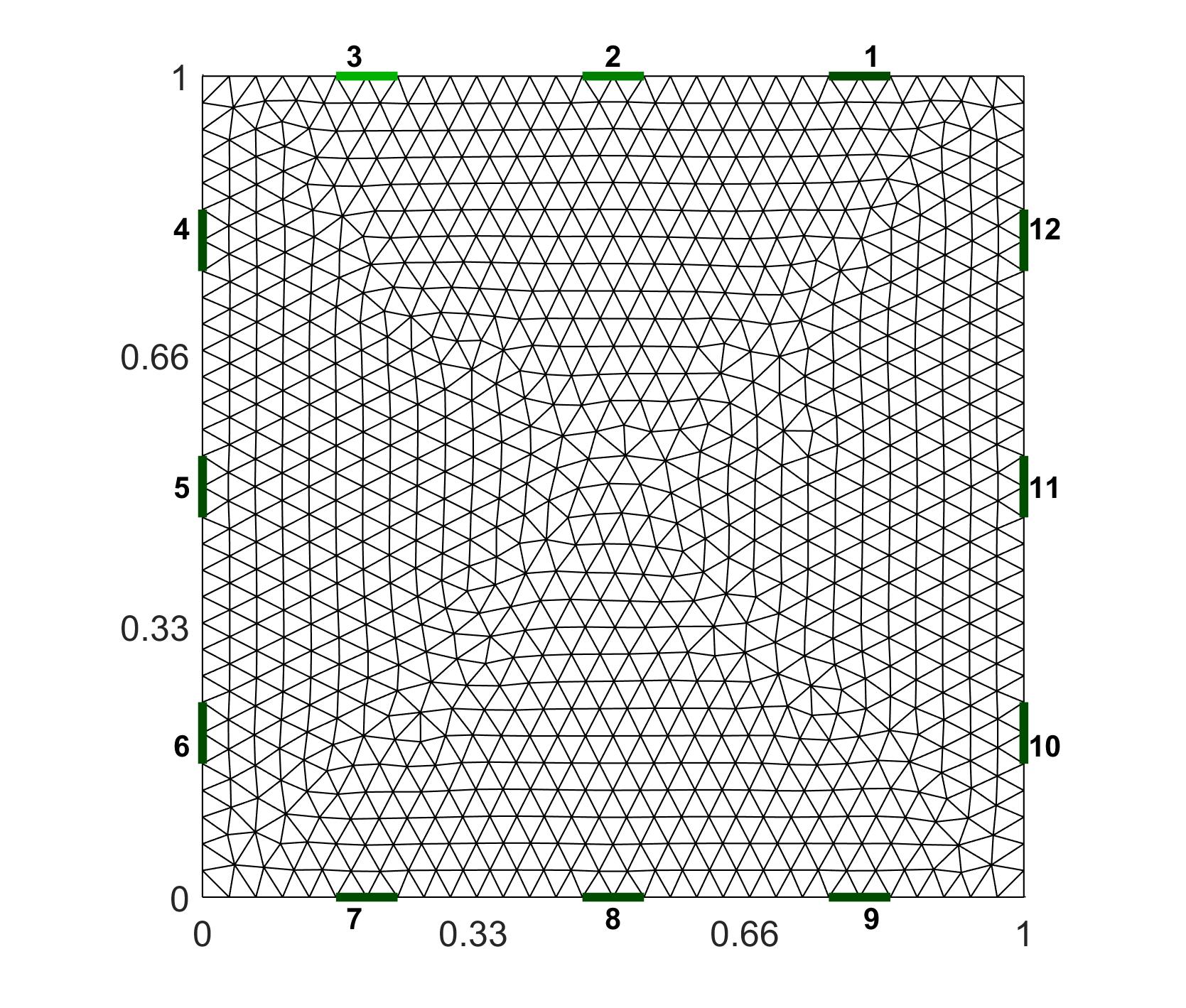} \\
\end{tabular}
\centering
\begin{tabular}{m{0.5cm} m{5.0cm}}
\textbf{\large { (c) } } \hspace{-4mm} & \includegraphics[width=57mm,trim=4.5cm 2cm 0cm 2cm, clip=true]{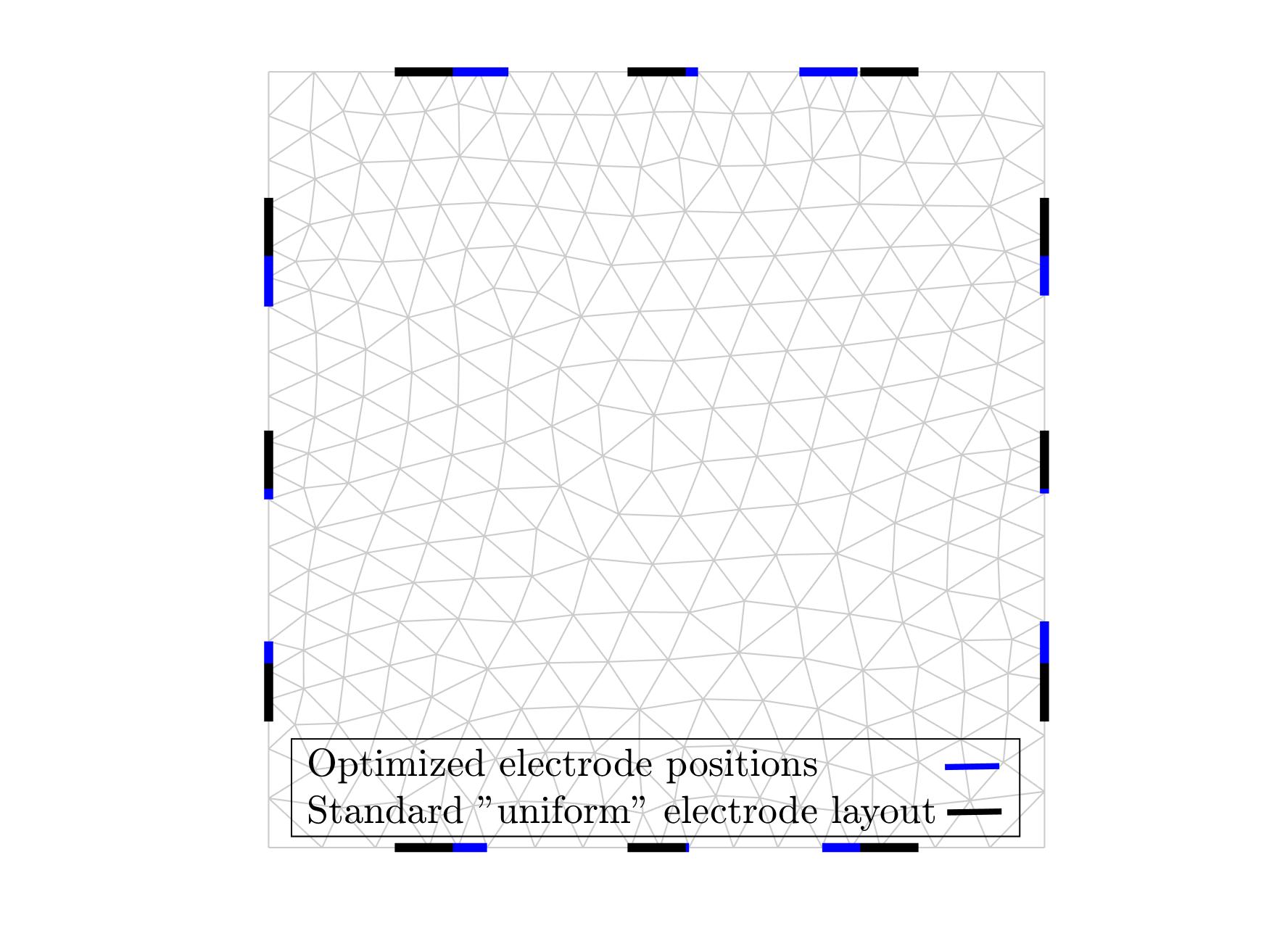}
\end{tabular}
\caption{1$\times$1 meshes for the (a) optimized electrode positions and (b) ``standard'' (evenly-spaced) electrode positions and (c) an overlay comparing the uniform electrode positions (black) plotted atop the optimized electrode positions (blue). Coarse and fine meshes used in the quality analysis are shown on the left- and right-hand sides, respectively.}
\label{E1optE}
\end{figure}

\begin{figure}[h!]
\centering
\includegraphics[width=8cm]{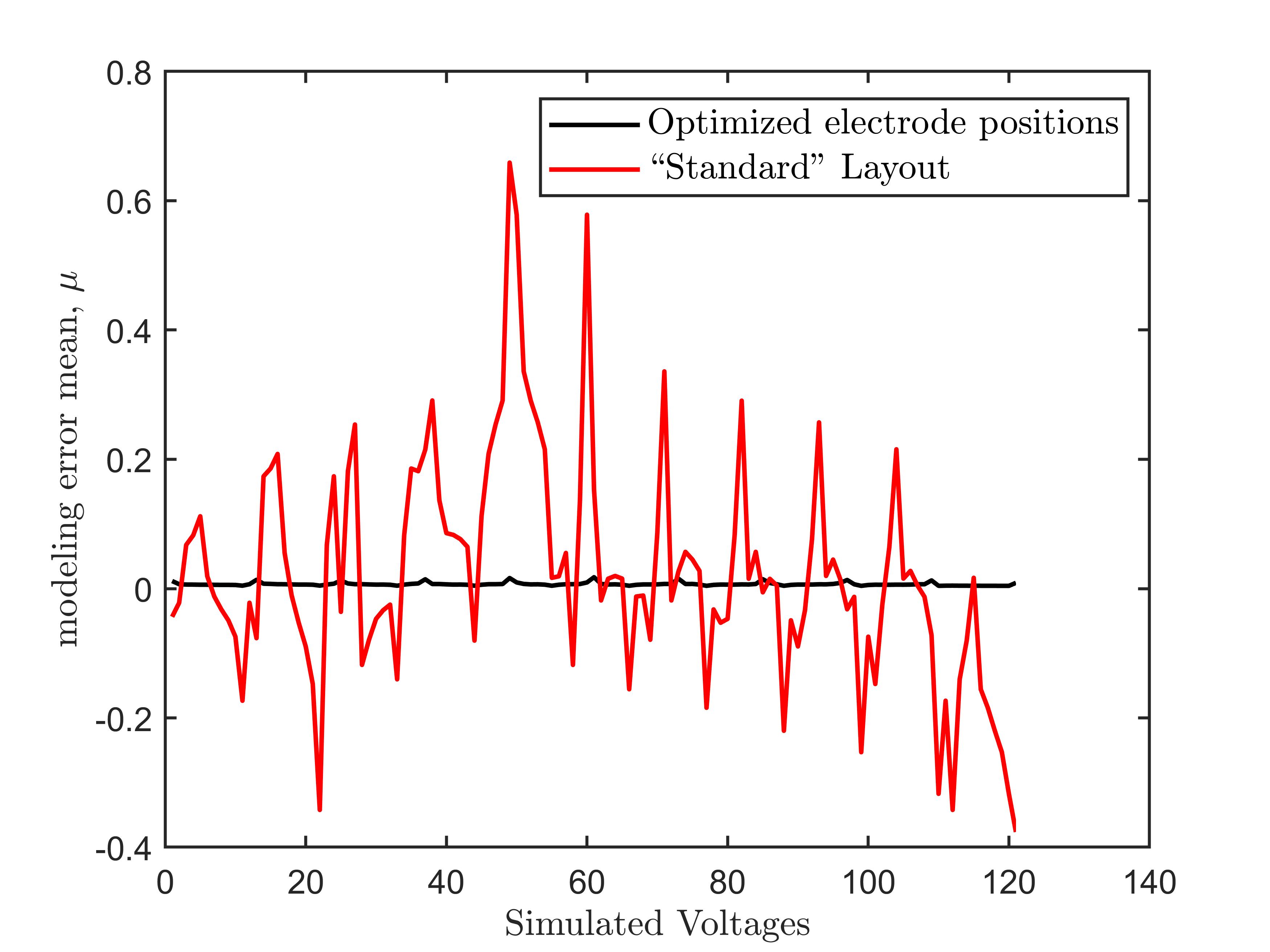}
\caption{Comparison of quality metrics $\mu$ for the optimized electrode positions (black) and ``standard'' positions (red) for the 1x1 example.}
\label{E1qual}
\end{figure}

Upon training $Q_{\bar{\Theta}, \bar{E}}$, the optimized electrode configuration was computed by inputting $\Theta = [1,0]^\mathrm{T}$ into $Q_{\bar{\Theta}, \bar{E}}$.
The optimized configuration is shown in Fig. \ref{E1optE}(a).
On first glance, it appears that the optimized electrode locations do not vary significantly in comparison to the uniform electrode layout shown in Fig. \ref{E1optE}(b).
However, to investigate this closer, we plot the uniform layout atop the optimized layout, as shown in Fig. \ref{E1optE}(c).
In Fig. \ref{E1optE}(c), we can clearly see notable differences between the uniform and optimized positions; we also observe that the optimized electrode positions are close to, but not exactly evenly spaced.
These small perturbations in the optimized positions appear to be random and are roughly 0.05 in magnitude.
The perturbations can be viewed as errors since, from the symmetry of the geometry, optimized electrode spacing should be uniform -- whereas, e.g., electrode 1 is clearly too close to electrode 2.
The source of errors likely results from over-fitting and/or the distance from the objective vector $\Theta$ to the fitted data (i.e. $\Theta = [1,0]^\mathrm{T}$ is too far from the trained values of $\kappa$ and $\beta$).
At this point, a comparison of the optimized and uniform electrode configurations' quality would be useful to determine the optimization approach's effectiveness.
This comparison, showing the mean modeling errors for $N_\sigma$ samples using coarse and fine meshes shown in Fig \ref{E1optE}(a,b), is provided in Fig. \ref{E1qual}.

The results of the electrode positioning quality analysis, shown in Fig. \ref{E1qual}, are highly unexpected.
The use of optimized electrode positioning resulted in a drastic reduction in mean modeling errors compared to the ``standard'' uniform layout.
In fact, the cumulative sum of mean errors for the uniform layout $||\mu_S||_1$ was approximately 18 times higher than the cumulative sum of mean errors for the optimized layout $||\mu_O||_1$, i.e. $||\mu_S||_1/||\mu_O||_1 \approx 18$.
This indicates that, despite the small electrode positioning differences between the optimized and ``standard'' uniform layout, the electrode position optimization approach was highly effective.
Moreover, this result demonstrates the ill-conditioned nature of the electrode positioning problem -- i.e., small changes in electrode positions result in large changes in modeling errors.

 {Following the previous analysis of the mean modeling errors resulting from discretization, we investigate the effects electrode position optimization has on Hessian and resistivity matrix conditioning using the metrics described in Eqs. \ref{qualH} - \ref{qualR}.
Based on 200 random blob-like conductivity samples, different from the samples used in training the network, we find the mean Hessian condition numbers to be $\bar{\kappa}_{H} = 1.98\times 10^{23}$ and $\bar{\kappa}_{H} = 6.56\times 10^{23}$ for the optimized and ``standard" layouts, respectively.
The mean condition numbers of the resistivity matrices are $\bar{\kappa}_{R}=1.43\times 10^{10}$ and $\bar{\kappa}_{R}=1.57\times 10^{10}$ for the optimized and ``standard" layouts, respectively.
It is interesting to note that a reduction in $\bar{\kappa}_{R}$ of approximately 9\% contributed to a decrease in $\bar{\kappa}_{H}$ of approximately 30\%, which is further evidence for the ``snowball effect" noted earlier (ill-conditioned Finite Element matrices contributing to ill-conditioned Hessians).
Nonetheless, we observe that the condition numbers for the Hessian and resistivity matrices are reduced when employing the optimized electrode layouts.
This result demonstrates the potential of the electrode position optimization approach for reducing the ill-posedness of the EIT inverse problem when employed.
}

\subsection{Example 2: 2$\times$1 rectangle}
\label{rectGeo}
\begin{figure}[h]
\centering
\begin{tabular}{m{0.03cm} m{5.0cm} m{0.05cm} m{5.0cm} m{0.03cm} m{5.0cm} }
\textbf{(a)} & \includegraphics[width=52mm,trim=6cm 10cm 5.5cm 10cm, clip=true]{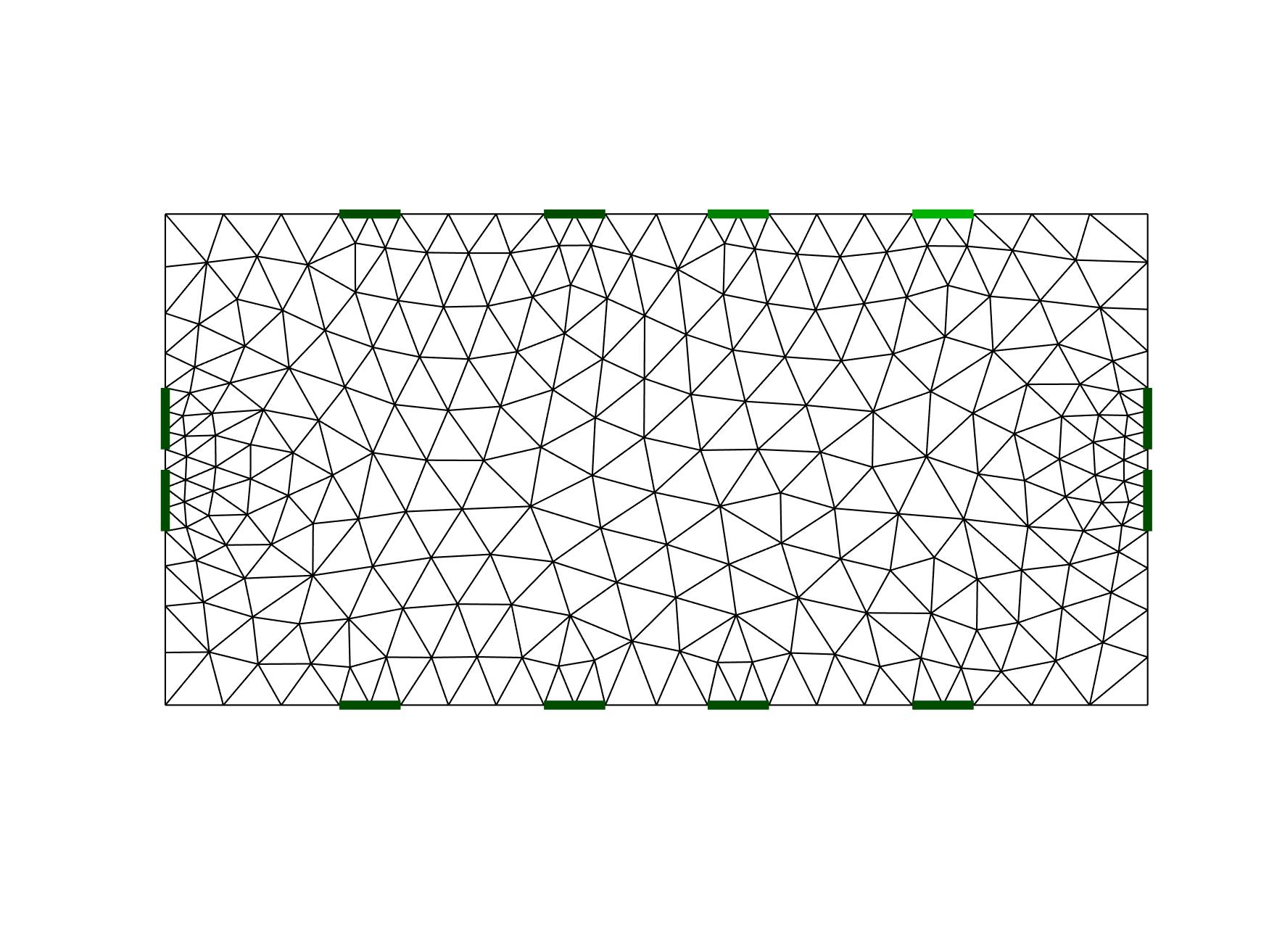}&
\textbf{(b)} & \includegraphics[width=52mm,trim=6cm 10cm 5.5cm 10cm, clip=true]{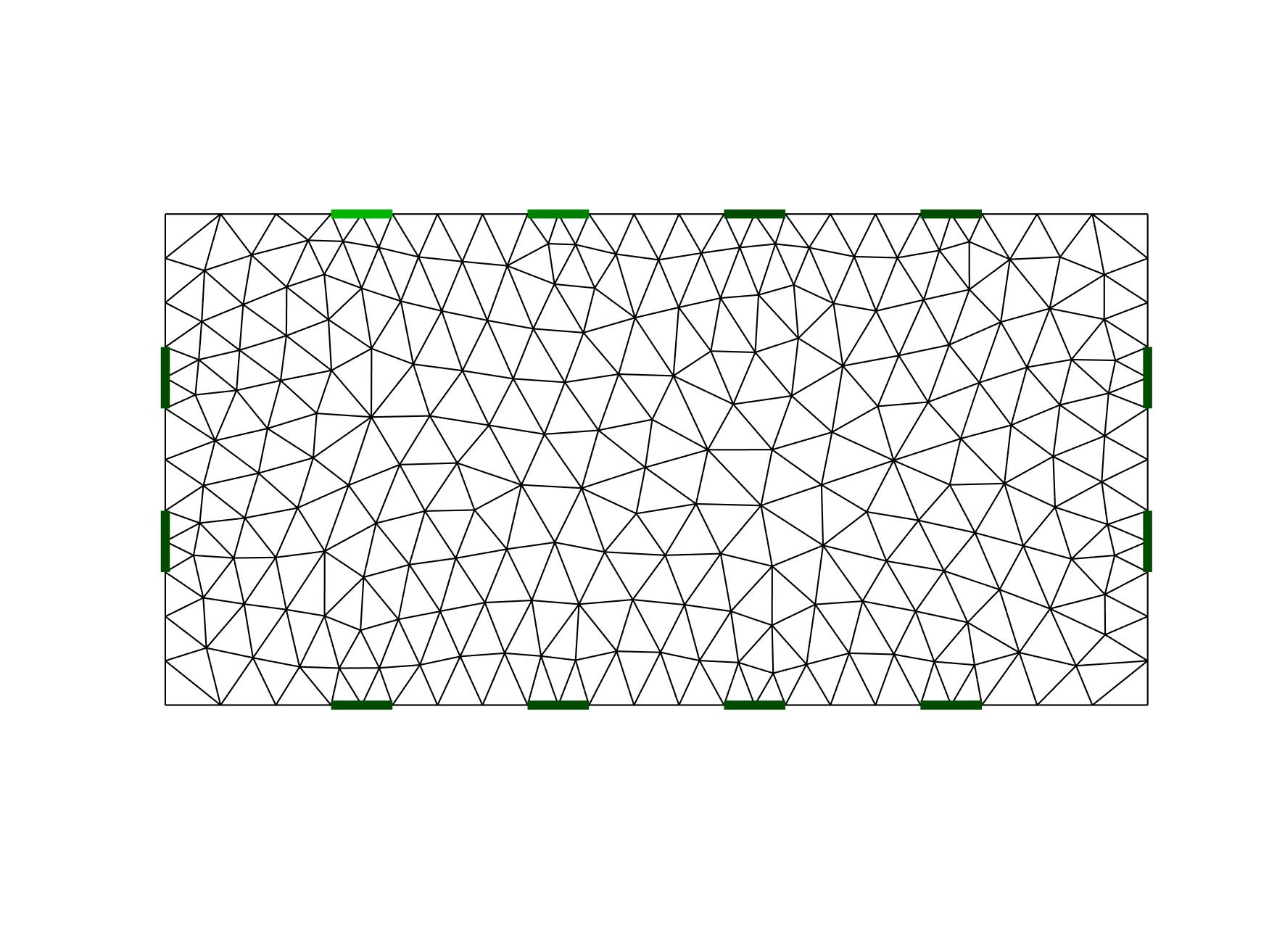} &
\textbf{(c)} & \includegraphics[width=52mm,trim=6cm 10cm 5.5cm 10cm, clip=true]{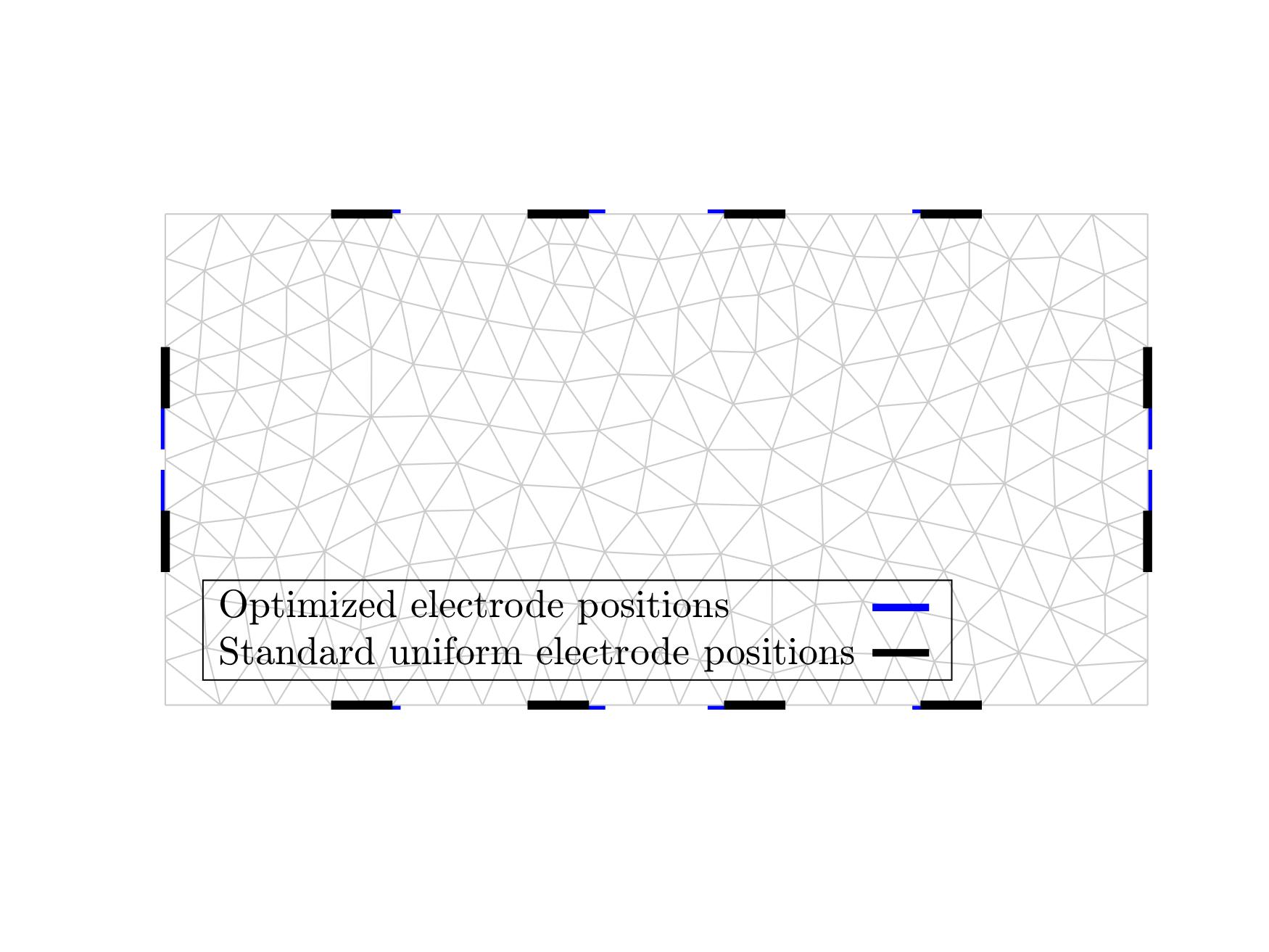} \\
\end{tabular}
\caption{2$\times$1 coarse meshes for the (a) optimized electrode positions and (b) ``standard'' (evenly-spaced) electrode positions and (c) an overlay comparing the uniform electrode positions (black) plotted atop the optimized electrode positions (blue).}
\label{E2optE}
\end{figure}

In this example, we study a 2$\times$1 rectangle and again consider $k=12$ electrodes.
However, in this example, we assume the short sides have 2 electrodes, whereas the longer sides have 4 electrodes.
We have followed the same procedure discussed in the previous subsection and for the purposes of conciseness and to avoid repetition, we immediately show the optimized electrode configuration in Fig. \ref{E2optE}(a).
The comparative uniformly-spaced ``standard'' layout with electrodes located at the one fifth and one third points on the long and short sides, respectively, is shown in Fig. \ref{E2optE}(b).
The differences in electrode positions are shown in Fig. \ref{E2optE}(c) for direct comparison.


Upon immediate visual inspection and comparison of the optimized and ``standard'' electrode positions shown in Fig. \ref{E2optE}(a,b), the largest visual difference is the placing of the short-side electrodes (since the central coordinates of the top and bottom side electrode differ by a maximum of $\approx$0.10 cm).
This is confirmed in the electrode position discrepancy plot in Fig. \ref{E2optE}(c).
On first thought, one may hypothesize that the number of elements differ significantly between the two configurations, which -- in itself -- may drive significant differences in the finite element matrix condition numbers.
However, the meshes vary by only approximately 20 nodes/elements, so the effect of element/node number is marginal.
As such, the locations of the electrodes on the short sides must play a major role in the optimality of the mesh in Fig. \ref{E2optE}(a).
Therefore, for this geometry, the movement of the end electrodes towards each other functions to increase the information in EIT data and reduce the ill-conditioning the and Hessian matrices -- one possible mechanism for this is likely the increasing distance between current injections.
Using $N_\sigma$ different trial conductivity samples, we found, on average, a reduction of $5 - 14\%$ in the 2-norm condition numbers of the optimized configuration's Hessian matrices compared to the uniform configuration -- which confirms the former hypothesis regarding the end electrode locations.

\begin{figure}[h!]
\centering
\includegraphics[width=8cm]{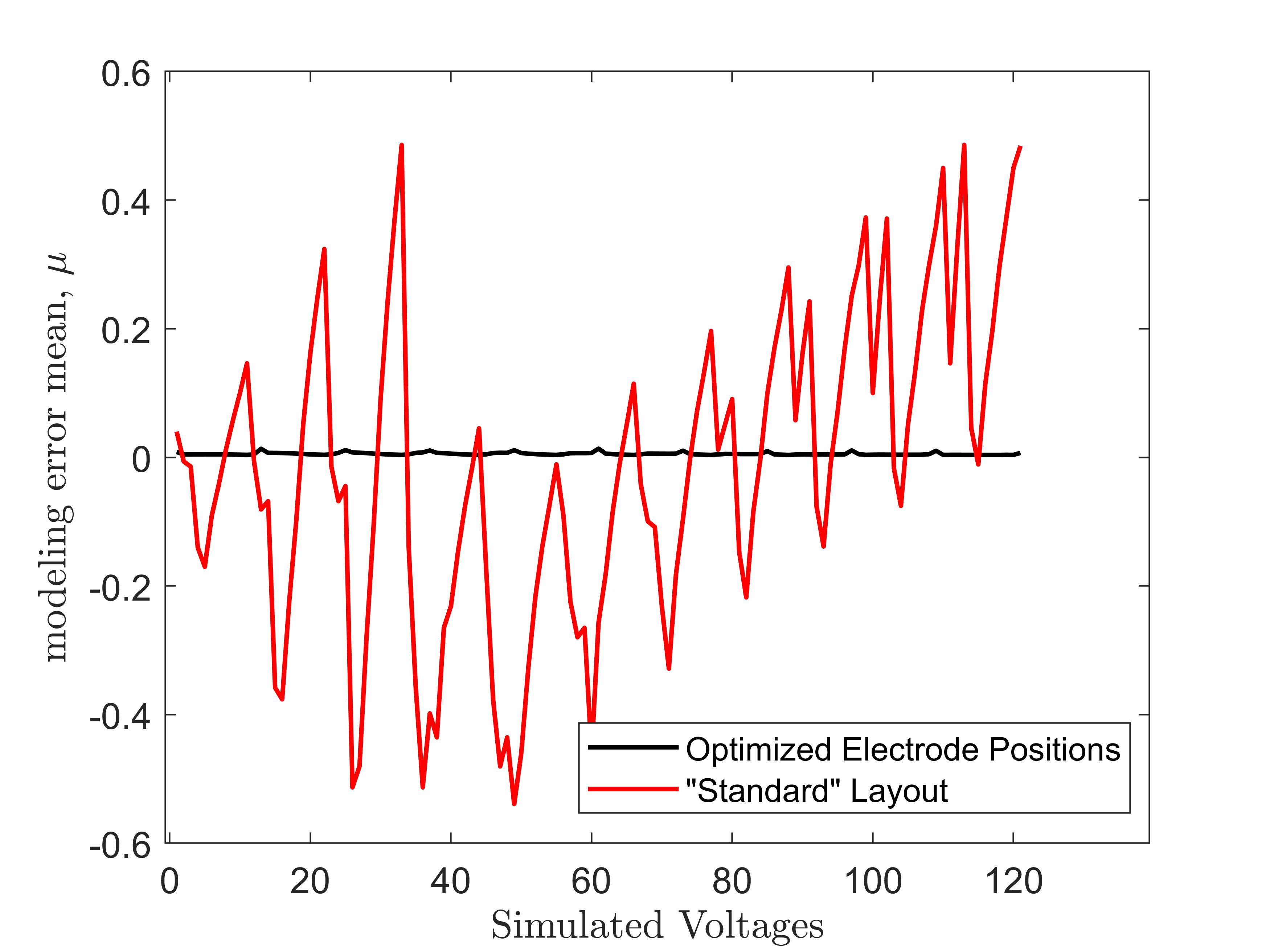}
\caption{Comparison of quality metrics $\mu$ for the optimized electrode positions (black) and ``standard'' positions (red) for the 2x1 example.}
\label{qualE2}
\end{figure}

What remains in this analysis is a quantification of the optimized electrode configuration quality in comparison to the uniform layout, which is shown in Fig. \ref{qualE2}.
A similar trend observed in the square example is also observed here.
In this example, the uniform layout mean error $||\mu_S||_1$ was approximately 30 times higher than the cumulative sum of mean errors for the optimized layout $||\mu_O||_1$, i.e. $||\mu_S||_1/||\mu_O||_1 \approx 30$.
This improvement in modeling error again confirms the effectiveness of the optimization approach.
Interestingly, we found a greater reduction in the mean error using optimized electrode positions in this example relative to the square example.
This may indicate that the use of optimized electrode configurations in slender geometries has a greater impact on the quality of EIT reconstructions than in geometries with an aspect ratio approaching 1 to 1.
This realization is intuitively logical since the optimal electrode configurations -- at least for randomized or homogeneous conductivity distributions -- for square or circular geometries are (or nearly are) uniform, as demonstrated in the previous example and also inferred from symmetry.
 {The amplified effects of electrode optimization in slender rectangular geometries is also supported here by a 73\% reduction in the the mean Hessian condition number compared to the ``standard'' electrode layout ($\bar{\kappa}_{H}=3.45\times 10^{24}$ and $\bar{\kappa}_{H}=1.25\times 10^{25}$ for the optimized and ``standard'' electrode layouts, respectively, computed using 200 blob-like conductivity samples).
Indeed, and while the reduction of the mean resistivity matrix condition number compared to the ``standard'' electrode layout is similar to the square example (approximately 9\% for both geometries), electrode optimization resulted in an additional 43\% reduction in the mean Hessian condition number relative to the square example.
}

\break

\subsection{Example 3:  {A right} triangle}
\begin{figure}[h]
\centering
\begin{tabular}{m{0.1cm} m{4.5cm} m{0.1cm} m{4.5cm} m{0.1cm} m{4.5cm} }
\textbf{(a)} & \includegraphics[width=55mm,trim=10cm 1cm 1cm 1cm, clip=true]{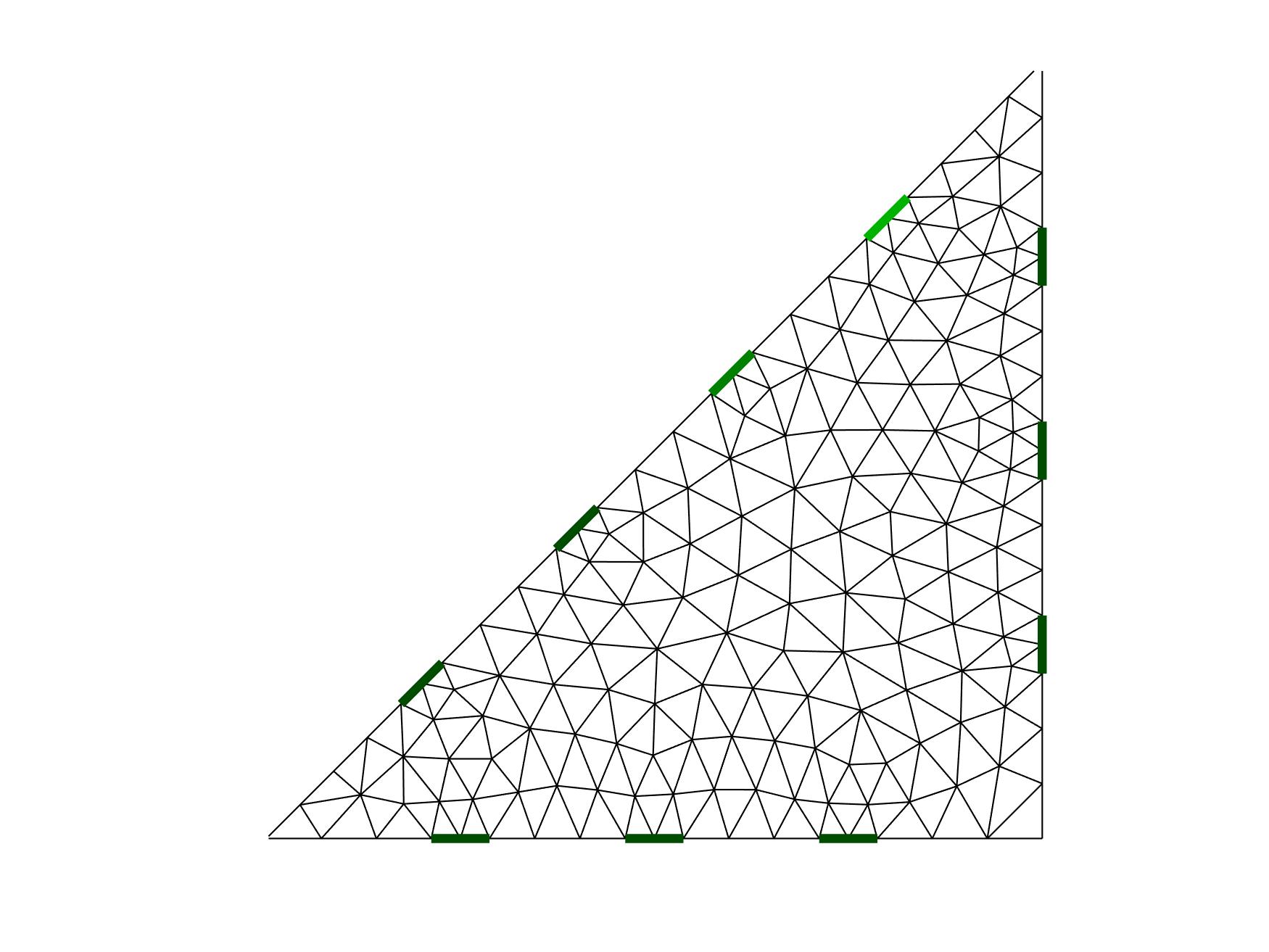} & \textbf{(b)} & \includegraphics[width=55mm,trim=10cm 1cm 1cm 1cm, clip=true]{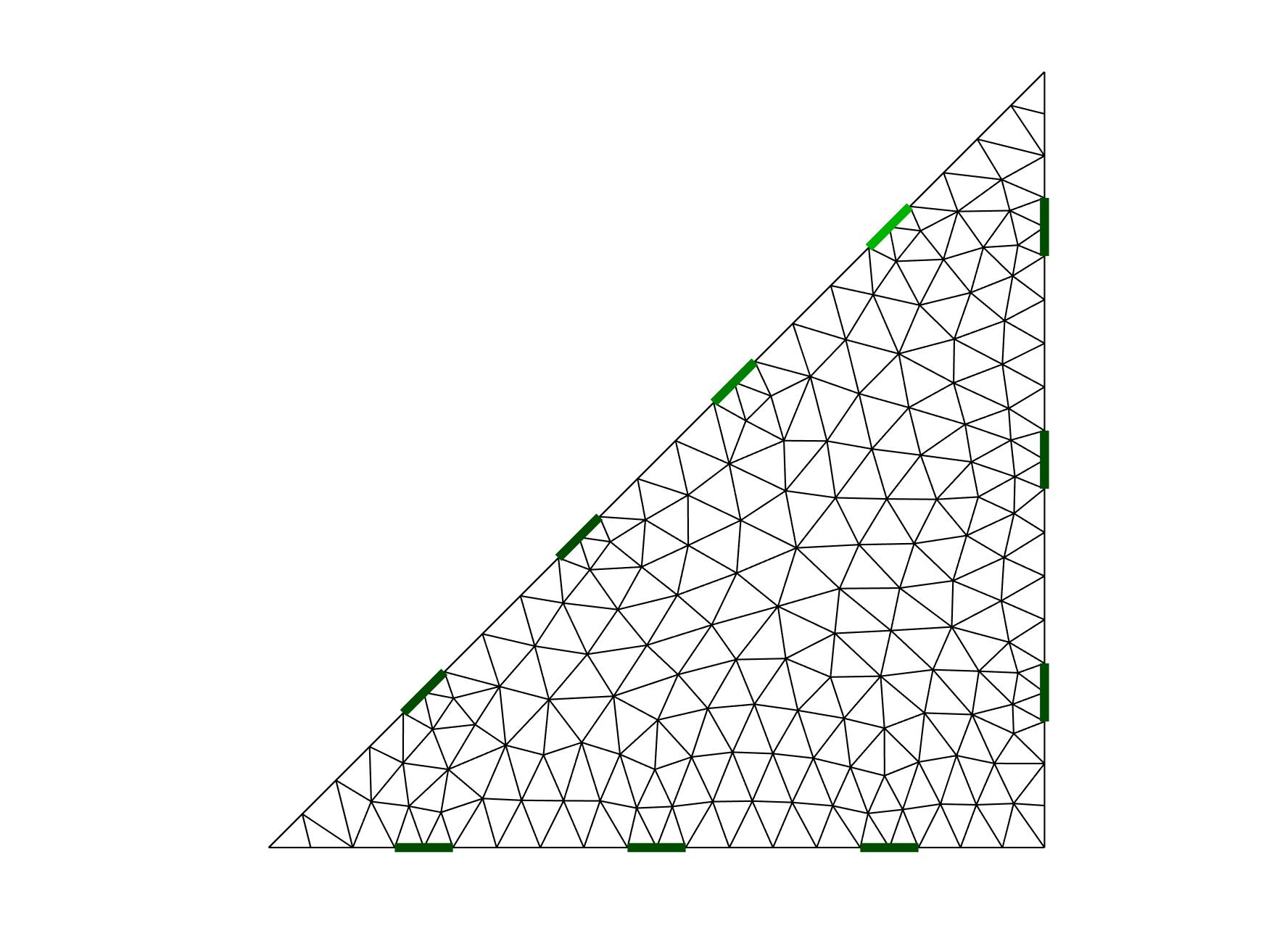} & \textbf{(c)} & \includegraphics[width=55mm,trim=10cm 1cm 1cm 1cm, clip=true]{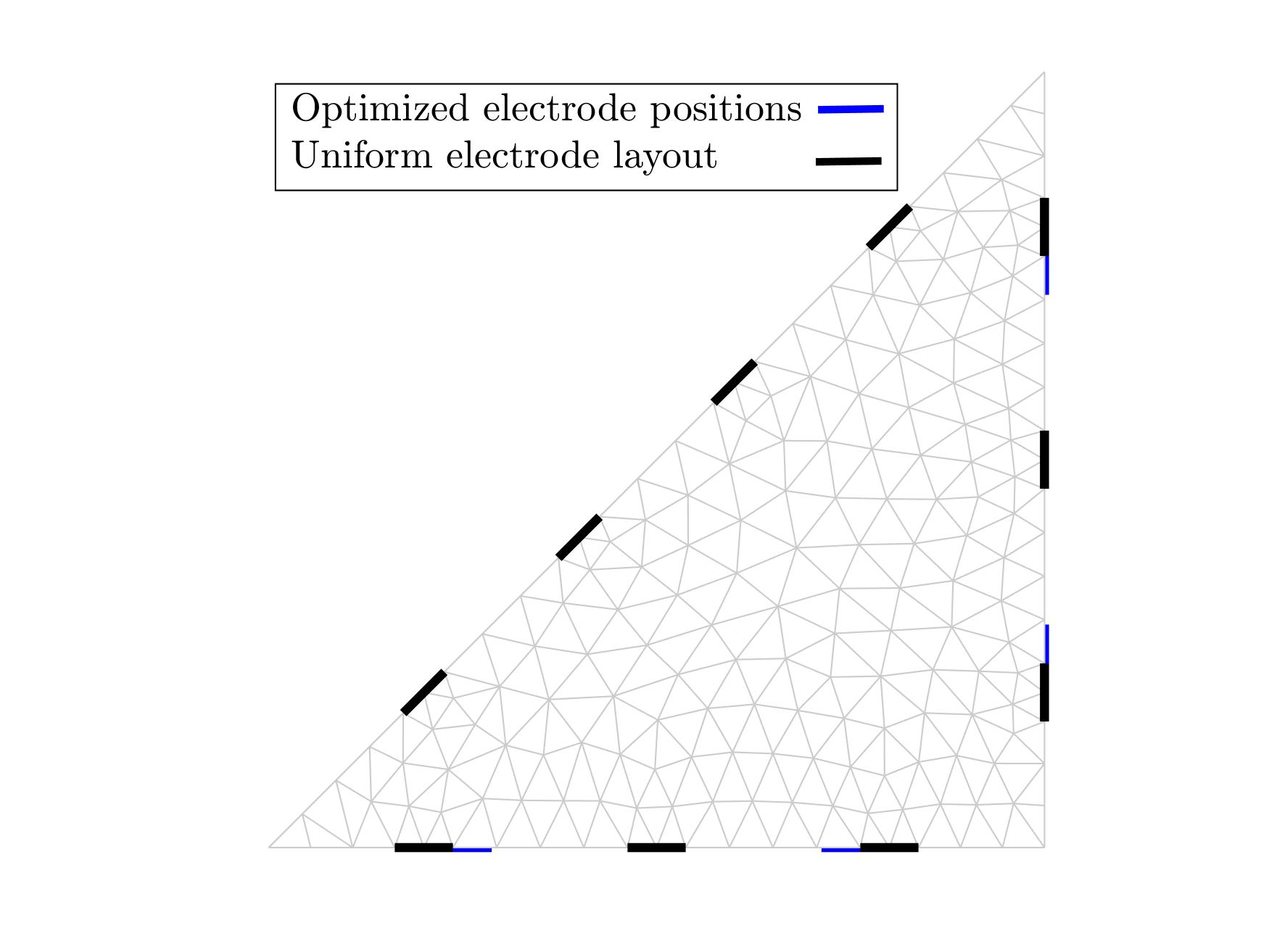} \\
\end{tabular}
\caption{  {$1 \times 1 \times \sqrt{2}$} triangle geometry coarse meshes for the (a) optimized electrode positions and (b) ``standard'' (evenly-spaced) electrode positions and (c) an overlay comparing the uniform electrode positions (black) plotted atop the optimized electrode positions (blue).}
\label{E3optE}
\end{figure}

This example explores electrode position optimization for a right triangle with side lengths  {$1 \times 1 \times \sqrt{2}$}.
Here, we assume that the hypotenuse has 4 electrodes and the other sides have three electrodes, for a total of $k = 10$ electrodes.
The uniform control electrode configuration used for comparison has electrode midpoints located at one fifth points along all sides.
The optimized, uniform control, and comparative electrode configurations are shown in Fig. \ref{E3optE}.


\begin{figure}[h]
\centering
\includegraphics[width=8cm]{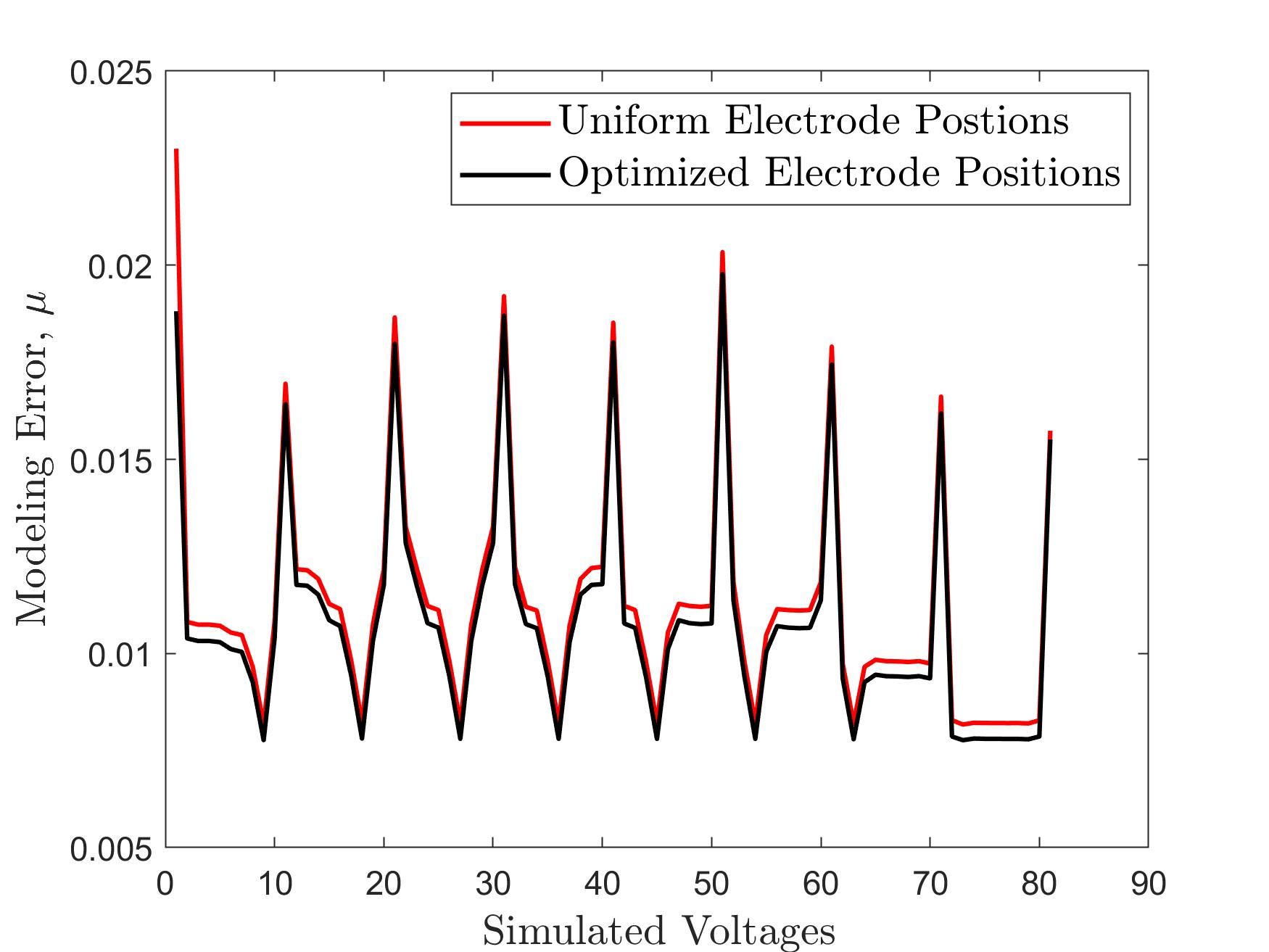}
\caption{Comparison of quality metrics $\mu$ for the optimized electrode positions (black) and uniform control positions (red) for the triangular geometry.}
\label{qualE3}
\end{figure}

We immediately notice that the optimized electrode positions on the diagonal match nearly exactly with those on the control triangle.
On the bottom and right-hand sides, we observe that the middle optimized electrode positions are the same as the control, while the others are slightly different.
In fact, the optimized electrodes on the bottom and right-hand sides are located at one quarter, one half, and three quarters the side length within a precision of 0.01.
Given the small differences, it would be interesting to determine if the resulting modeling errors for each configuration also vary by only a small amount.
To investigate this, the mean error plots for both electrode configurations are shown in Fig. \ref{qualE3}.

In this case, we clearly see that the mean errors of the optimized electrode positions are only slightly lower than the errors from the control triangle.
For this example, the cumulative ratio of the electrode quality metrics is $||\mu_S||_1/||\mu_O||_1 \approx 1.07$.
In other words, the optimized configuration only outperformed the control configuration by approximately 7\% -- which again supports the effectiveness of the optimization approach, albeit the improvement is small.
However, this is an interesting observation, which may indicate that unlike the square and rectangular examples,  {voltages computed using} EIT forward models for triangular geometries are less sensitive to pertubations in electrode positions.
Intuitively, this result does make sense since the electric current is much more concentrated towards the center in the triangle.
Therefore, small changes in the electrode positions do not affect the electric fields as drastically as in the previous examples.
We can therefore  {establish} that the effectiveness of the optimization regime is significantly influenced  {by} the target geometry.

 {In contrast to the former observations, the mean Hessian condition number computed over 200 randomized blob-like samples was 45\% lower for the geometry with optimized electrode positions ($\bar{\kappa}_{H}=3.87\times 10^{25}$ and $\bar{\kappa}_{H}=7.04 \times 10^{25}$ for the optimized and uniform electrode layouts, respectively).
Therefore, despite the concentration of electric currents noted, the conditioning of the Hessian is nonetheless significantly influenced by even small changes in electrode positions.
Conversely, the mean condition number of the resistivity matrices for the optimized and uniform layouts differed by only 4\% ($\bar{\kappa}_{R}=9.51\times 10^{9}$ and $\bar{\kappa}_{R}=9.84 \times 10^{9}$ for the optimized and uniform electrode layouts, respectively).
Based on these realizations, and those from the previous two examples, we conclude that employing the electrode positioning algorithm resulted in reduction of all three quality metrics.
Moreover, it is worth noting that, in all cases, a rather modest reduction in the mean condition number of the resistivity matrix (4\% - 9\%) contributed to drastic reductions in the mean Hessian condition numbers (30\% - 73\%) -- all reductions, of course, are relative to reference layouts with uniform electrode distributions.
The latter is a key takeaway from this work and also reinforces the influence of electrode position optimization on the ill-conditioning/ill-posedness of EIT.}

\section{Discussion: curious geometries,  {reconstructions, distinguishability,} and future work}
\subsection{ {Curious geometries}}
To this point, this article has investigated rather simple geometries and shown the effectiveness of the deep learning based approach for optimizing electrode configurations for these cases.
But, can the algorithm handle more complex geometries and to what extent is the optimization approach still viable?

\begin{figure}[h!]
\centering
\begin{tabular}{m{0.05cm} m{5.0cm} m{0.05cm} m{5.0cm}}
\textbf{(a)} & ~~~~~~\includegraphics[width=103mm,trim=15cm 38cm 10cm 38cm, clip=true]{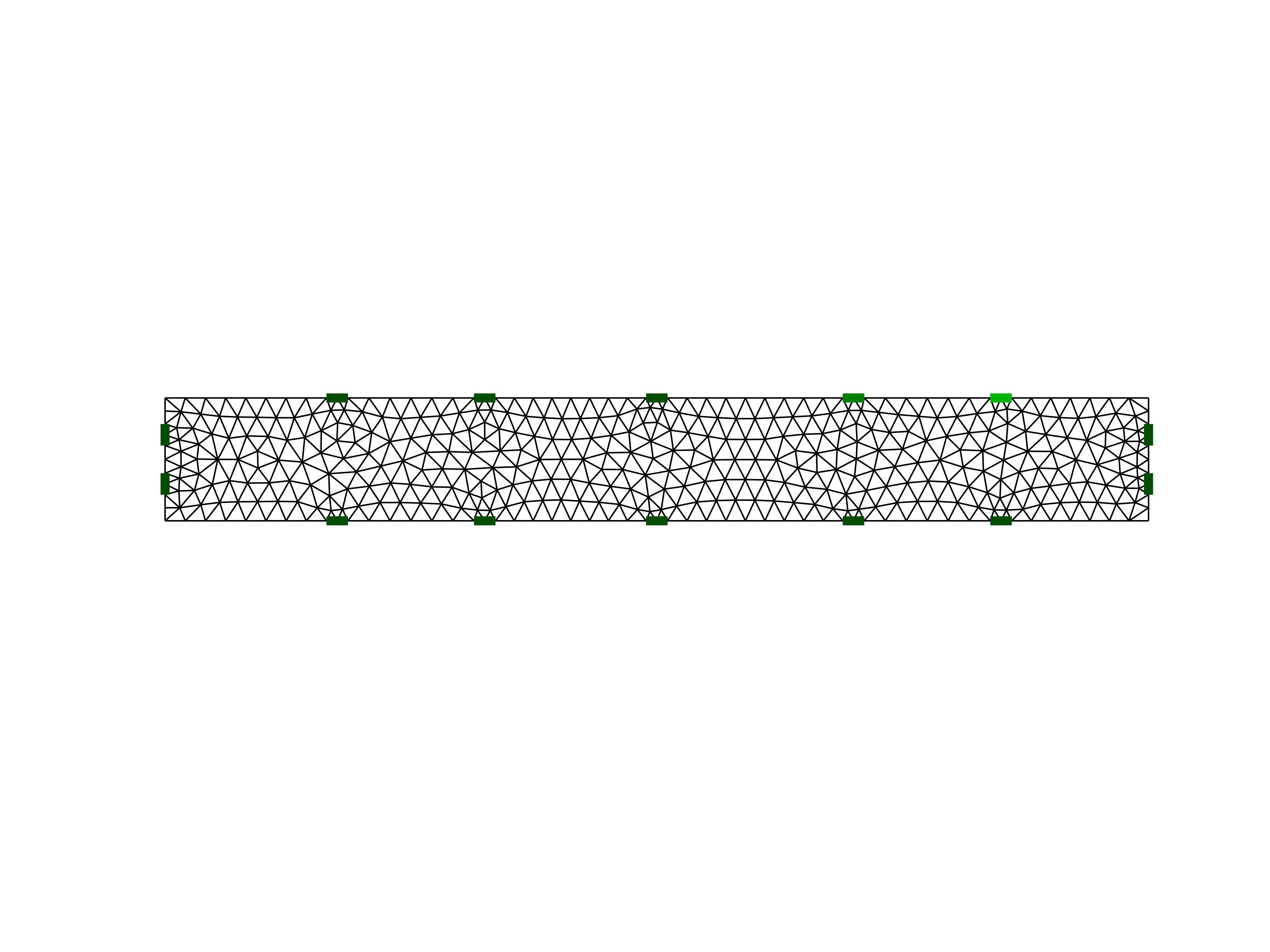} & \\
\textbf{(b)} & \includegraphics[width=55mm,trim=7cm 3cm 4cm 3cm, clip=true]{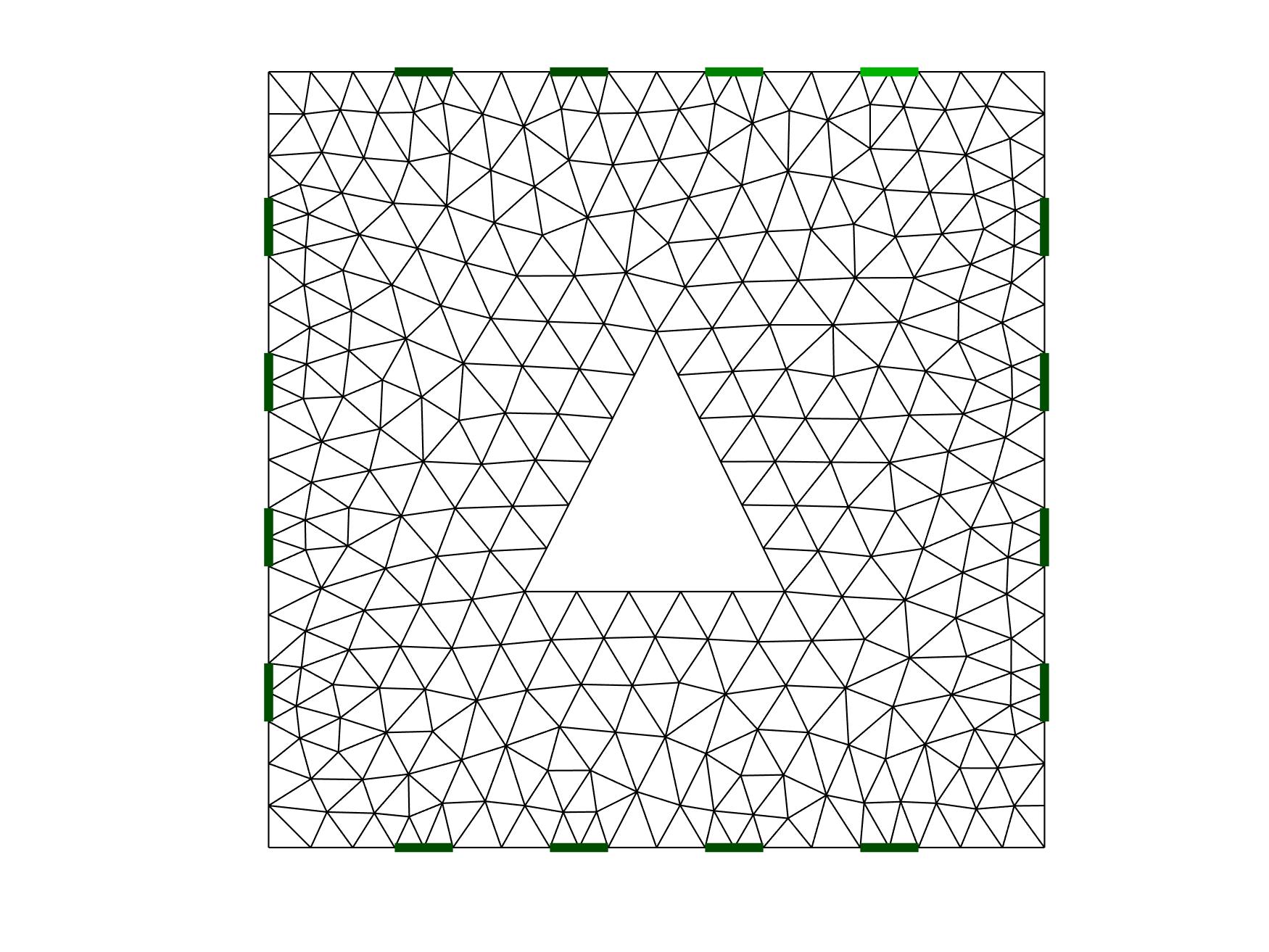} & \textbf{(c)} & \includegraphics[width=55mm,trim=7cm 3cm 4cm 3cm, clip=true]{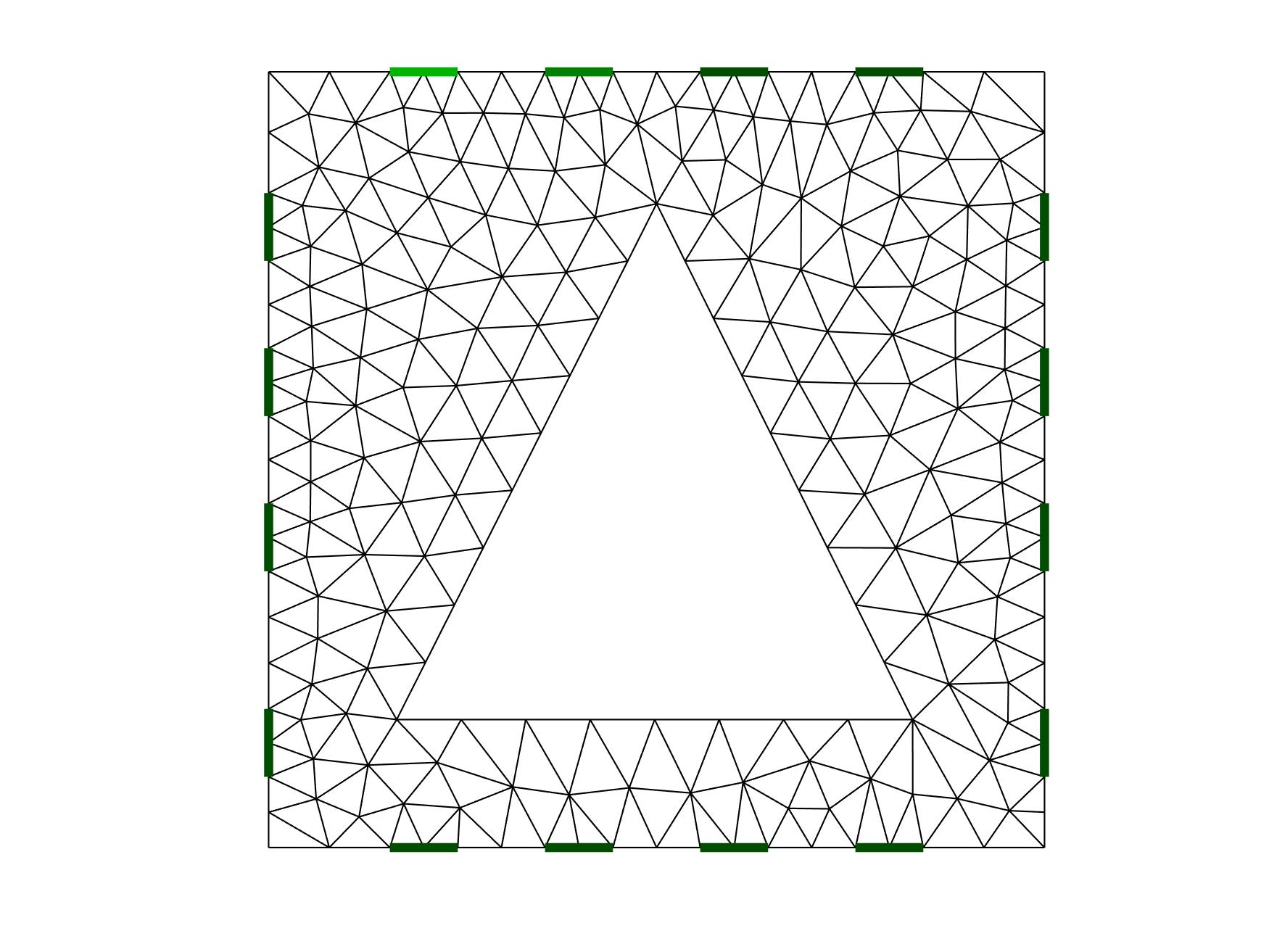} \\
\textbf{(d)} & \includegraphics[width=55mm,trim=7cm 3cm 4cm 3cm, clip=true]{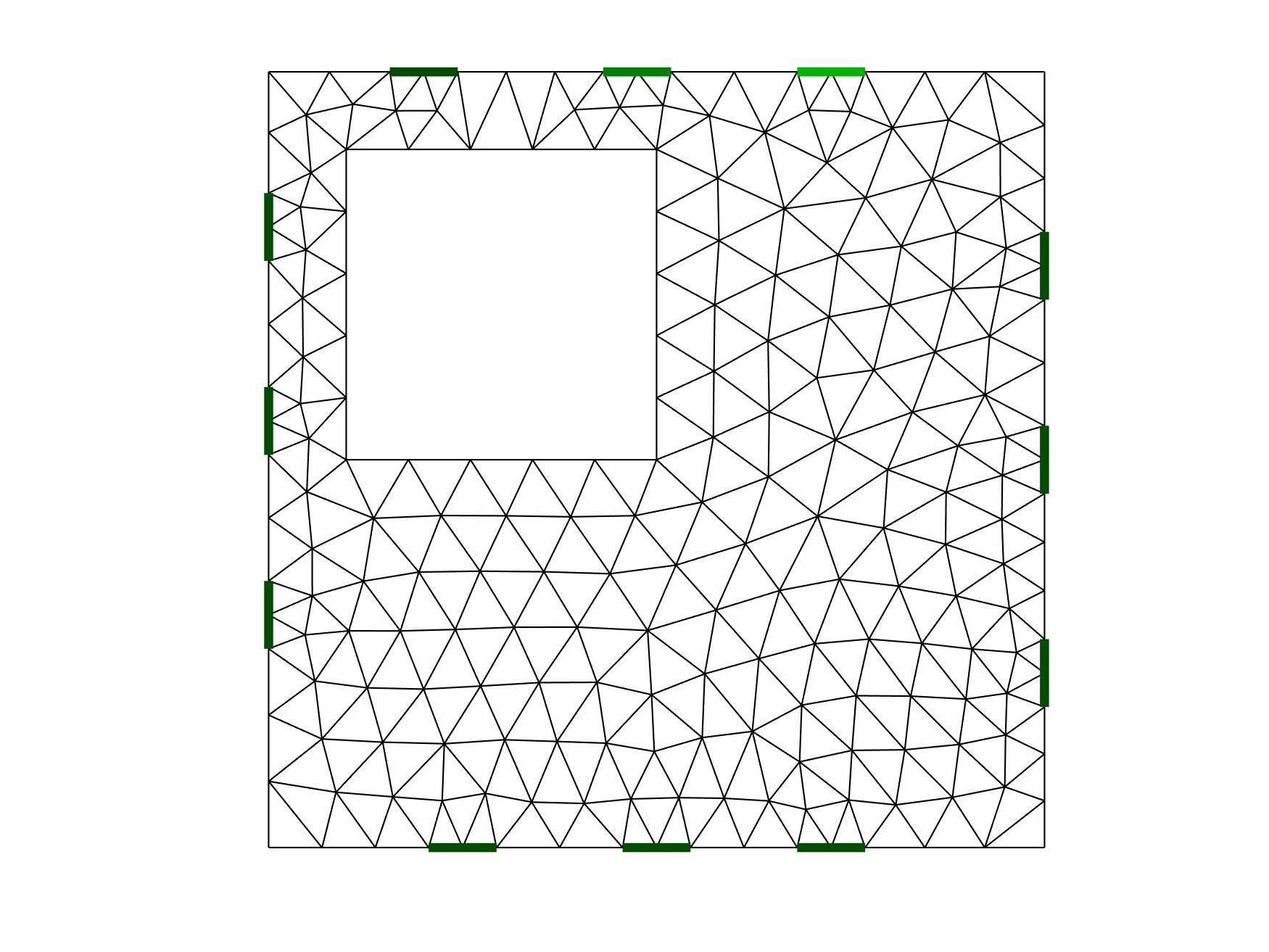} & \textbf{(e)} & \includegraphics[width=55mm,trim=7cm 3cm 4cm 3cm, clip=true]{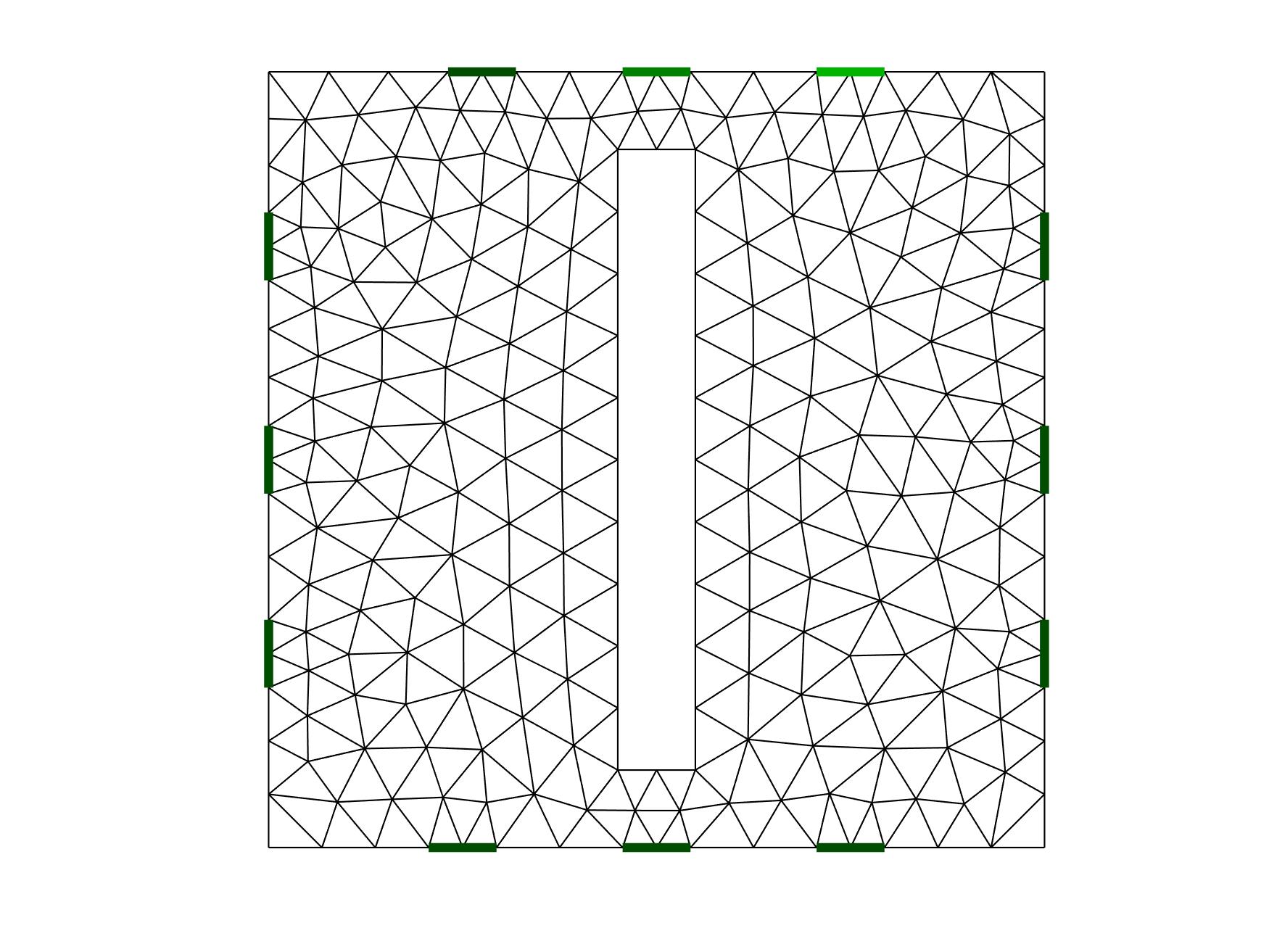} \\
\end{tabular}
\caption{Examples of optimized electrode positions using the deep learning based approach for various geometries.}
\label{odd}
\end{figure}

We began by investigating this query by considering rectangular cases where the aspect ratio $\gamma= \frac{\mathrm{width}}{\mathrm{height}}$ is large and found the deep learning approach highly sensitive to the value of the hyperparameter for $\gamma \gtrapprox 10$ due to the large amount of possible electrode location possibilities along the geometry's width.
Generally speaking, however, the number of electrode position possibilities truly stems from (a) the ratio of the perimeter to homogeneous element width and (b) the minimum/maximum element size.
In addition to this, when the aspect ratio becomes arbitrarily large and the geometry's height is arbitrarily small, the physics of the problem breaks down since the 2D geometry approaches a line.
Nonetheless, sensible optimized solutions are still attained up to around $\gamma = 8$ using the same hyperparameterr adopted throughout this work -- an example for $\gamma = 8$ is shown in Fig. \ref{odd}(a).

It is intuitively interesting to investigate optimal solutions for complex geometries, in particular those with geometrical discontinuities.
Take for example a square geometry with $k=16$ electrodes and an equilateral triangular hole in the center as shown in Fig. \ref{odd}(b).
Obviously, when the length of triangle's bottom side is equal to the square's width, we develop a circuit short in the current injection protocol and electrode measurements are meaningless.
On the other hand, when the triangle is arbitrarily small, the optimized electrode positions are the same as the unaltered square geometry.
But, optimized solutions for the intermediate range are certainly worth inquiry and are shown Fig. \ref{odd}(b,c).
Comparing the optimized positions in the triangular hole examples, we notice that the only significant difference in the electrode positions is a downward shift in the furthermost bottom side electrodes, which provides compensating information when the bottom side of the triangle approaches the square sides.
All in all, it is a bit counter-intuitive that such a large change in the hole size resulted in only a small change in the optimized positions -- on the other hand, the EIT forward model has a low sensitivity to changes far from the boundary.

Continuing this train of thought, we examine another square geometry with $k=12$ electrodes where a square hole is placed near the top left corner boundary.
We observe the optimized configuration for this geometry in Fig. \ref{odd}(d) and note that the electrodes on the top and left sides are localized near the hole, while the opposite side electrodes are spaced rather uniformly. 
This configuration is consistent with the previous statement regarding the sensitivity of the EIT forward model and demonstrates the tendency of electrodes to localize near non-conductive objects close to the boundary in optimized configurations.

The former example also begins to illuminate the effects of partial domain segmentation (i.e., splitting of different regions in the geometry).
Indeed, one could deduce that the L-shape in the top left-hand corner is locally segmented from the rest of the domain and that EIT information is improved when electrodes are localized near the segmented area.
However, what if the entire square domain is partially segmented?
We investigate this query in Fig. \ref{odd}(e) and immediately notice that the top and bottom side electrodes in the optimal configuration have again localized near the rectangular hole approaching the top and bottom boundary while the left and right sides, far from the hole, are roughly uniformly spaced.
This result again demonstrates that near-boundary effects have a large impact on the local positioning of the optimized electrodes and a small effect on electrodes far away.


\subsection{ {Reconstructions and distinguishability}}
 {Ultimately, a primary aim of optimizing electrode positions is to improve the quality of EIT reconstructions.
In this subsection we compare EIT reconstructions using optimized and non-optimized electrode configurations.
While there are numerous EIT reconstruction approaches, we utilize a least squares-based approach minimizing the following cost function
}

 {
\begin{equation}
\Psi( \sigma > 0 ) = ||L_n ( V_s - U(\sigma) )||^2 + R(\sigma)
\end{equation}
}

\noindent  {where $L_n $ is the Cholesky factor of the noise precision matrix $\Gamma_n^{-1}$ (i.e. $L_n^T L_n = \Gamma_n^{-1}$), $V_s$ are noisy simulated voltage measurements, $ R(\sigma) = ||L_\sigma (\sigma - \sigma_\mathrm{hom})||^2$ is a smoothness-promoting regularization term using the Cholesky factor $L_\sigma$ of the inverted covariance matrix described in Eq. \ref{PrCov}, and $\sigma_\mathrm{hom}$ is the best homogeneous estimate computed using $\sigma_\mathrm{hom} = \min ||L_n (V_s - U(\sigma))||^2$.
In solving the EIT problems, we utilize as Gauss-Newton optimization regime with a line-search and barrier functions to handle the constraints.
We would like to comment that the regime used herein is often referred to as and ``absolute imaging" approach, where we aim to reconstruct the ``absolute" values of $\sigma$ \cite{liu2015nonlinear}.}

\begin{figure}[h!]
\centering
\begin{tabular}{m{0.1cm} m{5.0cm} m{5.0cm}}
\textbf{(a)} & \includegraphics[width=55mm,trim=4cm 10cm 4cm 10cm, clip=true]{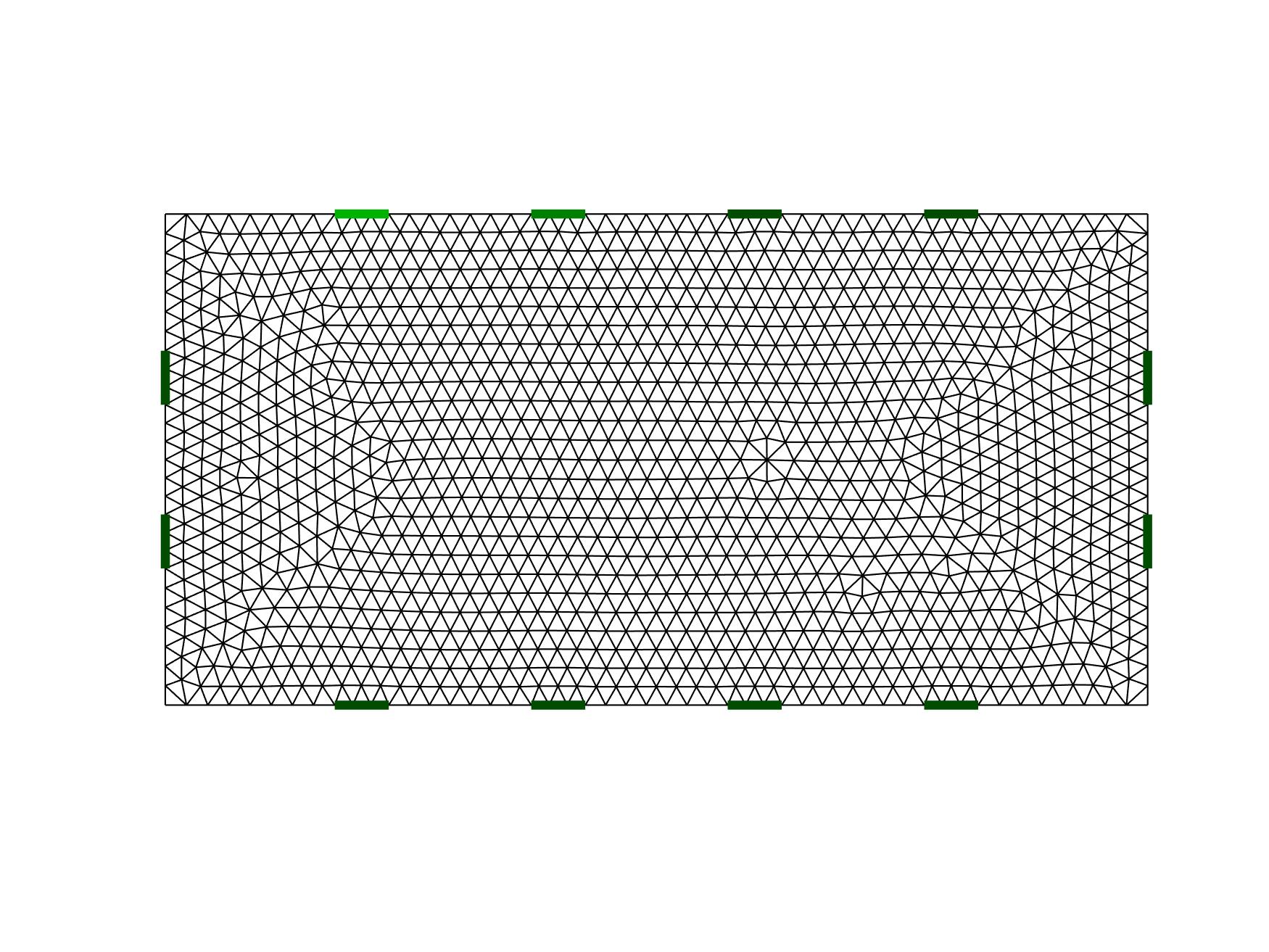} & \includegraphics[width=55mm,trim=4cm 10cm 4cm 10cm, clip=true]{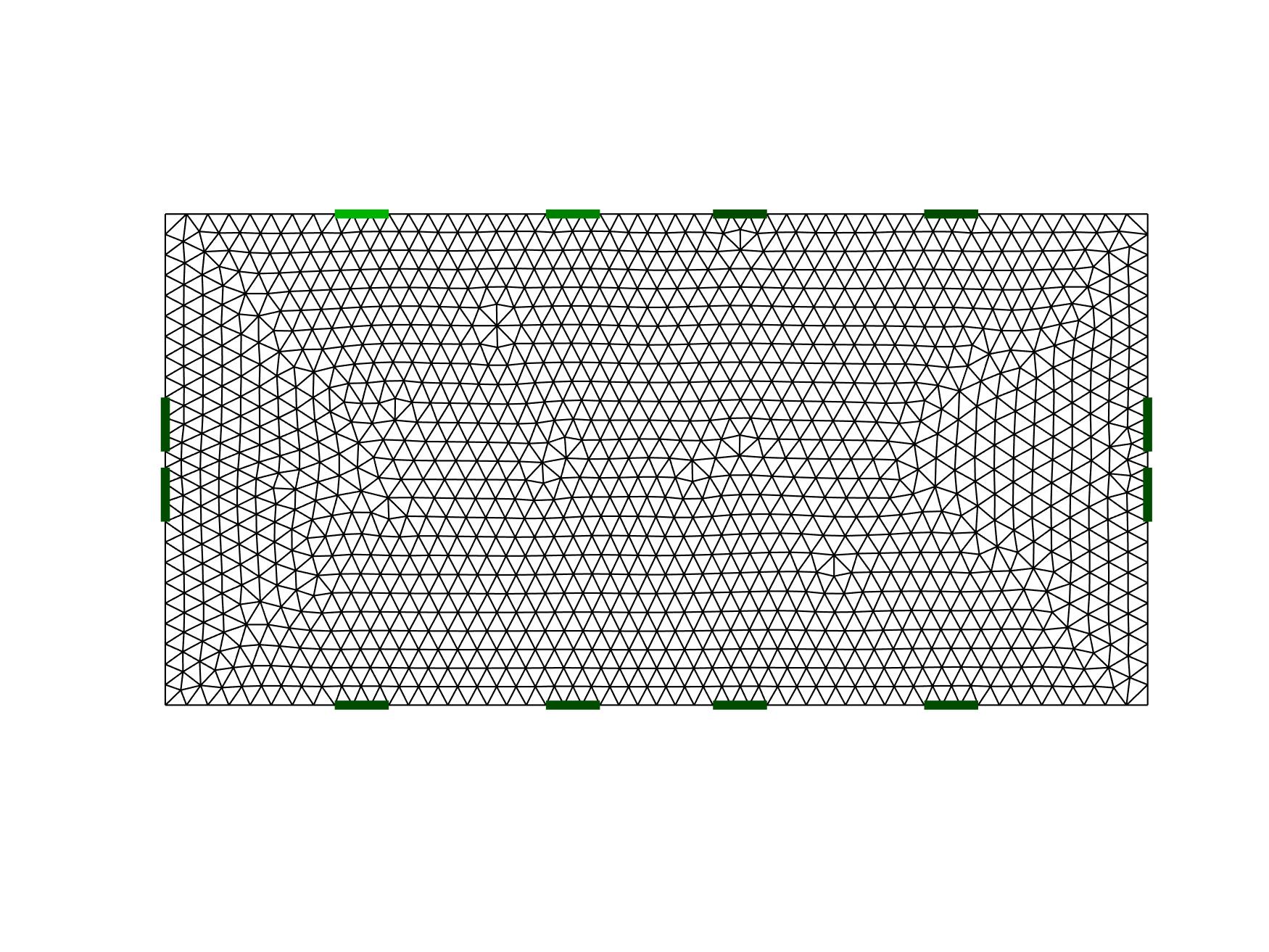} \\
\textbf{(b)} & \includegraphics[width=55mm,trim=4cm 10cm 4cm 10cm, clip=true]{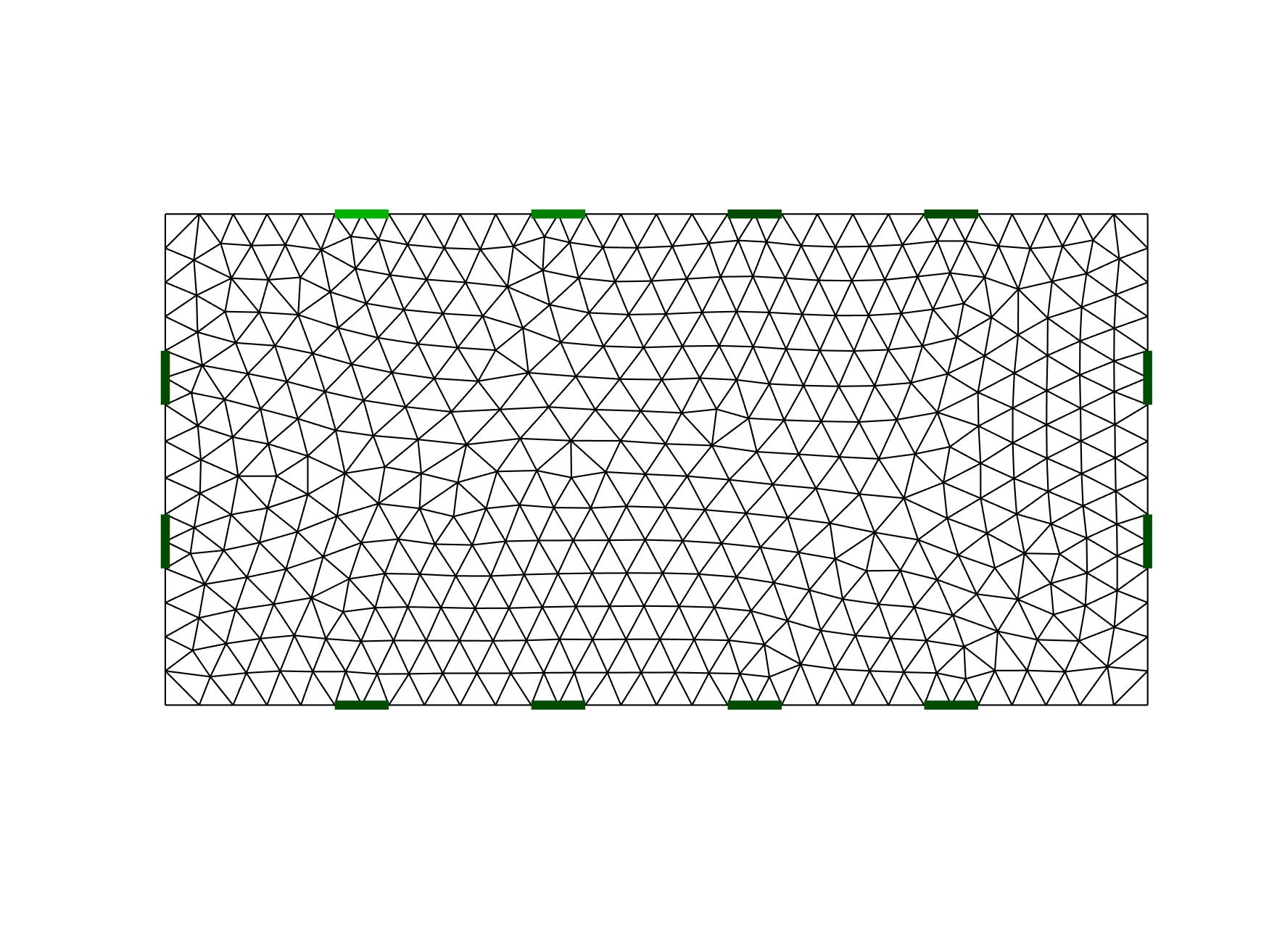} & \includegraphics[width=55mm,trim=4cm 10cm 4cm 10cm, clip=true]{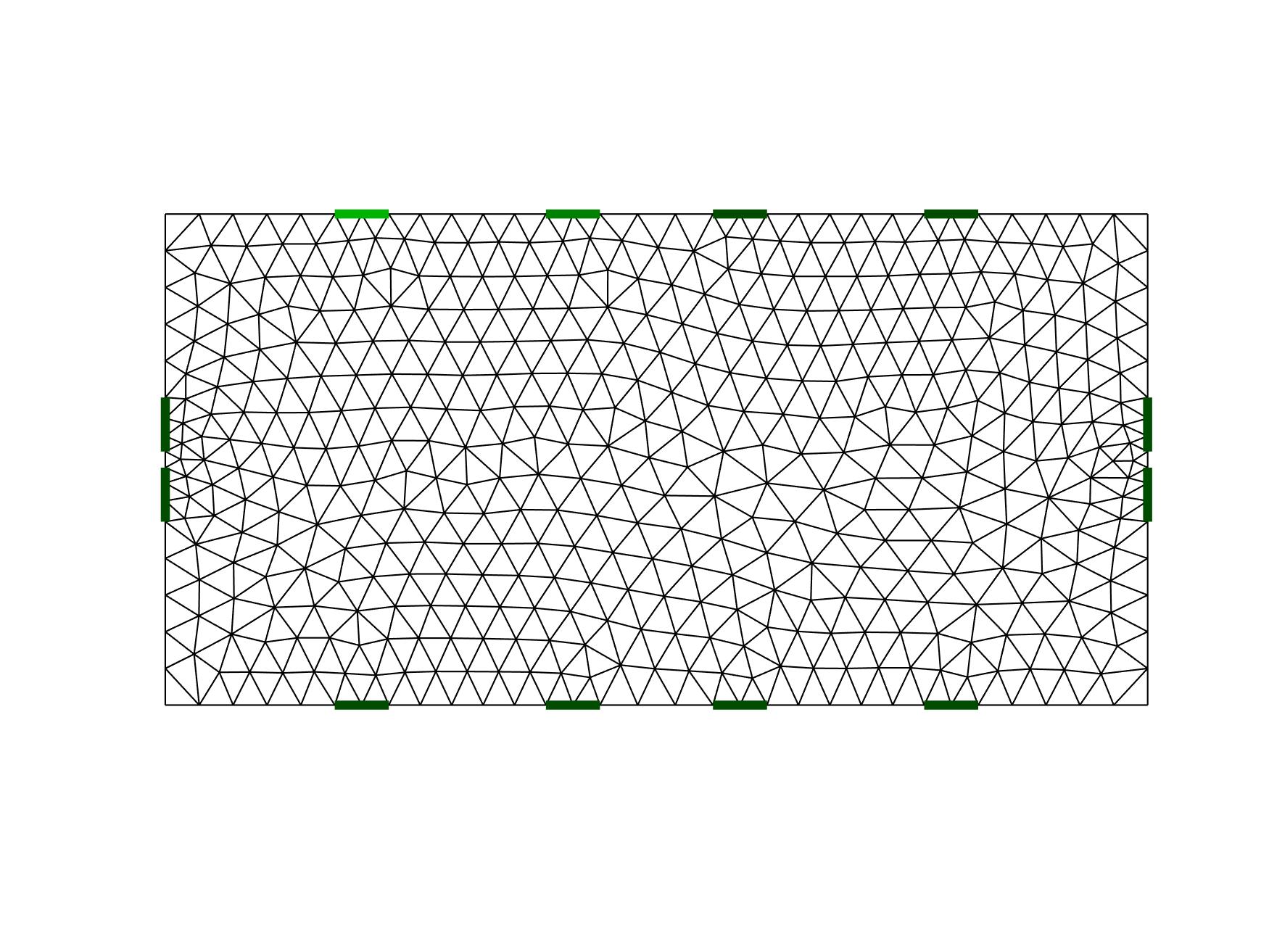} \\
\end{tabular}
\caption{ {Fine (a) and coarse (b) meshes used for EIT reconstructions to demonstrate the potential effectiveness of optimized electrode positions. In the left column, the electrode layout is a ``standard" layout and in the right column the electrode layout was optimized using the approach presented in this work.}}
\label{MESHRecon}
\end{figure}

 {To demonstrate the potential effectiveness of the optimized electrode positions in improving EIT reconstruction quality relative to a ``standard" layout, we test EIT on the rectangular geometries (and electrode layouts) described in section \ref{rectGeo}; however, we have scaled the geometries from centimeters to meters using 1:1 scaling.
To simulate the EIT data, we utilize the same stimulation and measurement pattern provided in example's description.
In generating the data, we consider two cases, (i) a randomized blob-like distribution and (ii) an ellipsoidal inclusion.
Using the simulated voltage data, we then add Gaussian noise $\eta$ with 1\%, 5\%, and 10\% standard deviation to the measurements.
In order to avoid an inverse crime, we use use different data simulation and inversion meshes.
For the fine simulation meshes, we generate meshes with maximum element dimension of 0.04m.
For the coarse inversion meshes, we generate meshes with a maximum element dimension of 0.075m, all meshes are shown in Fig. \ref{MESHRecon}.
}

\begin{figure}[h!]
\centering
\begin{tabular}{m{0.5cm} m{52mm} m{2mm}}
& \includegraphics[width=65mm,trim=3cm 13cm 3cm 10cm, clip=true]{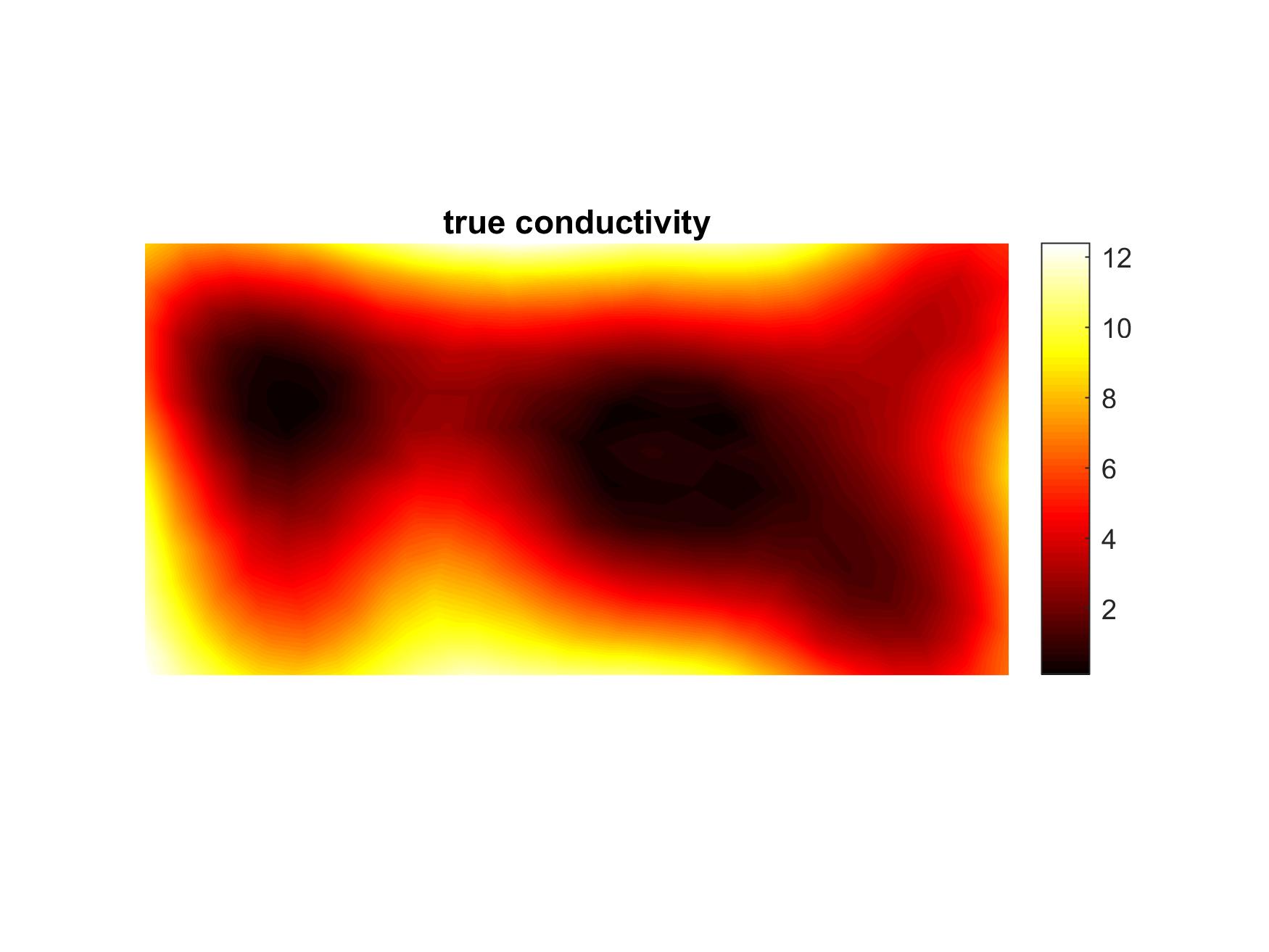} & ~~~~~~~[mS/cm] \\
\end{tabular}
\begin{tabular}{m{0.65cm} m{5.0cm} m{5.0cm}}
\textbf{$\eta$=1\%} & \includegraphics[width=50mm,trim=3cm 13cm 11.5cm 10cm, clip=true]{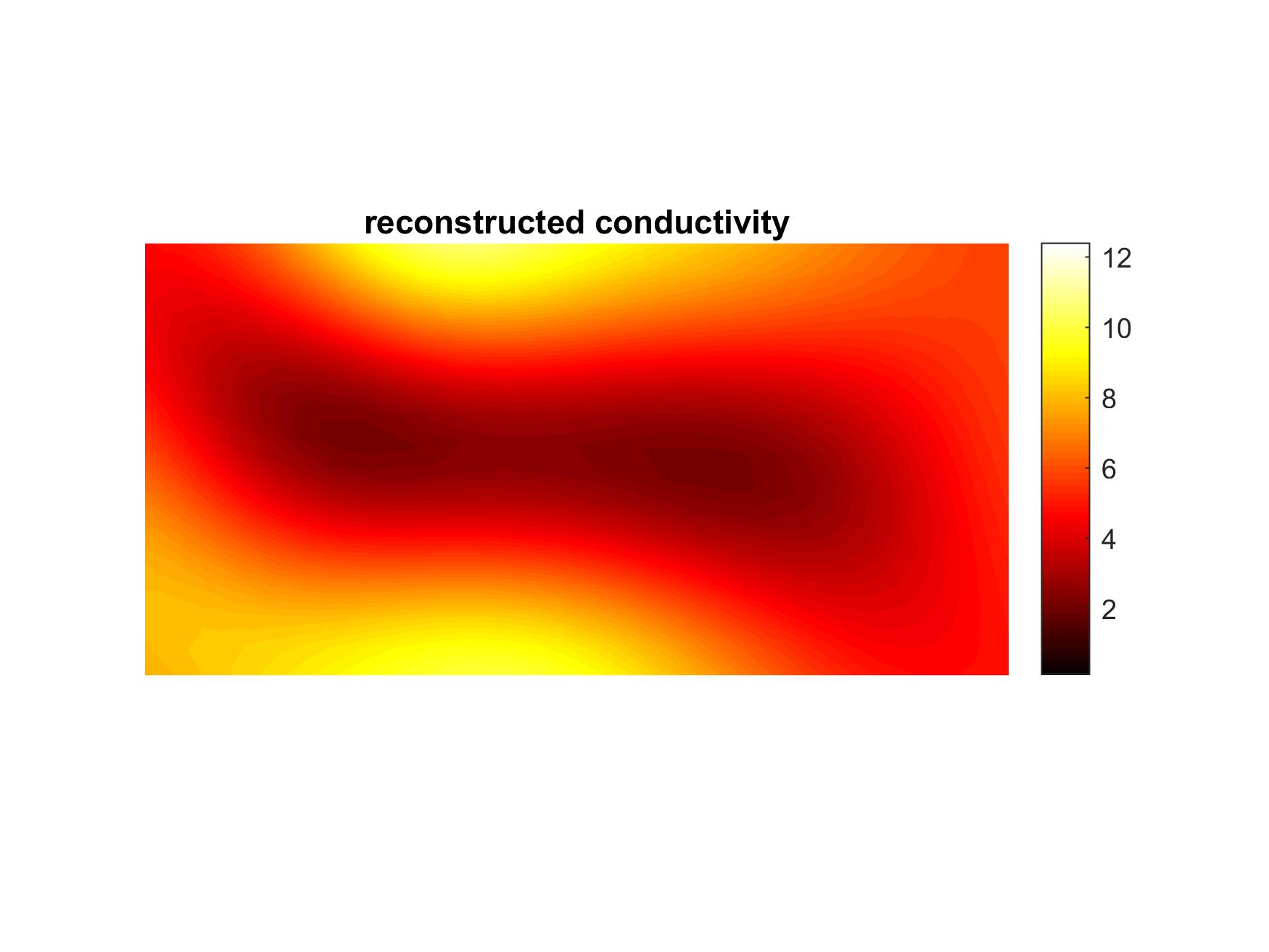} & \includegraphics[width=50mm,trim=3cm 13cm 11.5cm 10cm, clip=true]{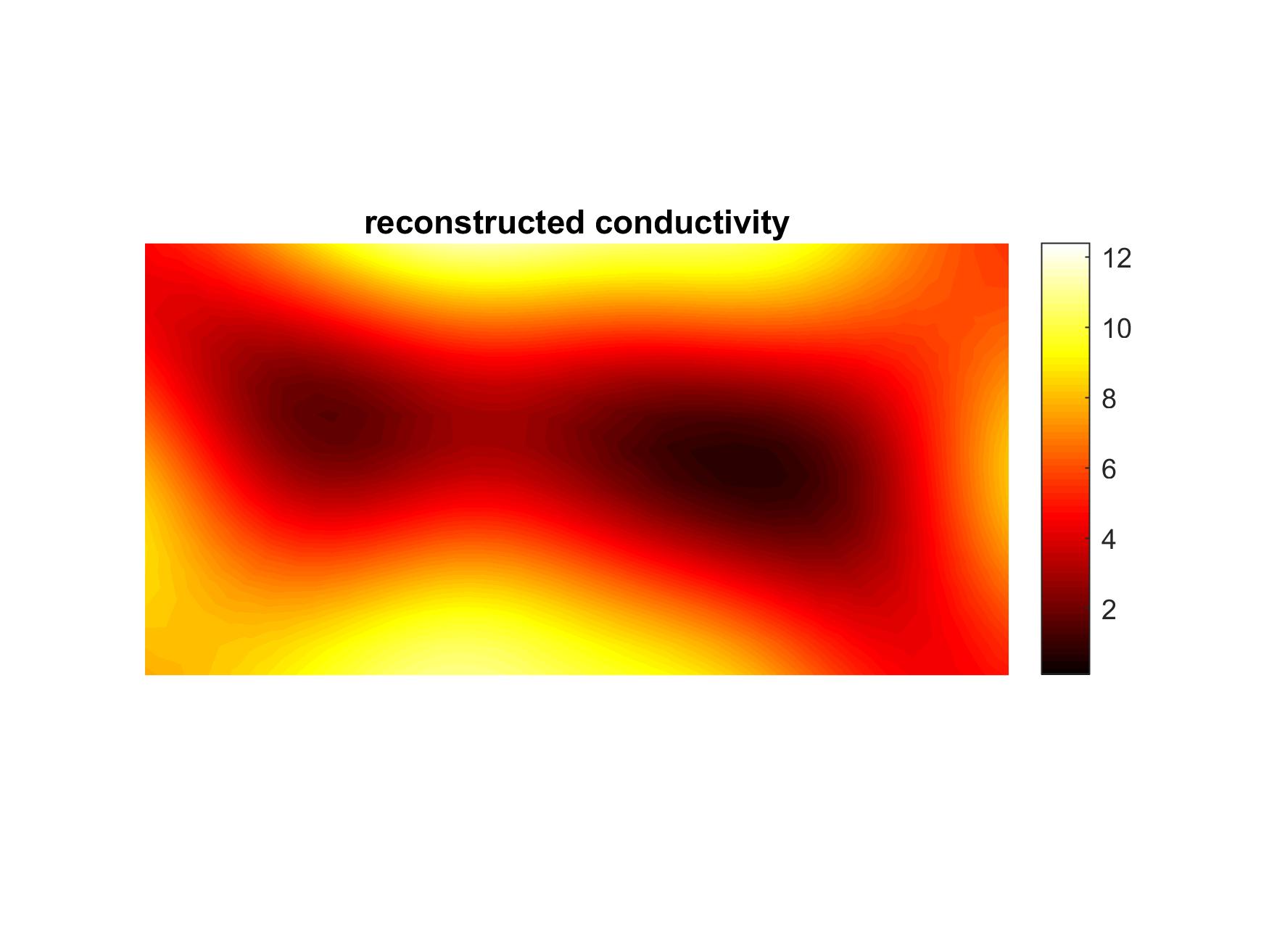} \\
\textbf{$\eta$=5\%} & \includegraphics[width=50mm,trim=3cm 13cm 11.5cm 10cm, clip=true]{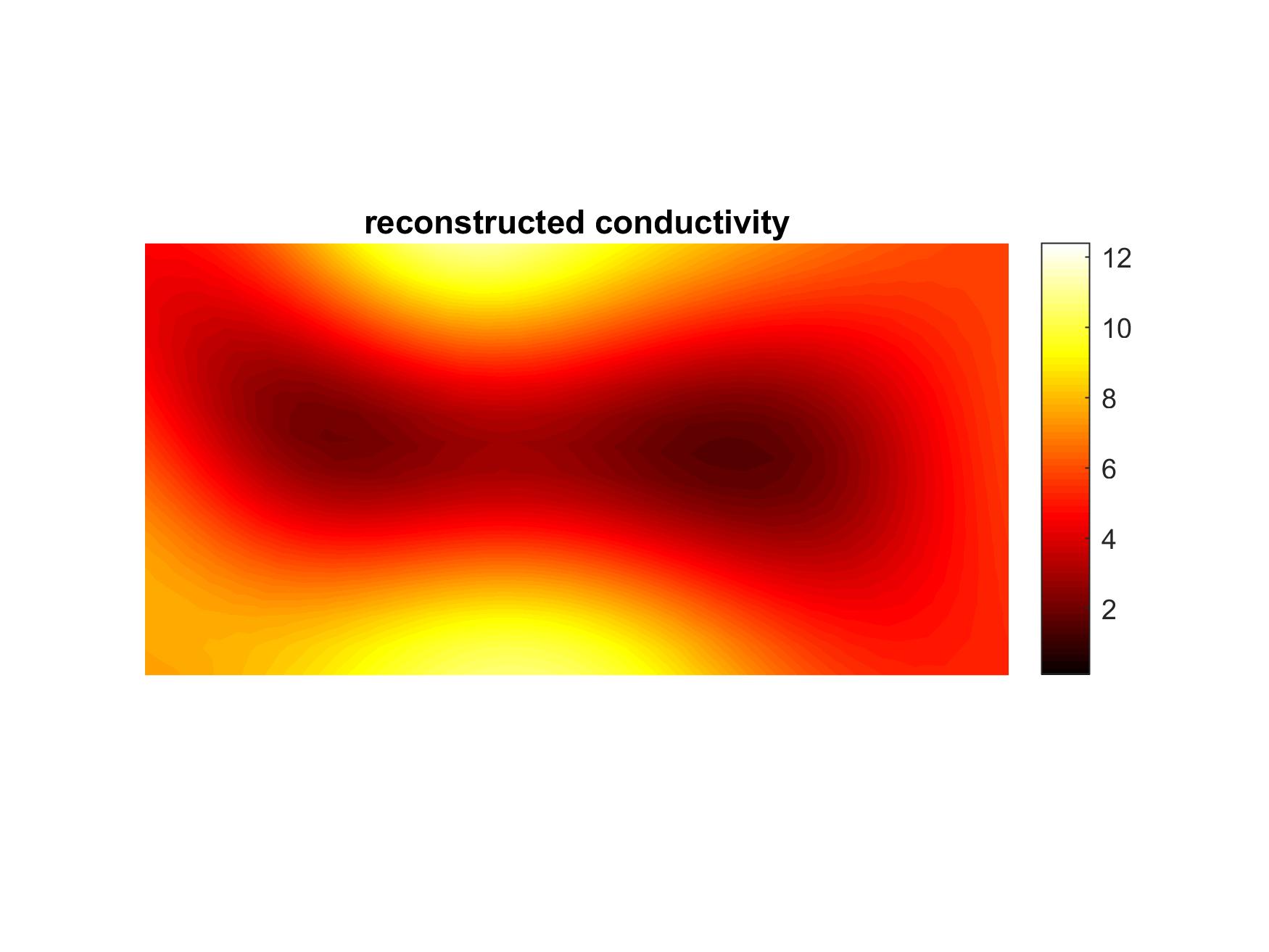} & \includegraphics[width=50mm,trim=3cm 13cm 11.5cm 10cm, clip=true]{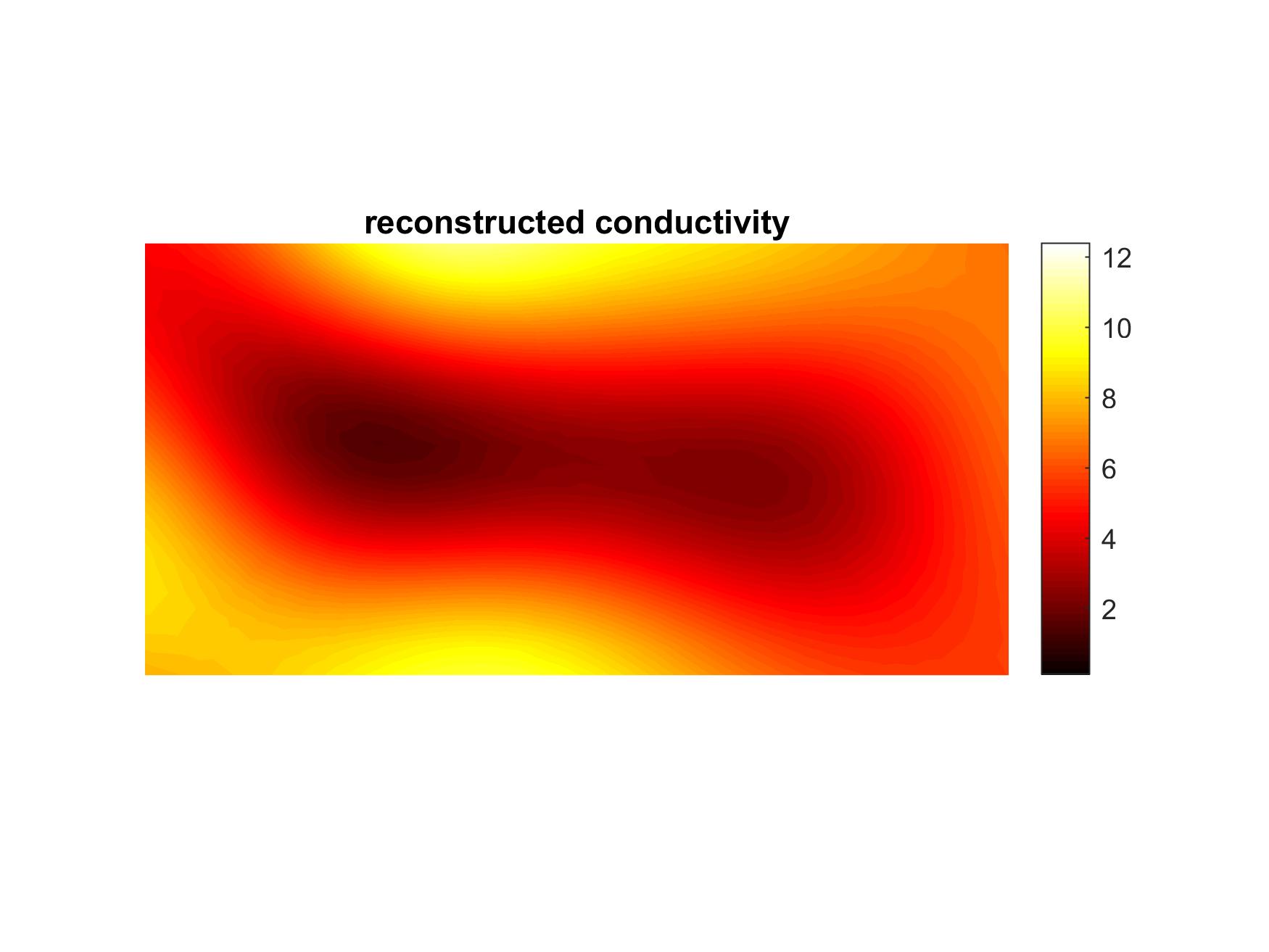} \\
\textbf{$\eta$=10\%} & \includegraphics[width=50mm,trim=3cm 13cm 11.5cm 10cm, clip=true]{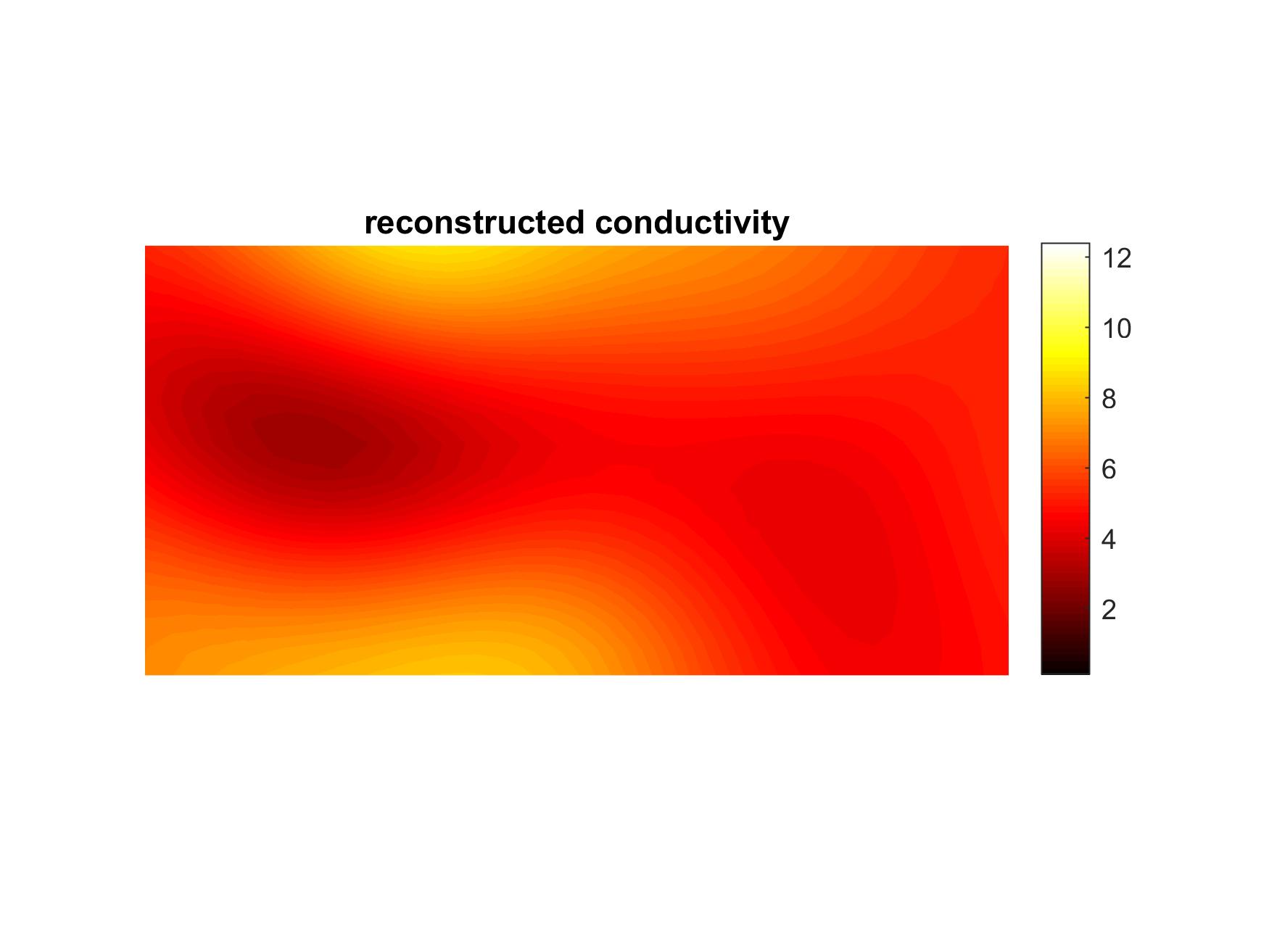} & \includegraphics[width=50mm,trim=3cm 13cm 11.5cm 10cm, clip=true]{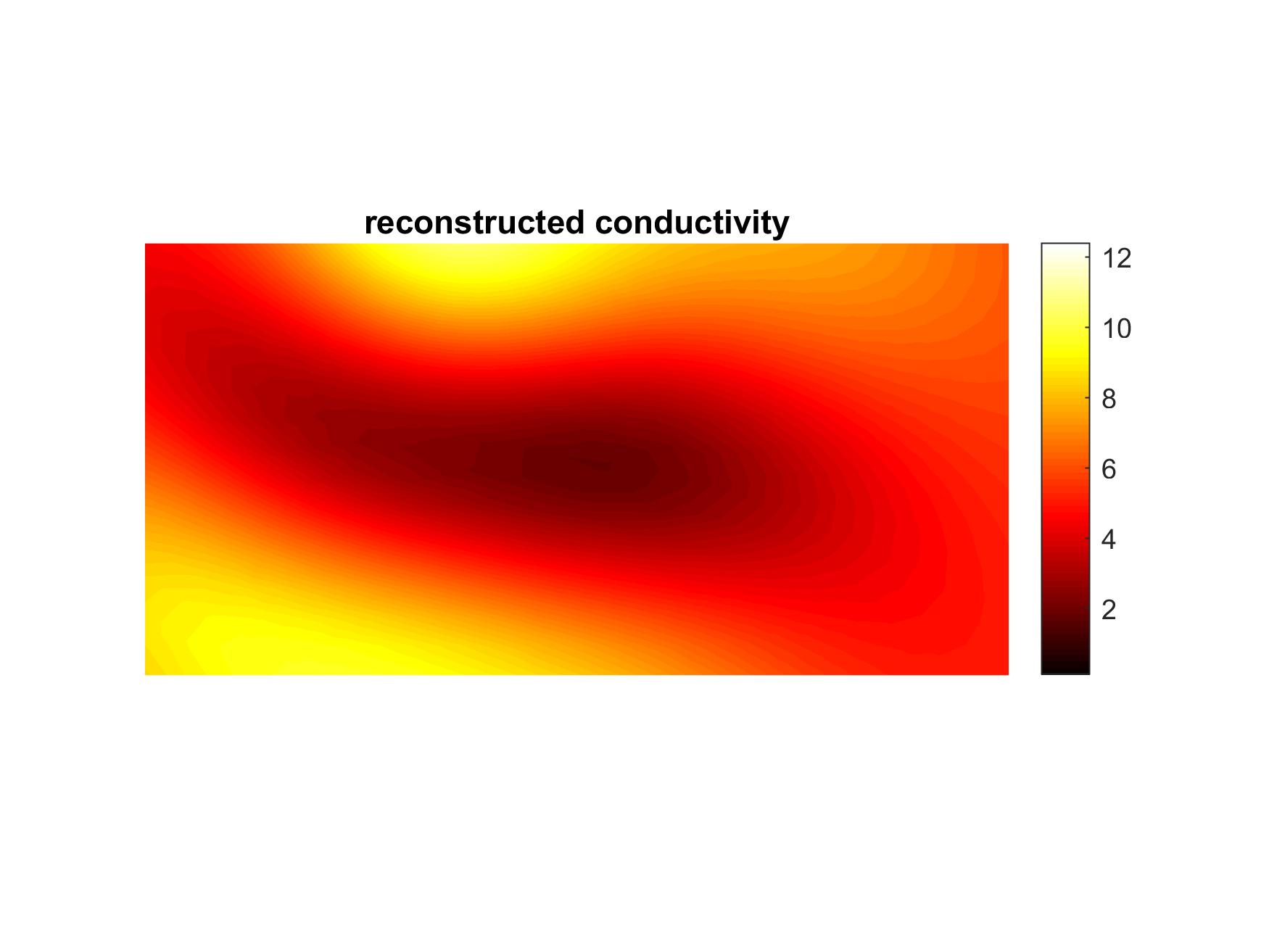} \\
\end{tabular}
\begin{tabular}{m{0.5cm} m{52mm} m{2mm}}
& \includegraphics[width=65mm,trim=3cm 13cm 3cm 10cm, clip=true]{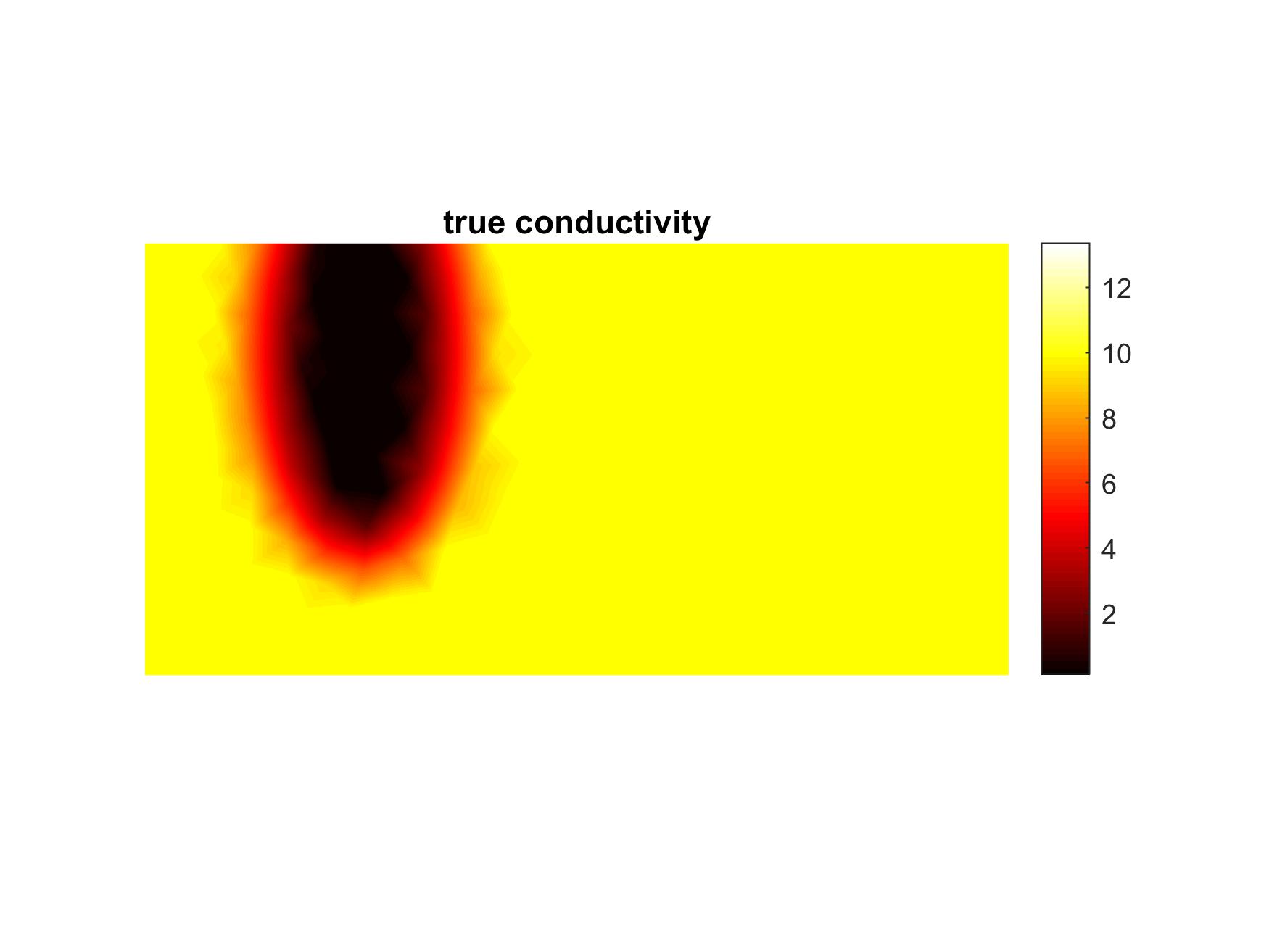}& ~~~~~~~[mS/cm]  \\
\end{tabular}
\begin{tabular}{m{0.65cm} m{5.0cm} m{5.0cm}}
\textbf{$\eta$=1\%} & \includegraphics[width=50mm,trim=3cm 13cm 11.5cm 10cm, clip=true]{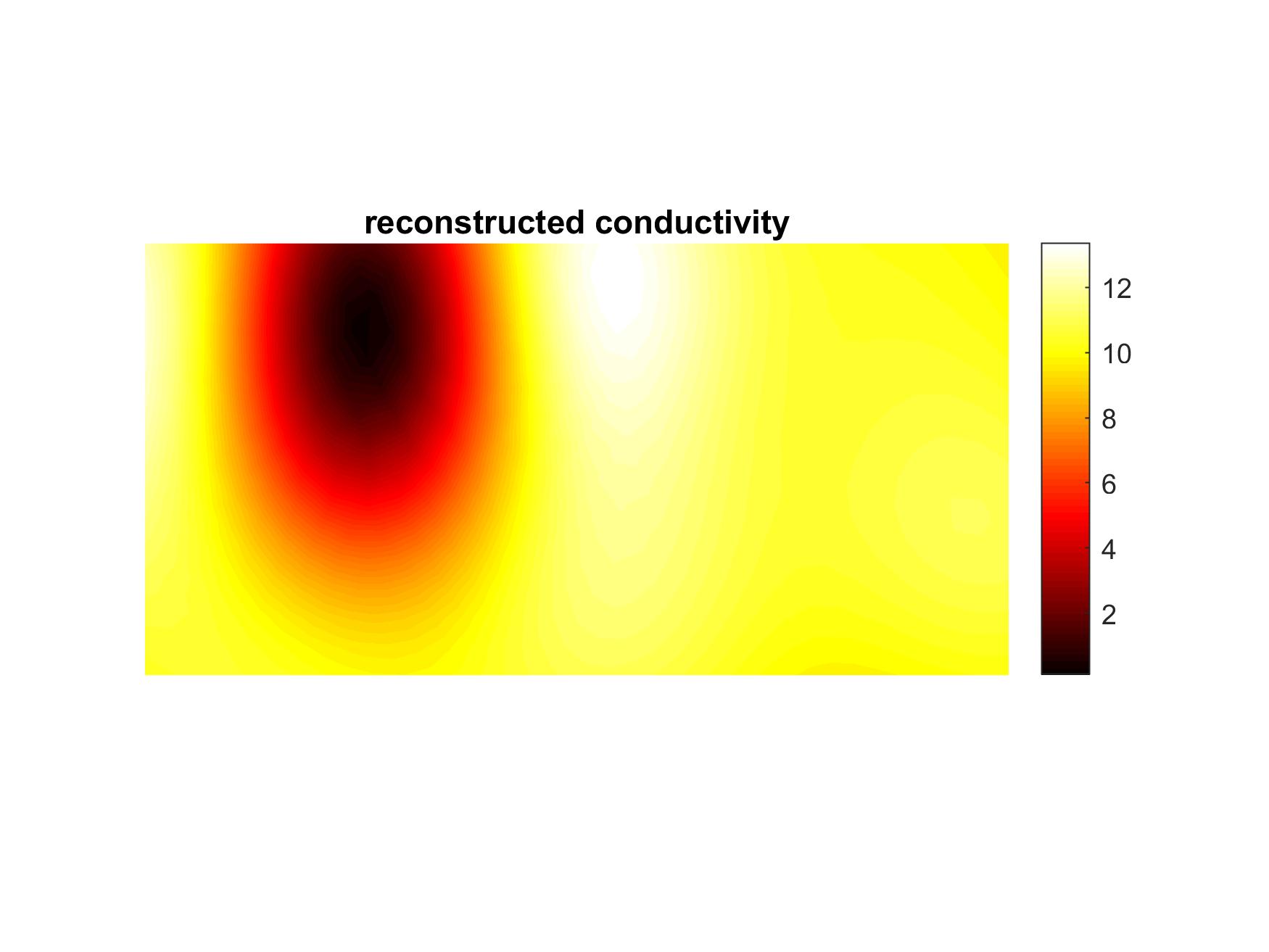} & \includegraphics[width=50mm,trim=3cm 13cm 11.5cm 10cm, clip=true]{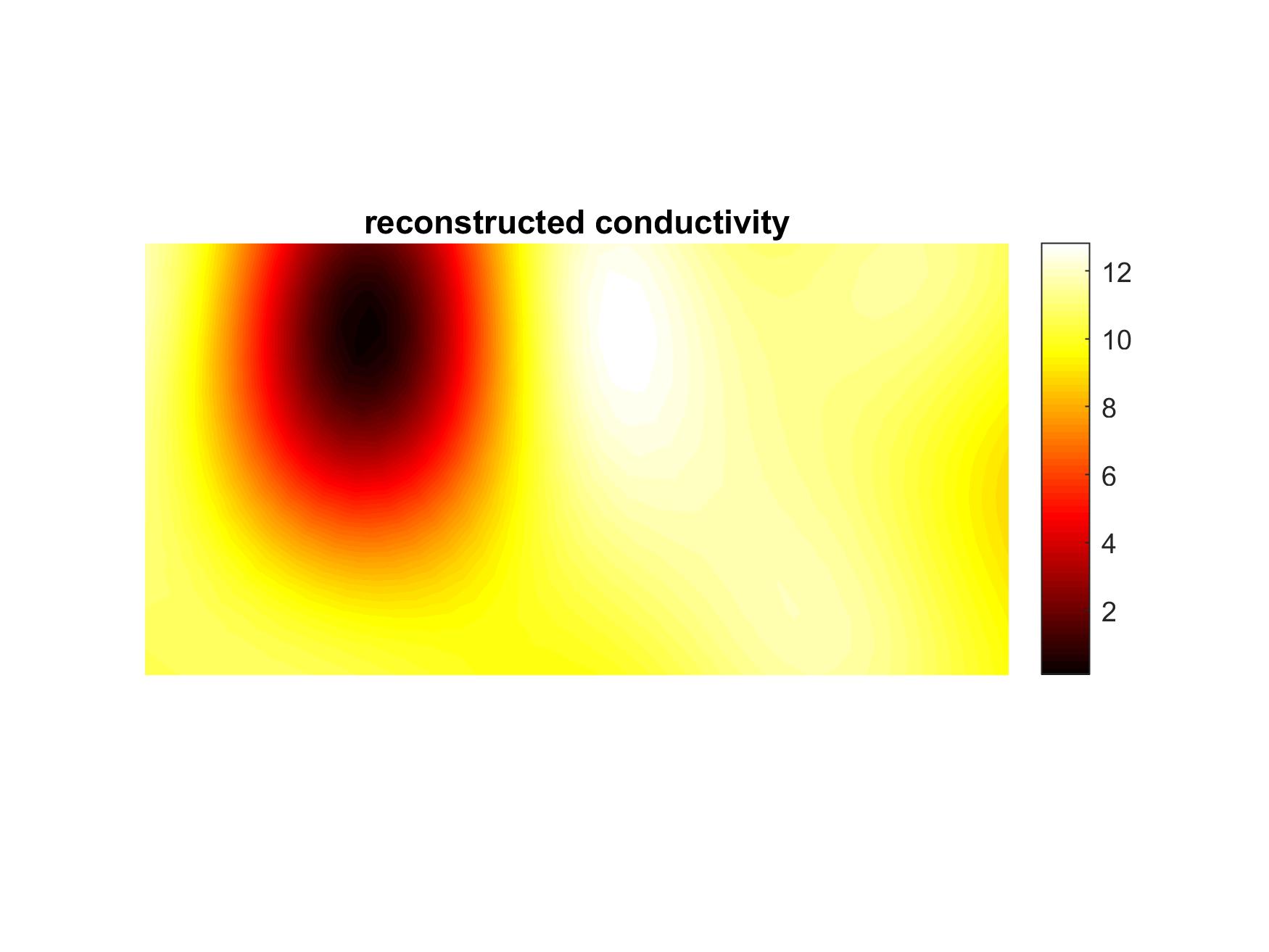} \\
\textbf{$\eta$=5\%} & \includegraphics[width=50mm,trim=3cm 13cm 11.5cm 10cm, clip=true]{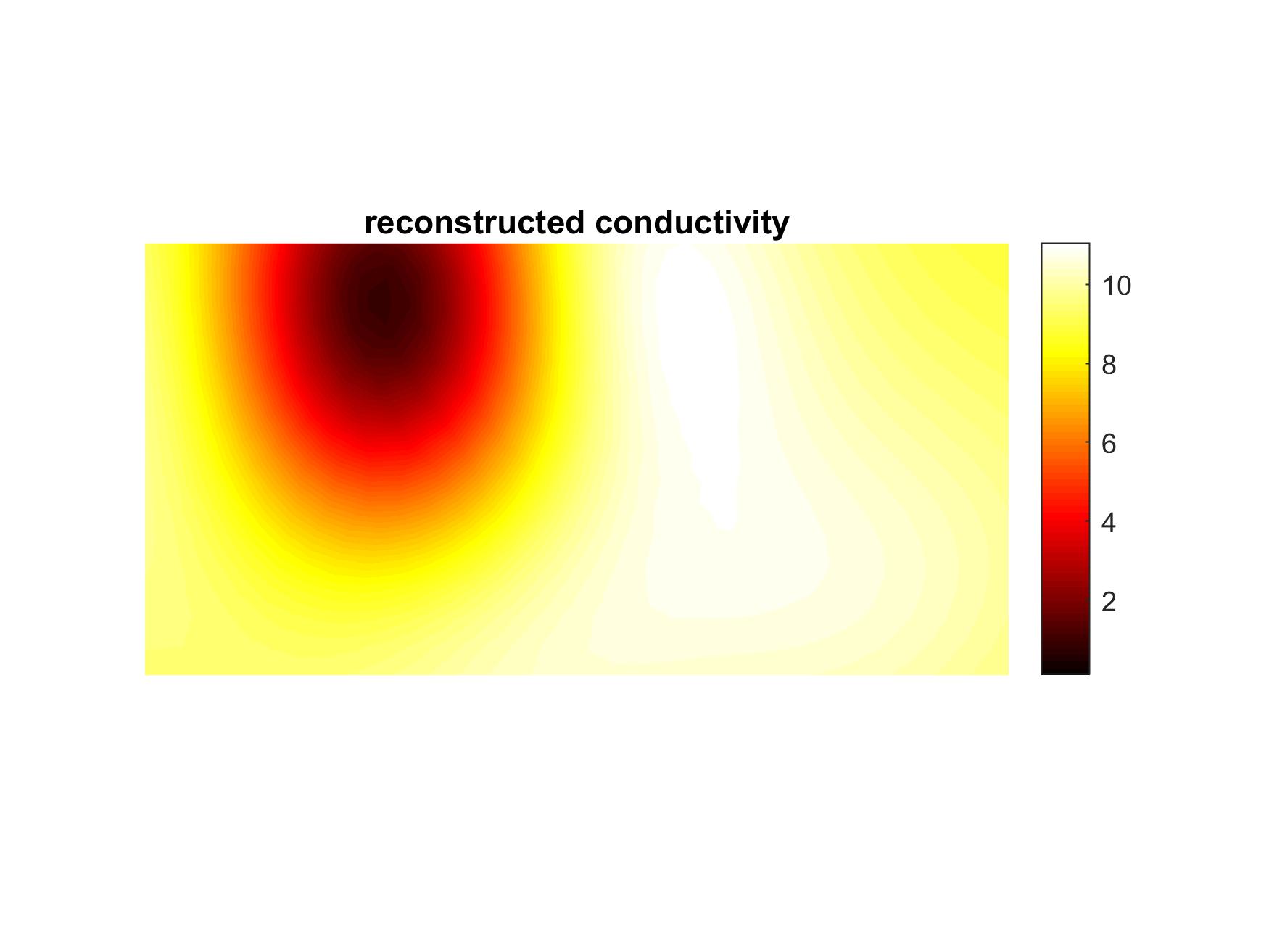} & \includegraphics[width=50mm,trim=3cm 13cm 11.5cm 10cm, clip=true]{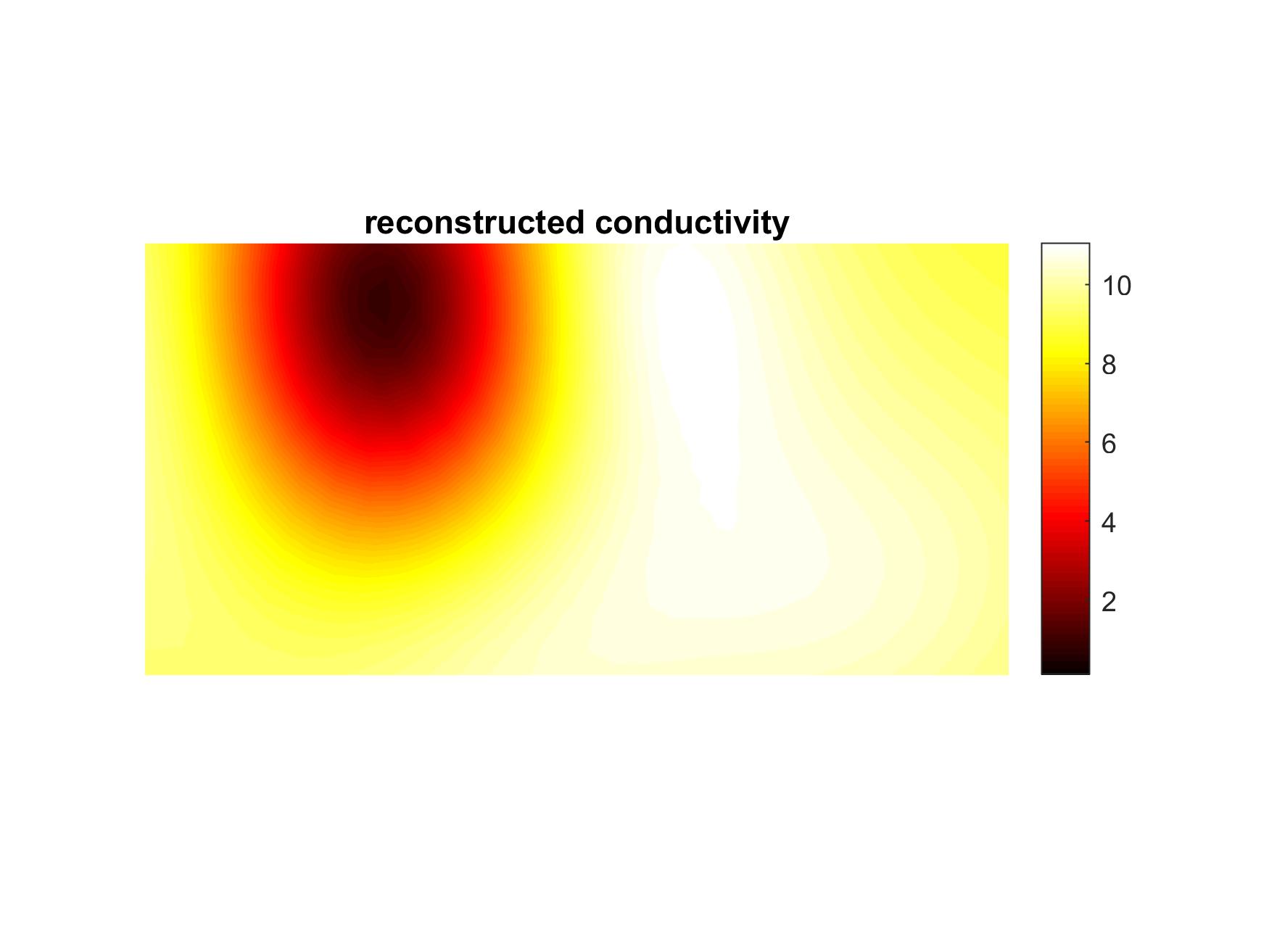} \\
\textbf{$\eta$=10\%} & \includegraphics[width=50mm,trim=3cm 13cm 11.5cm 10cm, clip=true]{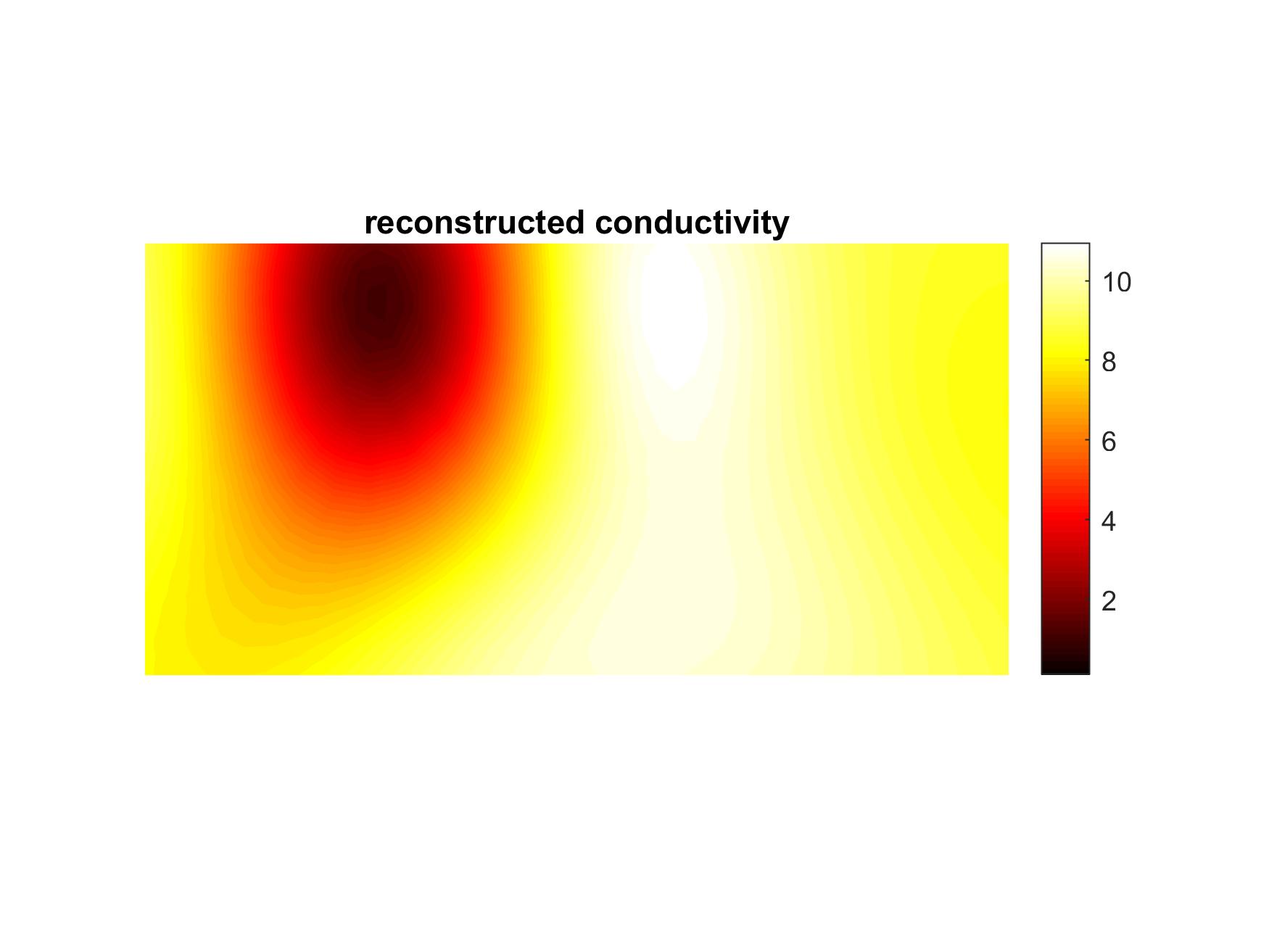} & \includegraphics[width=50mm,trim=3cm 13cm 11.5cm 10cm, clip=true]{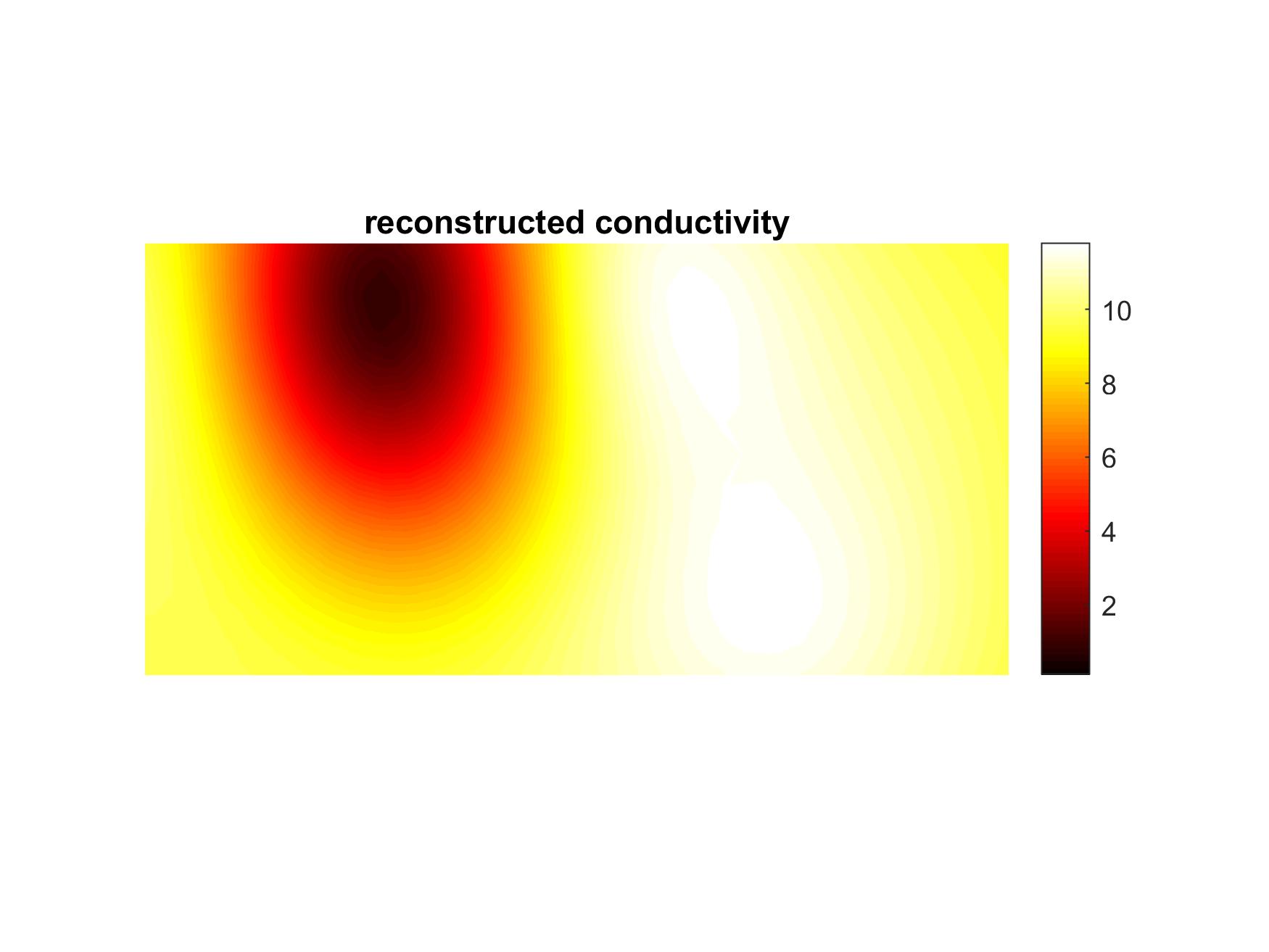} \\
& ~~~~~~~~~\textbf{``Standard" Layout} & ~~~~~~~~~~\textbf{Optimized Layout} \\
\end{tabular}
\caption{ {EIT reconstructions of a blob-like distribution (top half) and an ellipsoidal inclusion (bottom half) using a ``standard" electrode layout (left column) and an optimized layout (right column) considering Gaussian noise $\eta$ with 1\%, 5\%, and 10\% standard deviation added. True distributions are shown as a single row entry and reconstructions are reported in S/m on the same color bar for a given target distribution.}}
\label{reconz}
\end{figure}

 {The EIT reconstructions performed using the optimized and non-optimized electrode layouts are shown in Fig. \ref{reconz}, alongside the respective true conductivity distributions.
Beginning with the top half of Fig. \ref{reconz}, we immediately observe visual improvements in the reconstructions using the optimized electrode layout.
The relative improvements appear to be more distinguishable as noise increases.
To quantify these potential improvements, Table \ref{recontable} provides the root mean square errors (RMSEs) for all reconstructions reported herein, where RMSE = $\sqrt{\sum_{\mathrm{q}=1}^T (\hat{\sigma}_\mathrm{q,true} - \sigma_\mathrm{q})^2/T}$, $\hat{\sigma}_\mathrm{true}$ are the true values interpolated onto the coarse grid, and $T$ are the number of conductivity values estimated (in this case, nodal values).
The former result is confirmed as we observe from Table \ref{recontable} that the RMSE for the blob-like reconstructions increases at a higher rate, with respect the noise, in estimates computed using the ``standard" electrode layout.
This observation supports the earlier statements claiming that the proposed electrode optimization approach should improve reconstruction quality, in addition to the improvements in, e.g., Hessian and resistivity matrix conditioning.
}

 {Reconstructions shown in Fig. \ref{reconz} comparing ellipsoidal reconstructions are far less visually distinguishable than reconstructions of blob-like distributions.
These visual observations are also confirmed by RMSE values reported in Table \ref{recontable}, where reconstructions computed using the optimized electrode layout have only slightly lower error than the reconstructions computed using the non-optimized electrode layout.
This realization results from the fact that the prior model used promotes smoothness and is a less optimal choice for reconstructing a sharp inclusion than reconstructing the blob-like distribution.
This is a useful observation as it demonstrates that, while the use of optimal electrode layouts can improve reconstruction quality, it cannot completely compensate for the use of a poor prior model.
}

\begin{table}[h!]
\caption{ {Root mean square errors (RMSEs) of EIT reconstructions of a blob-like distribution and an ellipsoidal inclusion using optimized and non-optimized (``standard") electrode layouts considering Gaussian noise $\eta$ with 1\%, 5\%, and 10\% standard deviation added. }}
\resizebox{\textwidth}{!}{
\begin{tabular}{@{}lcccccc@{}}
\toprule
& Blobby, $\eta = 1\%$ & Blobby, $\eta =5\%$ & Blobby, $\eta = 10\%$ & Ellipsoidal, $\eta = 10\%$ & Ellipsoidal, $\eta = 5\%$ & Ellipsoidal, $\eta = 10\%$ \\ \midrule
RMSE (\%), optimized electrode layout & 6.0 & 7.4 & 9.8 & 6.0 & 8.0 & 9.7 \\
RMSE (\%),``standard" electrode layout & 7.6 & 9.1 & 14.1 & 6.0 & 8.3 & 11.0 \\ \bottomrule

\end{tabular}}
\label{recontable}
\end{table}
 {
In the previous examples, the effectiveness of employing an optimal electrode layout was demonstrated in the reduction of reconstruction errors relative to a ``standard" electrode layout.
While this result implies that the use of the electrode optimization approach improves the information contained in voltage measurements, we would like to quantify this using an independent means since there is some intrinsic bias in comparing reconstructions (as noted in section \ref{qualmetric1}).
For this, we employ the distinguishability criteria \cite{smyl2019invisibility,Isaacson1986}, where the distinguishability between some background conductivity $\sigma_1$ and target conductivity $\sigma_2$ is given by
}

 {
\begin{equation}
\delta = ||V_2 - V_1||^2 
\label{dist1}
\end{equation}
}
\noindent  {where $V_1$ and $V_2$ are voltage measurements corresponding to $\sigma_1$ and $\sigma_2$, respectively.
Noting that $\sigma_1$ and $\sigma_2$ are separated by a change in conductivity $\Delta \sigma$ (such that $\sigma_2 = \sigma_1 + \Delta \sigma$) and since we are using simulated data, Eq. \ref{dist1} may be rewritten as follows by recalling Eq. \ref{obeqr}:
}
 {
\begin{equation}
\delta = ||U(\sigma_1 + \Delta \sigma) - U(\sigma_1)||^2 
\label{dist2}
\end{equation}
}
\noindent  {assuming a fixed measurement/stimulation pattern, that the same discretization is used for $U$, and $U$ is the same for both simulations}

\begin{figure}[h]
\centering
\includegraphics[width=10cm]{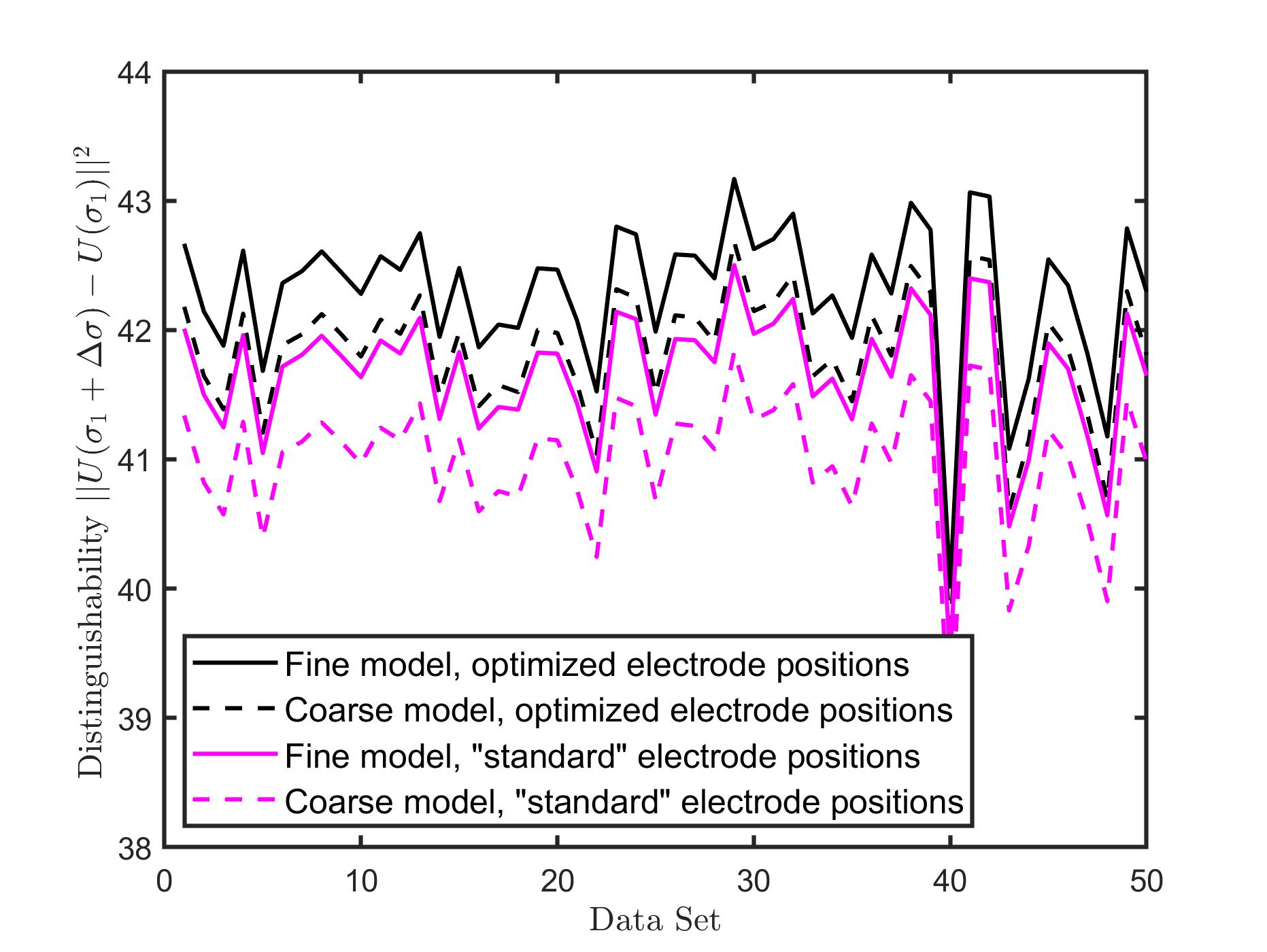}
\caption{ {Distingishability of simulated voltage measurements considering 50 random distributions of $\sigma_1$ and $\Delta \sigma$. Four discretizations are considered, including optimized and ``standard" electrode layouts with fine and coarse meshes.}}
\label{DISTINGUISH}
\end{figure}

 {Using the distinguishability model in Eq. \ref{dist2}, we can test the distinguishability of simulated voltage samples for the optimized and non-optimized electrode layouts.
We reinforce here that a computed distinguishability value $\delta$ is only valid if the simulated voltages $U(\sigma_1 + \Delta \sigma)$ and $U(\sigma_1)$ are computed using the same electrode layout and discretization, otherwise modeling error corruption may yield meaningless interpretations of $\delta$.
Given this, we compute $\delta$ using forward model simulations from (a) the fine discretization using the ``standard" electrode layout, (b) the coarse discretization using the ``standard" electrode layout, (c) the fine discretization using the optimized electrode layout, and (d) the coarse discretization using the optimized electrode layout.
For this, we generate 50 random blobby samples $\sigma_1$ ($1 < \sigma_1 <2 $) and 50 random blobby samples for $\Delta \sigma$ ($1 < \Delta \sigma <2 $).
These conductivity distributions were generated on a fine grid with a maximum node spacing of 0.01 m and interpolated onto the meshes in cases (a-d), so that the same conductivity distributions were used in all data simulations.
Using Eq. \ref{dist2}, the distinguishability between distributions $\sigma_1$ and $\sigma_2$ were computed and are shown in Fig. \ref{DISTINGUISH}.
}

 {
The results provided in Fig. \ref{DISTINGUISH} show a consistent improvement in distinguishability when employing the optimized electrode layouts.
An interesting finding here is that distinguishability of measurements using the optimized electrode layout with both the fine and coarse meshes is uniformly higher than the distinguishability of measurements using the ``standard" electrode layout with both fine and coarse meshes.
This indicates that the sensitivity of EIT measurements to changes in conductivity are higher when using the optimized electrode positions than when employing the ``standard" electrode layout irrespective of using fine or coarse meshes (at least within the discretization sizes tested herein).
}

\subsection{ {Future work}}
What remains to be discussed is the effects of over-fitting and regularization of the neural network, elements which are closely related.
We noted that a constant regularization term was used throughout this work, which was shown to be reasonably robust but certainly not optimal for every optimization problem.
The effects of this were not pronounced in the examples provided and were mainly manifested in small (seemingly random) perturbations in optimized electrode positions.
This was most obvious in Fig \ref{E1optE}(a), where the electrode spacings were not completely symmetric as expected.
Of course, the type and amount of training data also plays a role here; however, tailored changes in the regularization parameter for a given problem did fix this issue in preliminary trials, indicating that the choice of the regularization parameter had the most significant effect for the problems considered herein.

In future work, we look forward to improving the robustness of the deep learning approach to electrode position optimization.
 {One central challenge is the selection of the optimization criteria $\Theta = [\kappa,\beta,\dots]^T$.
Here, the use of $\kappa = 1$ and $\beta = 0$ proved reasonable despite the notable differences between $\Theta$ and training values for $\kappa$ and $\beta$.
Indeed, while training values for $\beta$ were $\mathcal{O}(10^{-1} - 10^2)$, training values for $\kappa$ ranged from $10^{19}-10^{26}$.
This may indicate that the trained network was actually predicting layouts that were not ``fully optimal" based on the distance between $\Theta$ and the training data.
Therefore, the selection of $\Theta$ used herein is probably not the best choice for all applications and additional work is required to improve $\Theta$, perhaps including information related to distinguishability, additional conditions numbers (e.g. $L_1$ condition numbers and resistivity matrix condition numbers), etc.}
 {Other} remaining challenges to address are centered around mitigating the effects of over-fitting and improving the robustness to a broader range of conductivity distributions (such as non-conductive cracks for non-destructive testing purposes).
For this, we look forward to integrating Bayesian regularized convolutional neural networks \cite{snoek2015scalable}, taking advantage of automatic regularization and the use of conductivity distributions in the training data.
In addition, we anticipate the inclusion of internal electrodes into the optimization approach to improve the sensitivity of EIT to inclusions far from the boundary.

 {We would like to conclude this section by commenting that the learning architecture used herein is quite simple (two hidden layers, full connectivity, etc.).
This choice was made to demonstrate that electrode position optimization is possible using basic networks -- thereby making the proposed approach accessible to those who may not be experts in deep learning.
We are presently extending implementation of the electrode position optimization approach to Keras/TensorFlow \cite{abadi2016tensorflow,chollet2015keras} in order to improve the approach's robustness by testing features such as dropout, different optimization regimes, different performance metrics, and regularization techniques.
In eventuality, we hope this will enable 3D electrode position optimization using deep learning.}

\section{Conclusions}
In this article, we proposed a straightforward deep learning based approach for optimizing electrode positions used in EIT sensing.
{In the broad sense, optimal electrode positions are dependent on both the domain geometry and conductivity.
Yet, while the geometry can be known with a high degree of certainty, the purpose of EIT is to reconstruct an \emph{unknown} conductivity distribution.
In the proposed electrode position optimization approach, information on the expected structure of the target conductivity is incorporated into the data used to train the neural network used in predicting optimal electrode positions.
The inclusion of such structural data can be viewed as prior information used to improve the quality of information contained in electrode data.}

To evaluate the quality of the optimized electrode configurations, simple statistical metric{s were} developed and applied to compare optimized configurations with ``standard" uniformly distributed electrode layouts.
The effectiveness of the optimization approach was demonstrated in a suite of examples, where reductions in cumulative modeling errors  {(resulting from discretization)} up to a factor of 30 were observed when employing optimized electrode positions relative to the ``standard" configurations.
 {In addition to this, reductions in the ill-conditioning of the resistivity and Hessian matrices of approximately (4\% - 9\%) and (30\% - 73\%), respectively, were noted.}
 {Further, it is found that the use of optimized electrode positions computed using the approach derived herein can reduce errors in EIT reconstructions as well as improve the distinguishability of EIT measurements.}
In the analysis, it was found that the following items notably influenced the solutions from of the optimization approach (in no particular order):

\begin{itemize}
\item The aspect ratio of the target geometry (if rectangular).
\item The regularization parameter used in training the neural network.
\item Localized domain segmentation.
\item Near-boundary effects (such as holes).
\item The condition number of the finite element  {resistivity} matrix.
\end{itemize}

\noindent Moreover, in future work, we aim to address robustness issues related to over-fitting and regularization of the neural network.
To do this, we aim to include a Bayesian regularized neural network in order to take advantage of automatic regularization and the inclusion of conductivity distributions directly in the training data.
These frameworks will be extended to include internal electrodes to improve the sensitivity of EIT to inclusions far from the boundary.

\section*{Acknowledgments}
DS would like to acknowledge the support of the Department of Civil and Structural Engineering at the University of Sheffield. 
DL was supported by the National Natural Science Foundation of China under Grant No. 61871356.

\bibliographystyle{IEEEtran}
\bibliography{bibliography}

\begin{thebibliography}{10}
\providecommand{\url}[1]{#1}
\csname url@samestyle\endcsname
\providecommand{\newblock}{\relax}
\providecommand{\bibinfo}[2]{#2}
\providecommand{\BIBentrySTDinterwordspacing}{\spaceskip=0pt\relax}
\providecommand{\BIBentryALTinterwordstretchfactor}{4}
\providecommand{\BIBentryALTinterwordspacing}{\spaceskip=\fontdimen2\font plus
\BIBentryALTinterwordstretchfactor\fontdimen3\font minus
  \fontdimen4\font\relax}
\providecommand{\BIBforeignlanguage}[2]{{%
\expandafter\ifx\csname l@#1\endcsname\relax
\typeout{** WARNING: IEEEtran.bst: No hyphenation pattern has been}%
\typeout{** loaded for the language `#1'. Using the pattern for}%
\typeout{** the default language instead.}%
\else
\language=\csname l@#1\endcsname
\fi
#2}}
\providecommand{\BIBdecl}{\relax}
\BIBdecl

\bibitem{hassan2019failure}
H.~Hassan and T.~N. Tallman, ``Failure prediction in self-sensing
  nanocomposites via genetic algorithm-enabled piezoresistive inversion,''
  \emph{Structural Health Monitoring}, p. (in press), 2019.

\bibitem{tallman2016}
T.~N. Tallman and K.~Wang, ``Damage and strain identification in
  multifunctional materials via electrical impedance tomography with
  constrained sine wave solutions,'' \emph{Structural Health Monitoring},
  vol.~15, no.~2, pp. 235--244, 2016.

\bibitem{tallman2017inverse}
T.~Tallman, S.~Gungor, G.~Koo, and C.~Bakis, ``On the inverse determination of
  displacements, strains, and stresses in a carbon nanofiber/polyurethane
  nanocomposite from conductivity data obtained via electrical impedance
  tomography,'' \emph{Journal of Intelligent Material Systems and Structures},
  pp. 1--13, 2017.

\bibitem{loh2009}
K.~J. Loh, T.-C. Hou, J.~P. Lynch, and N.~A. Kotov, ``Carbon nanotube sensing
  skins for spatial strain and impact damage identification,'' \emph{Journal of
  Nondestructive Evaluation}, vol.~28, no.~1, pp. 9--25, 2009.

\bibitem{Smyl2019SHM}
D.~Smyl, S.~Bossuyt, W.~Ahmad, A.~Vavilov, and D.~Liu, ``An overview of 38
  least squares--based frameworks for structural damage tomography,''
  \emph{Structural Health Monitoring}, p. 1475921719841012, 2019.

\bibitem{smyl2018detection}
D.~Smyl, M.~Pour-Ghaz, and A.~Sepp{\"a}nen, ``Detection and reconstruction of
  complex structural cracking patterns with electrical imaging,'' \emph{NDT \&
  E International}, vol.~99, pp. 123--133, 2018.

\bibitem{seppanen2009state}
A.~Sepp{\"a}nen, A.~Voutilainen, and J.~Kaipio, ``State estimation in process
  tomography—reconstruction of velocity fields using eit,'' \emph{Inverse
  Problems}, vol.~25, no.~8, p. 085009, 2009.

\bibitem{liuMMC2019}
D.~Liu and J.~Du, ``A moving morphable components based shape reconstruction
  framework for electrical impedance tomography,'' \emph{IEEE transactions on
  medical imaging}, vol.~38, no.~12, pp. 2937--2948, 2019.

\bibitem{mellenthin2018ace1}
\BIBentryALTinterwordspacing
M.~M. Mellenthin, J.~L. Mueller, E.~D. L.~B. de~Camargo, F.~S. de~Moura,
  T.~B.~R. Santos, R.~G. Lima, S.~J. Hamilton, P.~A. Muller, and M.~Alsaker,
  ``The ace1 electrical impedance tomography system for thoracic imaging,''
  \emph{IEEE Transactions on Instrumentation and Measurement}, 2018. [Online].
  Available: \url{DOI:10.1109/TIM.2018.2874127}
\BIBentrySTDinterwordspacing

\bibitem{liu2019nonstationary}
\BIBentryALTinterwordspacing
D.~Liu, D.~Smyl, and J.~Du, ``Nonstationary shape estimation in electrical
  impedance tomography using a parametric level set-based extended kalman
  filter approach,'' \emph{IEEE Transactions on Instrumentation and
  Measurement}, 2019. [Online]. Available: \url{DOI:10.1109/TIM.2019.2921441}
\BIBentrySTDinterwordspacing

\bibitem{adler2006uses}
A.~Adler and W.~R. Lionheart, ``Uses and abuses of eidors: an extensible
  software base for eit,'' \emph{Physiological measurement}, vol.~27, no.~5, p.
  S25, 2006.

\bibitem{yao2017application}
J.~Yao and M.~Takei, ``Application of process tomography to multiphase flow
  measurement in industrial and biomedical fields: A review,'' \emph{IEEE
  Sensors Journal}, vol.~17, no.~24, pp. 8196--8205, 2017.

\bibitem{wei2016super}
K.~Wei, C.-H. Qiu, and K.~Primrose, ``Super-sensing technology: Industrial
  applications and future challenges of electrical tomography,''
  \emph{Philosophical Transactions of the Royal Society A: Mathematical,
  Physical and Engineering Sciences}, vol. 374, no. 2070, p. 20150328, 2016.

\bibitem{holder2004electrical}
D.~Holder, \emph{Electrical impedance tomography: methods, history and
  applications}.\hskip 1em plus 0.5em minus 0.4em\relax CRC Press, 2004.

\bibitem{tan2019image}
C.~Tan, S.~Lv, F.~Dong, and M.~Takei, ``Image reconstruction based on
  convolutional neural network for electrical resistance tomography,''
  \emph{IEEE Sensors Journal}, vol.~19, no.~1, pp. 196--204, 2019.

\bibitem{wei2019dominant}
Z.~Wei, D.~Liu, and X.~Chen, ``Dominant-current deep learning scheme for
  electrical impedance tomography,'' \emph{IEEE Transactions on Biomedical
  Engineering}, 2019.

\bibitem{hamilton2018deep}
S.~J. Hamilton and A.~Hauptmann, ``Deep d-bar: Real-time electrical impedance
  tomography imaging with deep neural networks,'' \emph{IEEE transactions on
  medical imaging}, vol.~37, no.~10, pp. 2367--2377, 2018.

\bibitem{mueller2012}
J.~L. Mueller and S.~Siltanen, \emph{Linear and nonlinear inverse problems with
  practical applications}.\hskip 1em plus 0.5em minus 0.4em\relax SIAM, 2012.

\bibitem{kaipio2004posterior}
J.~Kaipio, A.~Sepp{\"a}nen, E.~Somersalo, and H.~Haario, ``Posterior covariance
  related optimal current patterns in electrical impedance tomography,''
  \emph{Inverse Problems}, vol.~20, no.~3, p. 919, 2004.

\bibitem{lionheart2001generalized}
W.~R. Lionheart, J.~Kaipio, and C.~N. McLeod, ``Generalized optimal current
  patterns and electrical safety in eit,'' \emph{Physiological measurement},
  vol.~22, no.~1, p.~85, 2001.

\bibitem{hyvonen2014optimizing}
N.~Hyvonen, A.~Seppanen, and S.~Staboulis, ``Optimizing electrode positions in
  electrical impedance tomography,'' \emph{SIAM Journal on Applied
  Mathematics}, vol.~74, no.~6, pp. 1831--1851, 2014.

\bibitem{horesh2010optimal}
L.~Horesh, E.~Haber, and L.~Tenorio, ``Optimal experimental design for the
  large-scale nonlinear ill-posed problem of impedance imaging,''
  \emph{Large-Scale Inverse Problems and Quantification of Uncertainty}, pp.
  273--290, 2010.

\bibitem{FAN2019109119}
Y.~Fan and L.~Ying, ``Solving electrical impedance tomography with deep
  learning,'' \emph{Journal of Computational Physics}, p. 109119, 2019.

\bibitem{hamilton2019beltrami}
S.~J. Hamilton, A.~H{\"a}nninen, A.~Hauptmann, and V.~Kolehmainen,
  ``Beltrami-net: domain independent deep d-bar learning for absolute imaging
  with electrical impedance tomography (a-eit),'' \emph{Physiological
  measurement}, 2019.

\bibitem{tan2018image}
C.~Tan, S.~Lv, F.~Dong, and M.~Takei, ``Image reconstruction based on
  convolutional neural network for electrical resistance tomography,''
  \emph{IEEE Sensors Journal}, vol.~19, no.~1, pp. 196--204, 2018.

\bibitem{sLiu2019}
S.~{Liu}, H.~{Wu}, Y.~{Huang}, Y.~{Yang}, and J.~{Jia}, ``Accelerated
  structure-aware sparse bayesian learning for three-dimensional electrical
  impedance tomography,'' \emph{IEEE Transactions on Industrial Informatics},
  vol.~15, no.~9, pp. 5033--5041, 2019.

\bibitem{rymarczyk2018non}
T.~Rymarczyk, G.~K{\l}osowski, and E.~Koz{\l}owski, ``A non-destructive system
  based on electrical tomography and machine learning to analyze the moisture
  of buildings,'' \emph{Sensors}, vol.~18, no.~7, p. 2285, 2018.

\bibitem{borgerding2016onsager}
M.~Borgerding and P.~Schniter, ``Onsager-corrected deep learning for sparse
  linear inverse problems,'' in \emph{2016 IEEE Global Conference on Signal and
  Information Processing (GlobalSIP)}.\hskip 1em plus 0.5em minus 0.4em\relax
  IEEE, 2016, pp. 227--231.

\bibitem{antholzer2019deep}
S.~Antholzer, M.~Haltmeier, and J.~Schwab, ``Deep learning for photoacoustic
  tomography from sparse data,'' \emph{Inverse problems in science and
  engineering}, vol.~27, no.~7, pp. 987--1005, 2019.

\bibitem{han2019k}
Y.~Han, L.~Sunwoo, and J.~C. Ye, ``k-space deep learning for accelerated mri,''
  \emph{IEEE transactions on medical imaging}, 2019.

\bibitem{mccann2017convolutional}
M.~T. McCann, K.~H. Jin, and M.~Unser, ``Convolutional neural networks for
  inverse problems in imaging: A review,'' \emph{IEEE Signal Processing
  Magazine}, vol.~34, no.~6, pp. 85--95, 2017.

\bibitem{vauhkonen99}
P.~Vauhkonen, M.~Vauhkonen, T.~Savolainen, and J.~Kaipio, ``{Three-dimensional
  electrical impedance tomography based on the complete electrode model},''
  \emph{IEEE T. Biomedical Eng.}, vol.~46, no.~9, pp. 1150--1160, 1999.

\bibitem{borcea2002electrical}
L.~Borcea, ``Electrical impedance tomography,'' \emph{Inverse problems},
  vol.~18, no.~6, p. R99, 2002.

\bibitem{Arridge2019}
S.~Arridge and A.~Hauptmann, ``Networks for nonlinear diffusion problems in
  imaging,'' \emph{arXiv preprint arXiv:1811.12084}, 2018.

\bibitem{huang2003learning}
G.-B. Huang, ``Learning capability and storage capacity of two-hidden-layer
  feedforward networks,'' \emph{IEEE Transactions on Neural Networks}, vol.~14,
  no.~2, pp. 274--281, 2003.

\bibitem{nawi2007}
N.~M. Nawi, R.~Ransing, and M.~Ransing, ``An improved conjugate gradient based
  learning algorithm for back propagation neural networks,''
  \emph{International Journal of Computational Intelligence}, vol.~4, no.~1,
  pp. 46--55, 2007.

\bibitem{nissinen2007bayesian}
A.~Nissinen, L.~Heikkinen, and J.~Kaipio, ``The bayesian approximation error
  approach for electrical impedance tomography—experimental results,''
  \emph{Measurement Science and Technology}, vol.~19, no.~1, p. 015501, 2007.

\bibitem{smyl2019less}
D.~Smyl and D.~Liu, ``Less is often more: Applied inverse problems using
  hp-forward models,'' \emph{Journal of Computational Physics}, p. 108949,
  2019.

\bibitem{liu2015nonlinear}
D.~Liu, V.~Kolehmainen, S.~Siltanen, and A.~Sepp{\"a}nen, ``A nonlinear
  approach to difference imaging in eit; assessment of the robustness in the
  presence of modelling errors,'' \emph{Inverse Problems}, vol.~31, no.~3, p.
  035012, 2015.

\bibitem{smyl2019invisibility}
D.~Smyl and D.~Liu, ``Invisibility and indistinguishability in structural
  damage tomography,'' \emph{Measurement Science and Technology}, vol.~31,
  no.~2, p. 024001, 2019.

\bibitem{Isaacson1986}
D.~Isaacson, ``Distinguishability of conductivities by electric current
  computed tomography,'' \emph{IEEE transactions on medical imaging}, vol.~5,
  no.~2, pp. 91--95, 1986.

\bibitem{snoek2015scalable}
J.~Snoek, O.~Rippel, K.~Swersky, R.~Kiros, N.~Satish, N.~Sundaram, M.~Patwary,
  M.~Prabhat, and R.~Adams, ``Scalable bayesian optimization using deep neural
  networks,'' in \emph{International conference on machine learning}, 2015, pp.
  2171--2180.

\bibitem{abadi2016tensorflow}
M.~Abadi, P.~Barham, J.~Chen, Z.~Chen, A.~Davis, J.~Dean, M.~Devin,
  S.~Ghemawat, G.~Irving, M.~Isard \emph{et~al.}, ``Tensorflow: A system for
  large-scale machine learning,'' in \emph{12th $\{$USENIX$\}$ Symposium on
  Operating Systems Design and Implementation ($\{$OSDI$\}$ 16)}, 2016, pp.
  265--283.

\bibitem{chollet2015keras}
F.~Chollet \emph{et~al.}, ``Keras,'' \url{https://github.com/fchollet/keras},
  2015.

\end{thebibliography}

\end{document}